\documentclass{aastex63}   

\usepackage{array}      
\usepackage{multirow}   
\usepackage{xspace}		
\usepackage{comment}    

\newcommand{\ie}{{i.e.}\xspace}

\submitjournal{Astronomical Journal May 2023}

\shorttitle{DESI Optical Corrector}
\shortauthors{Miller et al.}

\begin{document}
 
\title{The Optical Corrector for the Dark Energy Spectroscopic Instrument}


\correspondingauthor{Timothy N.\ Miller}
\email{tmiller@ssl.berkeley.edu}

\author[0009-0005-2445-7544]{Timothy N.\ Miller}
\affiliation{University of California, Space Sciences Laboratory, Berkeley, CA 94720, USA}

\author{Peter Doel}
\affiliation{Department of Physics \& Astronomy, University College London, Gower Street, London, WC1E 6BT, UK}

\author{Gaston Gutierrez}
\affiliation{Fermi National Accelerator Laboratory, P.\ O.\ Box 500, Batavia, IL 60510, USA}

\author{Robert Besuner}
\affiliation{University of California, Space Sciences Laboratory, Berkeley, CA 94720, USA}

\author{David Brooks}
\affiliation{Department of Physics \& Astronomy, University College London, Gower Street, London, WC1E 6BT, UK}

\author{Giuseppe Gallo}
\affiliation{Fermi National Accelerator Laboratory, P.\ O.\ Box 500, Batavia, IL 60510, USA}

\author{Henry Heetderks}
\affiliation{University of California, Space Sciences Laboratory, Berkeley, CA 94720, USA}

\author{Patrick Jelinsky}
\affiliation{University of California, Space Sciences Laboratory, Berkeley, CA 94720, USA}

\author[0000-0003-4207-7420]{Stephen M.\ Kent}
\affiliation{Fermi National Accelerator Laboratory, P.\ O.\ Box 500, Batavia, IL 60510, USA}

\author{Michael Lampton}
\affiliation{University of California, Space Sciences Laboratory, Berkeley, CA 94720, USA}

\author[0000-0003-1887-1018]{Michael E.\ Levi}
\affiliation{Lawrence Berkeley National Laboratory, One Cyclotron Road, Berkeley, CA 94720, USA}

\author{Ming Liang}
\affiliation{National Optical Astronomy Observatory, Tucson, AZ 85719, USA}

\author[0000-0002-1125-7384]{Aaron Meisner}
\affiliation{National Optical Astronomy Observatory, Tucson, AZ 85719, USA}

\author{Michael~J.~Sholl}
\affiliation{Space Exploration Technologies Corp., 1 Rocket Rd, Hawthorne, CA 90250, USA}

\author{Joseph Harry Silber}
\affiliation{Lawrence Berkeley National Laboratory, One Cyclotron Road, Berkeley, CA 94720, USA}

\author{David Sprayberry}
\affiliation{NSF's National Optical-Infrared Astronomy Research Laboratory, 950 N. Cherry Avenue, Tucson, AZ 85719, USA}

\author{Jessica Nicole Aguilar}
\affiliation{Lawrence Berkeley National Laboratory, One Cyclotron Road, Berkeley, CA 94720, USA}

\author{Axel de la Macorra}
\affiliation{Instituto de F\'{\i}sica, Universidad Nacional Aut\'{o}noma de M\'{e}xico,  Cd. de M\'{e}xico  C.P. 04510,  M\'{e}xico}

\author{Daniel Eisenstein}
\affiliation{Center for Astrophysics $|$ Harvard \& Smithsonian, 60 Garden Street, Cambridge, MA 02138, USA}

\author[0000-0003-2371-3356]{Kevin Fanning}
\affiliation{The Ohio State University, Columbus, 43210 OH, USA}

\author[0000-0002-3033-7312]{Andreu Font-Ribera}
\affiliation{Institut de F\'{i}sica d'Altes Energies (IFAE), The Barcelona Institute of Science and Technology, Campus UAB, 08193 Bellaterra Barcelona, Spain}

\author{Enrique Gaztañaga}
\affiliation{Institute of Space Sciences, ICE-CSIC, Campus UAB, Carrer de Can Magrans s/n, 08913 Bellaterra, Barcelona, Spain}

\author[0000-0003-3142-233X]{Satya Gontcho A Gontcho}
\affiliation{Lawrence Berkeley National Laboratory, One Cyclotron Road, Berkeley, CA 94720, USA}

\author{Klaus Honscheid}
\affiliation{The Ohio State University, Columbus, 43210 OH, USA}

\author{Jorge Jimenez}
\affiliation{Institut de F\'{i}sica d'Altes Energies (IFAE), The Barcelona Institute of Science and Technology, Campus UAB, 08193 Bellaterra Barcelona, Spain}

\author[0000-0003-0201-5241]{Dick Joyce}
\affiliation{NSF's National Optical-Infrared Astronomy Research Laboratory, 950 N. Cherry Avenue, Tucson, AZ 85719, USA}

\author{Robert Kehoe}
\affiliation{Department of Physics, Southern Methodist University, 3215 Daniel Avenue, Dallas, TX 75275, USA}

\author[0000-0003-3510-7134]{Theodore Kisner}
\affiliation{Lawrence Berkeley National Laboratory, One Cyclotron Road, Berkeley, CA 94720, USA}

\author[0000-0001-6356-7424]{Anthony Kremin}
\affiliation{Lawrence Berkeley National Laboratory, One Cyclotron Road, Berkeley, CA 94720, USA}

\author[0000-0003-1838-8528]{Martin Landriau}
\affiliation{Lawrence Berkeley National Laboratory, One Cyclotron Road, Berkeley, CA 94720, USA}

\author[0000-0001-7178-8868]{Laurent Le Guillou}
\affiliation{Sorbonne Universit\'{e}, CNRS/IN2P3, Laboratoire de Physique Nucl\'{e}aire et de Hautes Energies (LPNHE), FR-75005 Paris, France}

\author{Christophe Magneville}
\affiliation{IRFU, CEA, Universit\'{e} Paris-Saclay, F-91191 Gif-sur-Yvette, France}

\author[0000-0002-4279-4182]{Paul Martini}
\affiliation{The Ohio State University, Columbus, 43210 OH, USA}

\author{Ramon Miquel}
\affiliation{Institut de F\'{i}sica d'Altes Energies (IFAE), The Barcelona Institute of Science and Technology, Campus UAB, 08193 Bellaterra Barcelona, Spain}

\author[0000-0002-2733-4559]{John Moustakas}
\affiliation{Department of Physics and Astronomy, Siena College, 515 Loudon Road, Loudonville, NY 12211, USA}

\author[0000-0001-6590-8122]{Jundan Nie}
\affiliation{National Astronomical Observatories, Chinese Academy of Sciences, A20 Datun Rd., Chaoyang District, Beijing, 100012, P.R. China}

\author[0000-0002-0644-5727]{Will Percival}
\affiliation{Department of Physics and Astronomy, University of Waterloo, 200 University Ave W, Waterloo, ON N2L 3G1, Canada}

\author{Claire Poppett}
\affiliation{Lawrence Berkeley National Laboratory, One Cyclotron Road, Berkeley, CA 94720, USA}

\author[0000-0001-7145-8674]{Francisco Prada}
\affiliation{Instituto de Astrof\'{i}sica de Andaluc\'{i}a (CSIC), Glorieta de la Astronom\'{i}a, s/n, E-18008 Granada, Spain}

\author{Graziano Rossi}
\affiliation{Department of Physics and Astronomy, Sejong University, Seoul, 143-747, Korea}

\author{David Schlegel}
\affiliation{Lawrence Berkeley National Laboratory, One Cyclotron Road, Berkeley, CA 94720, USA}

\author{Michael Schubnell}
\affiliation{University of Michigan, Ann Arbor, MI 48109, USA}

\author[0000-0002-6588-3508]{Hee-Jong Seo}
\affiliation{Department of Physics \& Astronomy, Ohio University, Athens, OH 45701, USA}

\author{Ray Sharples}
\affiliation{Centre for Advanced Instrumentation, Department of Physics, Durham University, South Road, Durham DH1 3LE, UK}

\author[0000-0003-1704-0781]{Gregory Tarl\'{e}}
\affiliation{University of Michigan, Ann Arbor, MI 48109, USA}

\author{Mariana Vargas-Maga\~na}
\affiliation{Instituto de F\'{\i}sica, Universidad Nacional Aut\'{o}noma de M\'{e}xico,  Cd. de M\'{e}xico  C.P. 04510,  M\'{e}xico}

\author[0000-0002-4135-0977]{Zhimin Zhou}
\affiliation{National Astronomical Observatories, Chinese Academy of Sciences, A20 Datun Rd., Chaoyang District, Beijing, 100012, P.R. China}

\author{the DESI Collaboration}

\begin{abstract}
The Dark Energy Spectroscopic Instrument (DESI) is currently measuring the spectra of 40\,million galaxies and quasars, the largest such survey ever made to probe the nature of cosmological dark energy.
The 4-meter Mayall telescope at Kitt Peak National Observatory has been adapted for DESI, including the construction of a $3.2^\circ$ diameter prime focus corrector that focuses astronomical light onto a 0.8-meter diameter focal surface with excellent image quality over the DESI bandpass of 360-980nm.  The wide-field corrector includes six lenses, as large as 1.1-meters in diameter and as heavy as 237\,kilograms, including two counter-rotating wedged lenses that correct for atmospheric dispersion over Zenith angles from 0 to $60^\circ$.  The lenses, cells, and barrel assembly all meet precise alignment tolerances on the order of tens of microns.  The barrel alignment is maintained throughout a range of observing angles and temperature excursions in the Mayall dome by use of a hexapod, which is itself supported by a new cage, ring, and truss structure. 
In this paper we describe the design, fabrication, and performance of the new corrector and associated structure, focusing on how they meet DESI requirements.  In particular we describe the prescription and specifications of the lenses, design choices and error budgeting of the barrel assembly, stray light mitigations, and integration and test at the Mayall telescope.  We conclude with some validation highlights that demonstrate the successful corrector on-sky performance, and list some lessons learned during the multi-year fabrication phase.
\end{abstract}


\section{Introduction} \label{sec:intro}

The Dark Energy Spectroscopic Instrument (DESI) has begun its spectroscopic survey of 40\,million astronomical objects in order to answer cosmological questions about the nature of dark energy in the universe, with the goal of performing the most precise measurement of the expansion history of the universe ever obtained \citep{levi13, DESIcollab2016a}.  The survey will observe four classes of galactic targets: Bright Galaxy Survey targets \citep[BGS,][]{Hahn2022}, luminous red galaxies \citep[LRG,][]{Zhou2023}, emission
line galaxies \citep[ELG,][]{Raichoor2023}, and quasi-stellar objects \citep[QSO,][]{Chaussidon2023}.  These observations are accompanied by a Milky Way survey program \citep[MWS,][]{Cooper2022}.  The specific target choices are based on legacy imaging surveys \citep[][; Schlegel et al, in preparation]{Zou2017, Dey2019}.  With the galactic observations, DESI will calculate the Baryon Acoustic Oscillation (BAO) scale to determine the varying distance-vs-redshift relationship over the lifetime of the universe, from recent times to approximately redshift 3.5.  In addition to the expansion history and dark energy, DESI will also measure the growth of cosmic structure, provide new information on the sum of the neutrino masses, study the scale dependence of primordial density fluctuations from inflation, and test potential modifications to the general theory of relativity \citep{DESIcollab2016a}.  

DESI began its installation in 2018 at the Kitt Peak National Observatory on the 4-meter Mayall telescope near Tucson, Arizona \citep{DESIcollab2016b}, and started its main survey in May 2021 after a survey validation phase (DESI Collaboration et al, in preparation).  The first data release is expected in 2023 (DESI Collaboration et al, in preparation).  DESI aims to obtain approximately an order of magnitude increase in sample size compared to previous observations, 40\,million targets, and yet complete its ambitious survey in only five years, a fraction of previous survey durations.  For comparison, the closest similar survey, the Sloan Digital Sky Survey (SDSS), measured approximately four million spectra over nearly 20 years \citep{ahumada20}.  To achieve this, DESI takes advantage of the large collecting area of the Mayall 4-meter telescope in order to obtain images in only 1000\,seconds (about 17 minutes).  DESI is also capable of measuring 5020 spectra simultaneously, a significant multiplexing improvement over other surveys.  Furthermore, DESI's robotic positioners and software allow changing to the next set of fields to be done within a few minutes, allowing measurements to be taken on a roughly 20-minute cadence.  The SDSS used a 2.5\,meter telescope and measured 1000 spectra simultaneously with a cadence between measurements of over an hour.  DESI also created customized pipelines and software, including an extensive spectroscopic reduction \citep{Guy2023}, redshift classifications using Redrock (Bailey et al, in preparation), efficient fiber assignments (Raichoor et al, in preparation), and optimized observation tilings (Schlafly et al, in preparation).

\begin{figure}
\centering 
\includegraphics[width=.8\textwidth]{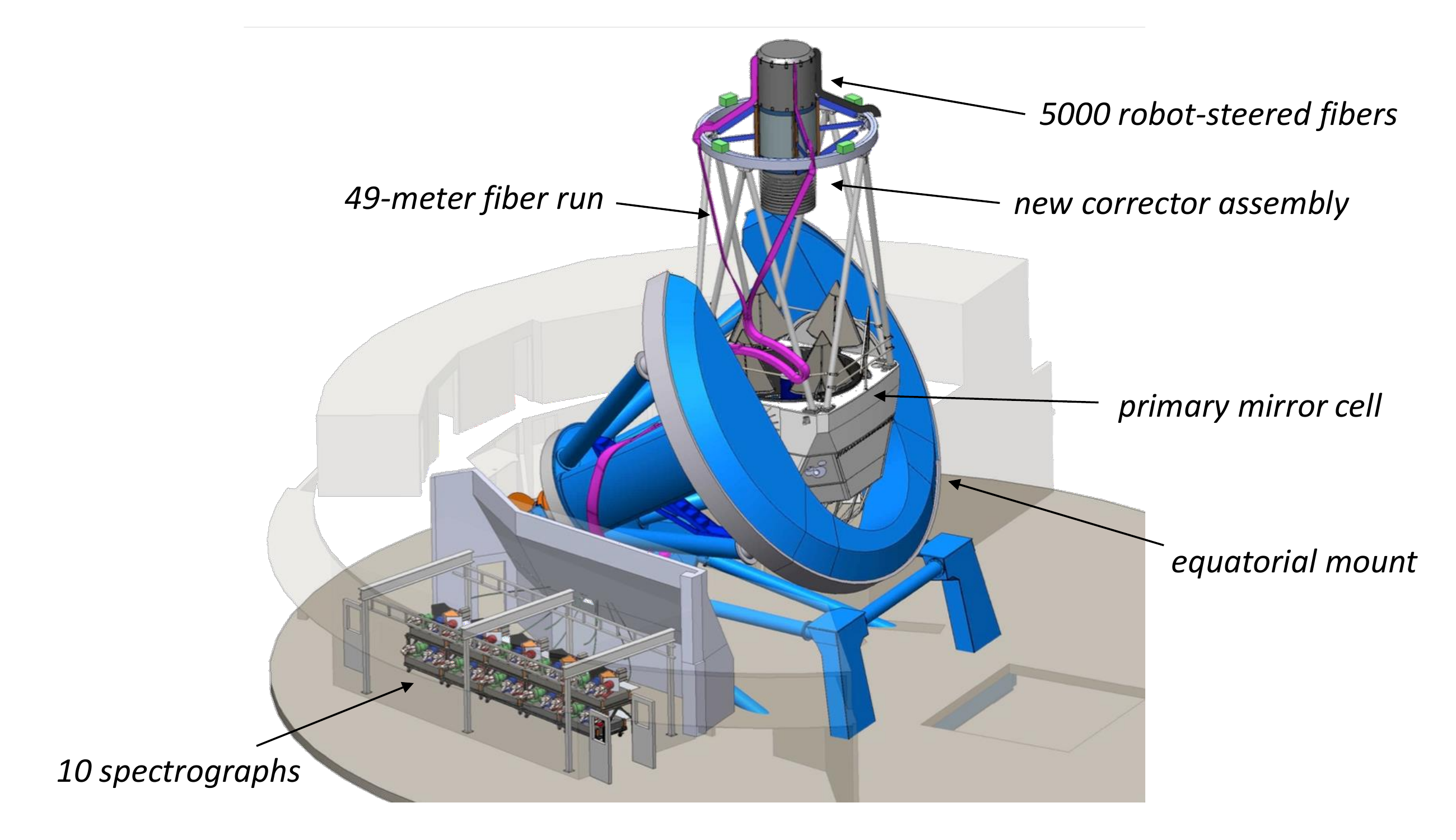}
\caption{Layout of the Mayall telescope and the DESI instrument. \citep[Image previously published in][]{miller18}}
\label{fig:DESIoverall} 
\end{figure}

Figure~\ref{fig:DESIoverall} shows a model of the Mayall telescope with the new instrumentation added for DESI.  First, there is a new wide-field corrector that achieves excellent imaging of the sky to the focal surface over a 3.2-degree diameter field of view.  A complex focal plane assembly is installed after the corrector which contains 5020 robotic fiber positioners, each of which can move accurately to capture focused light from individual targets in a single wide scene.  The positioners feed light into a 49-meter, high-efficiency fiber run that carries light from the telescope to a separate room, where an array of ten spectrographs performs spectroscopy on the astronomical light.  The telescope also contains a fiber view camera which views the focal plane assembly through the corrector lenses and provides positional feedback that allows the fiber positioners to be placed reliably within a handful of microns, thus minimizing light lost due to positioning errors.


The Mayall telescope was commissioned in 1973 and is the largest of the dozens of optical telescopes at Kitt Peak National Observatory (KPNO).  The heart of the Mayall is the 4-meter primary mirror, a concave hyperbolic surface coated with aluminum for high broadband reflectivity.  The primary mirror by itself does not produce good image quality at its focus over an extended field of view, and thus it requires a corrector, a group of lenses between the primary mirror and focal surface that corrects the wavefront error.  Over the years there have been several corrector assemblies installed.  The most recent corrector before DESI was the “Jacoby” or “Mosaic” corrector installed in 1998, a four-lens design which had a 51-arcmin diameter field of view \citep{jacoby98}.  To achieve the 3.2-degree diameter field of view needed for DESI, a new optical corrector was required.  Meeting those needs is not trivial: the prime-focus corrector must correct wavefront errors over a large field and a wide bandpass; its obscuration of the primary aperture must be minimized; the glasses used for the lenses must be available in very large sizes and high purity grades; and the design must minimize the number and complexity of the lenses.  To achieve its design goals, DESI took advantage of lessons learned from the Dark Energy Camera \citep[DECam,][]{flaugher15}, a similar prime focus corrector with five lenses which was built by a team that included future DESI members.  The resulting design for the DESI corrector and associated hardware met all of the requirements and achieves excellent performance, as demonstrated during the DESI commissioning phase.


This paper presents a full description of the DESI corrector and its support system.
Section 2 describes the optical design of the corrector, and the fabrication of the lenses.
Section 3 describes the mechanical design and fabrication of the mechanical elements of the corrector, including the lens cells, the barrel, the hexapod, the surrounding cage, and the “top end” consisting of the outer ring and connecting vanes.
Section 4 describes stray light in the corrector, specifically “best practices” that were made part of the design from the start, and high-level analysis that was performed to identify possible scatter issues and mitigate them.
Section 5 briefly describes the collection of all the corrector pieces at the Mayall, their integration into the telescope, and assembly-level testing.
Section 6 describes the successful on-sky performance of the integrated corrector.
Section 7 lists some lessons learned, gained over the course of the corrector design and build, and Section 8 provides a brief summary.
A series of companion papers describes the wider DESI instrumentation in greater depth: an instrument overview \citep{DESIcollab2022_overview}, the Focal Plane System \citep{silber2022}, the Fiber System (Poppett et al. in preparation), and the Spectrograph System (Jelinsky et al. in preparation).

\section{Corrector Optical Design and Fabrication}
\label{sec:corrdesign}

The corrector optics are the heart of the Mayall telescope, transporting incident light from astronomical targets into the DESI fiber system.  This section explains the requirements flowed to the optical design and describes the prescription and performance of the optical system that meets those requirements, including the atmospheric dispersion compensation scheme and the initial ghost analysis.  We also describe the fabrication process and verification testing of the lens blanks, polishing, and coating by their vendors.  Finally we address the maintenance of the primary mirror reflectivity.

\subsection{Requirements}

The needs of DESI place challenging requirements on the optical design of the corrector.  
The DESI instrument must be designed for high throughput in order to meet a top-level signal-to-noise (SNR) survey requirement \citep{DESIcollab2022_overview}, and most corrector requirements flow from that throughput need.  This includes requirements on image quality, atmospheric dispersion correction, and fiber coupling.

The DESI front end collects the light from each astronomical target, and focuses it into the 107-micron core of one of the fibers in the focal plane.  Therefore, the requirements on image quality are derived from throughput considerations: blur (optical PSF width) must be minimized in order to reduce light loss from rays missing the fiber core.  This applies across the entire field of view.  The corrector also must be highly achromatic over the DESI bandpass as well so that target light of different wavelengths focuses onto the fiber core with little residual blur.  Furthermore the corrector must meet its image quality requirements in the presence of atmospheric dispersion as a function of observing Zenith angle that would otherwise cause throughput loss at the fiber (see section \ref{sec:adc}). 

In addition, the fiber has a specific acceptance angle outside of which light does not transmit through the fiber and is scattered.  Therefore the corrector F/\# must match the fiber angle to maximize coupling throughput.  Furthermore the incident chief ray angle must be well-aligned to the fiber core normal angle, to keep the entire F/\# within the fiber acceptance angle.

Table~\ref{tab:corT2} lists the main optical requirements of the corrector. The rightmost column in the Table references what drives each requirement.  The design must take manufacturability into account: all of the subassemblies and components must be within the reasonable capabilities of their vendors.  For this reason, Table~\ref{tab:corT2} includes a requirement that limits the slope of the aspheric departure on lens surfaces to less than 30\,milliradians.  This value is based on vendor recommendations, and allows interferometric testing of the surfaces to be done with better than Nyquist fringe sampling.

The requirements in Table~\ref{tab:corT2} do acknowledge the challenges of meeting all needs, and therefore do allow some relaxation.  For example, the maximum zenith angle is limited, and a larger blur is accepted there, because of the difficulty in correcting atmospheric dispersion over a broad range.  Similarly, glass transmission is relaxed for the bluer wavelengths, given the natural lower transmission of DESI’s chosen glasses there.

\begin{deluxetable*}{lll}   
\tabletypesize{\footnotesize}
\tablecolumns{3}
\tablecaption{Prime focus corrector requirements \label{tab:corT2} }
\tablehead{
  \colhead{Requirement} &
  \colhead{Value} &
  \colhead{Note} }
\startdata
Optical design & Compatible with Mayall 4-m primary mirror & Mayall telescope selected for DESI. \\ [3pt]
Wavelength band & 360--980~nm &  Required for z range of QSO, LRG, and ELG targets \\ [3pt]
Zenith angle & 0 to $60^\circ$ &  Required for sky coverage \\ [3pt]
Design residual blur & Zenith & \\
(arcsec) & 360--450~nm: $<$0.4 mean, 0.60 max & 360--450 nm: required for QSO/LRG system throughput \\
         & 450--980~nm: $<$0.4  mean, 0.50  max & 450--980~nm: required for LRG/ELG system throughput \\
         & 60~deg from Zenith: & \\
         & 360--450~nm: 0.4 mean, 0.75 max & \\
         & 450--980~nm: 0.4 mean, 0.6 max & \\ [3pt]
Field of view & $>3^\circ$ diameter &  Required for Poisson statistics of targets \\ [3pt]
F/\# variation & F/3.7 to F/4.3 &  Required to match acceptance angle of fiber; \\
 & & F/\# variation is allowed, but must be constrained \\ [3pt]
Focal surface diameter & Sufficiently large to contain 5000 fiber positioners & Required to fit required number of positioners; \\
 & & verified by the focal plate layout simulator \\ [3pt]
Focal surface curvature & Convex radius greater than 3000~mm  & Required to allow dense packing of positioners \\ [3pt]
Aspheric departure slope & $<$30~mrad & Required for manufacturability; aspheres are permissible \\ 
 & & on one surface of smaller, non-ADC elements.  \\ [3pt]
Chief ray alignment & $<0.5^\circ$ average, $<1.0^\circ$ maximum & Required to keep incident light within acceptance angle of \\
 to focal surface normal & & fiber; required at all field points, and thus all varying F/\#s  \\ [3pt]
Glass transmission of  & $>75\%$ at 360 nm & Required for system throughput; does not include coatings \\ 
 combined elements    & $>88\%$ at 375 nm & \\
                     & $>94\%$ at 400 nm & \\
                     & $>95\%$ for $>450$ nm & \\ [3pt] 
Ghosting & $<1\%$ of DESI survey area contaminated by & Required to limit noise contribution to instrument SNR \\ 
 & ghost irradiance brighter than 0.01$\times$ night sky \\ [3pt]
Glass mass & $<900$ kg & Required to keep within allocation in top end mass budget \\
& & in order for telescope to support corrector \\ [3pt]
Lens diameter & $\leq 1175$ mm & Required to limit obscuration of incoming target light, \\ 
 & & and to keep lens manufacturable \\ [3pt]
\enddata
\tablecomments{Blurs are image full width at half maximum (FWHM), maximums are the largest value for any field position, and means are averaged over the entire field of view.}
\end{deluxetable*}

In the preliminary design stage of DESI, we approached optical vendors to get their feedback on what specifications they could achieve, and developed a working optical design that satisfied both the vendors' capabilities and the high-level corrector requirements in Table~\ref{tab:corT2}.  This gave us confidence that an optical design could be made to meet the requirements flowed down by the program.

\subsection{Optical Design Overview}
\label{sec:optdesign}

The DESI optical design is that of a prime-focus corrector: the large concave Mayall primary mirror reflects light towards a focus, and the optical aberrations are corrected by a lens assembly that achieves excellent imaging over a 3.2-degree field of view.  The lenses furthermore change the converging rays leaving the primary from F/2.8 to approximately F/3.9, as required to match the fibers at the focal surface and allow a reasonable pitch of focal plane positioners. 
 Table~\ref{tab:corrparams} summarizes the high-level optical parameters of the DESI corrector.


\begin{deluxetable}{ll}
\tabletypesize{\footnotesize}
\tablecolumns{3}
\tablecaption{DESI corrector optical design parameters \label{tab:corrparams} }
\tablehead{
  \colhead{Parameter} &
  \colhead{Value} }
\startdata
Primary mirror outer clear aperture &  3.797 meters diameter \\ [3pt]
Primary mirror inner clear aperture &  1.346 meters diameter \\ [3pt]
Field of view	& 3.2 degrees diameter, circular \\
	& 811.8 mm diameter at the focal plane \\ [3pt]
F/\#	& 3.68  on-axis \\
	& 3.86 average over FOV \\ [3pt]
Focal length & 13.920 meters on-axis \\
    & 14.572 meters average over FOV \\ [3pt]
Plate scale	& 67.48 microns arcsec$^{-1}$ on-axis \\
	& 70.65 microns arcsec$^{-1}$ average over FOV \\
\enddata
\tablecomments{The F/\#, focal length, and plate scale vary slightly across the field of view, see Figure~\ref{fig:anamorphic} for description.}
\end{deluxetable}

\begin{figure}
\centering 
\includegraphics[width=.8\textwidth]{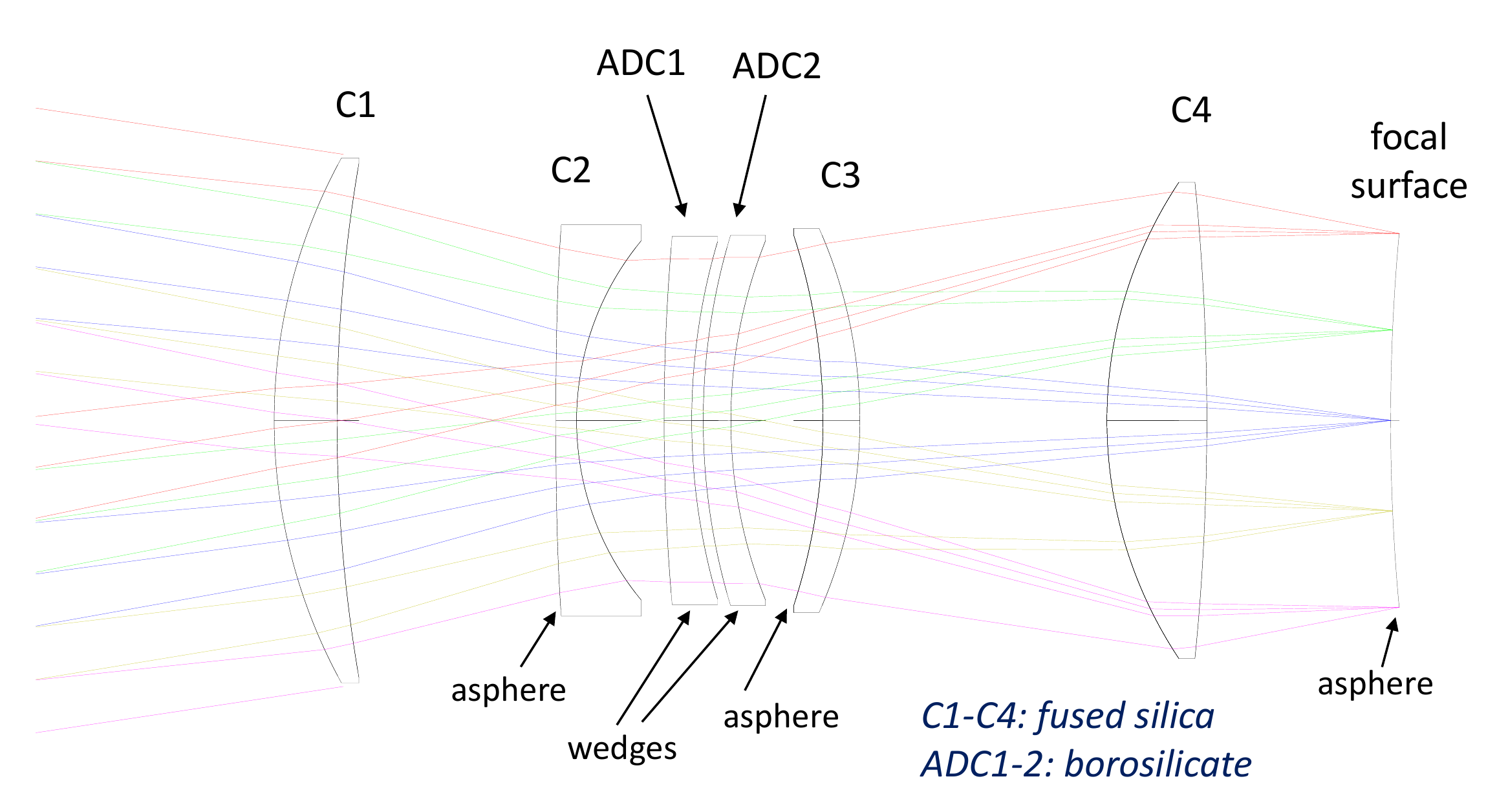}
\caption{Optical layout of DESI corrector lens assembly. \citep[Image previously published in][]{miller18}}
\label{fig:corF1} 
\end{figure}

Figure~\ref{fig:corF1} shows the optical layout of the DESI corrector lenses.  The assembly consists of six lenses, the largest of which is 1.1\,meters diameter, and the heaviest is 237\,kg.  The total mass of the lenses together is 864\,kg.  DESI's large field of view drives the lenses to be large, a typical feature of prime-focus corrector designs, although their sizes are still within the limits of available material supplies and optical fabrication capabilities.  The materials used are only fused silica and borosilicate, two common glasses chosen for their relative ease of availability in this size and excellent internal quality.  Table~\ref{tab:corT3} lists the physical characteristics of the lenses.


\begin{table}
\centering
\caption{Physical characteristics of the corrector lenses.}
\begin{tabular}{c|c|c|c|c} \hline 
Lens	&	Diameter	&	Total	&	Mass	&	Material	\\	
	&	(mm)	&	Thickness	&	(kg)	&		\\	
	&		&	(mm)	&		&		\\	\hline
C1	&	1140	&	186.0	&	201	&	Fused silica	\\	
C2	&	850	&	216.5	&	152	&	Fused silica	\\	
ADC1	&	800	&	119.7	&	102	&	N-BK7	\\	
ADC2	&	804	&	140.0	&	89	&	S-BSL7	\\	
C3	&	834	&	148.3	&	84	&	Fused silica	\\	
C4	&	1034	&	216.9	&	237	&	Fused silica	\\	\hline
\end{tabular}
\label{tab:corT3}
\end{table}

The design contains only two aspheric surfaces, one each on the C2 and C3 lenses, and all other surfaces are spherical.  The focal surface is also aspheric.  Two lenses function as an atmospheric dispersion corrector (ADC), slightly wedged lenses that can be independently rotated to counteract the effect of wavelength-dependent dispersion by the atmosphere when the telescope looks at off-zenith angles.

The physical size of the corrector unavoidably blocks some of the light reaching the primary mirror, reducing the transmission to only 77.5\% on-axis.  This transmission decreases slightly for off-axis light due to the three-dimensional shadow of the corrector, down to 76\% at a field of 1.45$^\circ$.  Furthermore, the design allows for some vignetting of the outermost fields at the C1 and C4 optical elements, in order to reduce the size and mass of those lenses, leading to a transmission of 72\% out at the 1.6$^\circ$ edge of the field of view.

The optical prescription of the corrector design is given in Table~\ref{tab:corT4}.  These values are the best final measurements of the lens radii, thicknesses and spacings, as measured by the lens vendors or the DESI team.  The lens spacings were reoptimized after the lenses were fabricated, based on knowledge of the lenses parameters as they were actually built.  The reoptimization also included knowledge of the actual index and dispersion of the glass material, obtained from melt data of the glass.  The table includes the refractive index Nd and Abbe number V for the C1, C2, and ADC1, ADC2 blanks.  Melt data for C3 and C4 were not requested - analysis showed it made little difference.  The table also includes the wedge of the ADC lenses, described as surface tilt; the tilts are on the inner surfaces of the pair, almost equal and opposite.  The spacings listed in the table refer to the separation from one element's rear surface to the front surface of the next element.  Finally, note that the primary mirror radius of curvature in the table is the corrected value discovered during DESI integration, as described in Section \ref{sec:onsiteint}. 

Table~\ref{tab:corT4} also includes the physical diameter of the optics.  The primary mirror is fitted with two masks that define its aperture, an outer mask with diameter 3797\,mm, and an inner mask with diameter 1346\,mm.  The corrector lenses mounted in their barrel assembly cause an obscuration of the incoming light, a circular shadow of roughly 1762\,mm diameter.

\begin{figure}
\centering 
\includegraphics[width=0.8\textwidth]{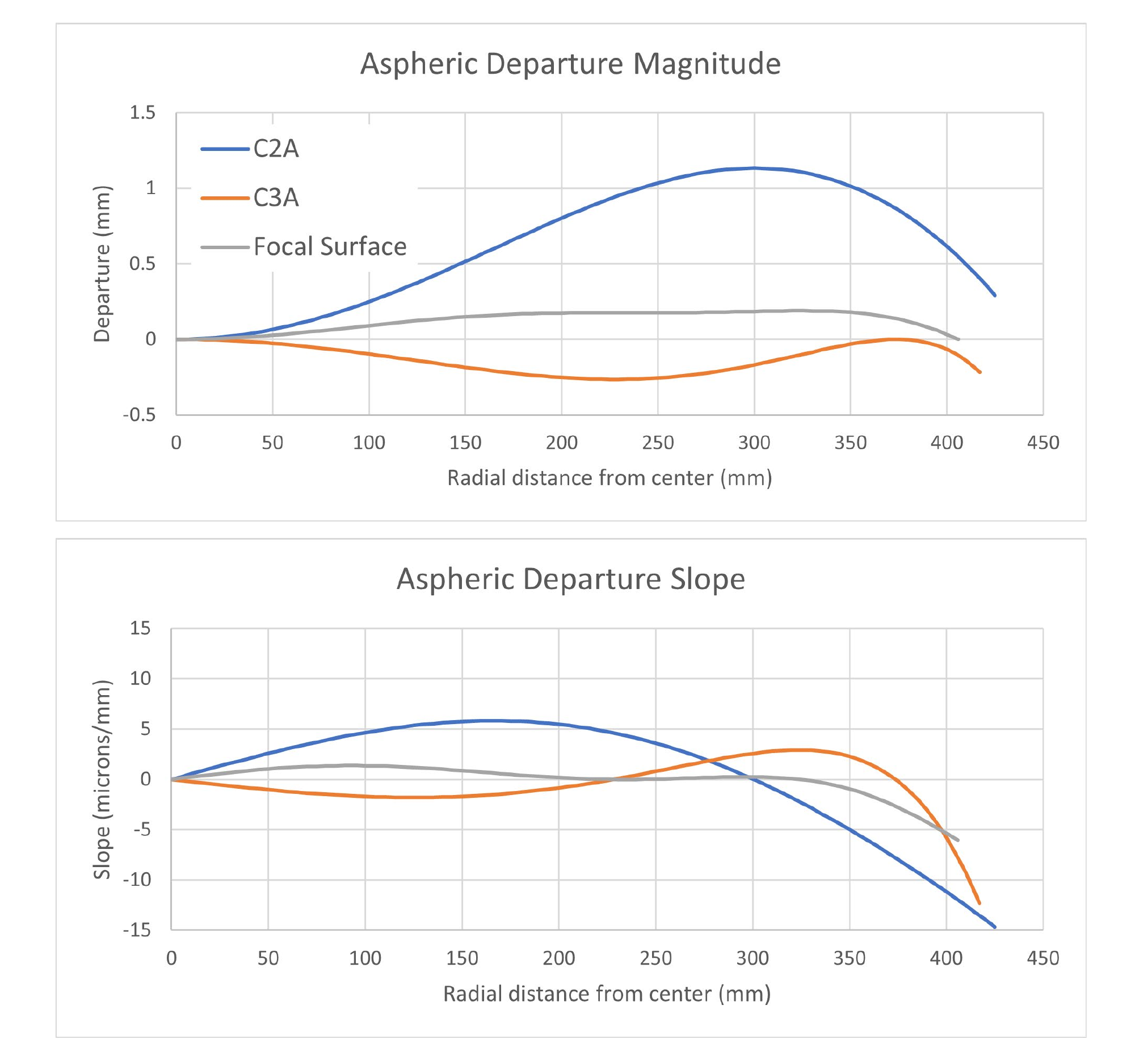} 
\caption{Description of the corrector aspheric surfaces. Top: aspheric departure of the surfaces. Bottom: slope of the aspheric departures.}
\label{fig:asphere}
\end{figure}

Finally, Table~\ref{tab:corT5} describes the aspheres in the optical design.  The primary mirror has a simple conic shape, but C2, C3 and the focal surface are 10th-order polynomial aspheres, described in the Table by a set of coefficients where the extra sag due to asphericity is $(A4)r^4 + (A6)r^6 + (A8)r^8 + (A10)r^{10}$.  Figure~\ref{fig:asphere} shows the aspheric departure from the base spherical curvature of these three surfaces.

Lenses with such polynomial aspheres are more difficult to manufacture and measure than simple conics, requiring smaller polishing tools and more complex metrology schemes.  Furthermore, the departures of these aspheres on C2 and C3 are substantial.  However, these were still within the capabilities of our chosen polishing vendor.  The focal surface is not an optical surface, rather it is a virtual curve to be matched with a mechanical assembly that can be precisely aligned and machined, so the aspheric departure is merely a bookkeeping exercise.  Allowing aspheric departure on the focal surface allows the optical design another degree of freedom with which to better control optical aberrations.


\begin{table}
\centering
\caption{DESI corrector optical prescription, as built.}
\begin{tabular}{c|c|c|c|c|c|c|c} \hline 
Surface	&	Radius	&	Center Thickness	&	Diameter	&	Tilt	&	Material	& Melt Data	&	Melt Data	\\	
	&	(mm)	&	or Spacing (mm)	&	(mm)	&	(deg)	&		&	Nd 	&	V	\\	\hline
Primary	&	-21318.000	&	8690.802	&	4000	&		&	Mirror	&		&		\\	
C1 front	&	-1184.990	&	136.590	&	1140	&		&	Fused silica	&	1.45850	&	67.825	\\	
C1 back	&	-3296.970	&	475.217	&	1140	&		&	                     	&		&		\\	
C2 front	&	-12615.960	&	44.720	&	850	&		&	Fused silica	&	1.45856	&	67.845	\\	
C2 back	&	-612.426	&	190.874	&	850  &		&	                     	&		&		\\	
ADC1 front	&	-4590.100	&	60.150	&	800	&		&	S-BSL7	&	1.51750	&	64.301	\\	
ADC1 back	&	-1371.000	&	24.850	&	800	&	0.2465	&	                     	&		&		\\	
ADC2 front	&	-1392.120	&	60.150	&	804	&	0.2501	&	N-BK7	&	1.51675	&	64.128	\\	
ADC2 back	&	-1049.580	&	199.257	&	804	&		&	                     	&		&		\\	
C3 front	&	1339.997	&	80.086	&	834	&		&	Fused silica	&	Nominal	&	Nominal	\\	
C3 back	&	1026.278	&	537.186	&	834	&		&	                     	&		&		\\	
C4 front	&	-934.200	&	216.920	&	1034	&		&	Fused silica	&	Nominal	&	Nominal	\\	
C4 back	&	5185.790	&	399.007	&	1034	&		&	                     	&		&		\\	
Focal surface	&	-4977.994	&	...	&	812	&		&		&	                     	&		\\	\hline
\end{tabular}
\label{tab:corT4}

\vspace{0.25in}

\centering
\caption{DESI corrector optical prescription: aspheric surfaces.}
\begin{tabular}{c|c|c|c} \hline 
Primary		&	C2 front		&	C3 front		&	Focal surface		\\	\hline
k =	-1.09763	&	A4 =	-1.7304552E-10	&	A4 =	1.2876124E-10	&	A4 =	-2.9648197E-10	\\	
		&	A6 =	1.5352565E-16	&	A6 =	-2.4723966E-16	&	A6 =	3.4523087E-15	\\	
		&	A8 =	3.6754102E-22	&	A8 =	4.4388319E-22	&	A8 =	-1.8041979E-20	\\	
		&	A10 =	-9.7376415E-28	&	A10 =	-7.3773076E-27	&	A10 =	3.2570782E-26	\\	\hline
\end{tabular}
\label{tab:corT5}
\end{table}


\subsection{Atmospheric Dispersion Compensator (ADC)} \label{sec:adc}

When DESI observes a target at off-Zenith angles, the layers of the atmosphere refracts the incoming light by an angle that varies as a function of wavelength, thus causing the PSF to blur in one direction.  This blur can be substantial, as much as several arcseconds over the full DESI bandpass, when viewing at our $60^\circ$ Zenith angle design limit.  For comparison, the fiber core diameter is approximately 1.5\,arcseconds and so this blur would cause a substantial light loss.  Therefore DESI incorporated an Atmospheric Dispersion Compensator (ADC) into its optical design to recover excellent imaging performance.

One common approach in ADC designs is to use dual rotating two-glass Amici prisms \citep{jacoby98} or wedged doublets \citep{Azais2016}.  However, the DESI ADC is the first to use two single wedged lenses to perform the compensation.  This simplification removes the need for two separate materials, thus avoiding either transmission losses from the extra surfaces or else the complexity and risk of bonding the optics.

\begin{figure}
\centering
\includegraphics[height=3in]{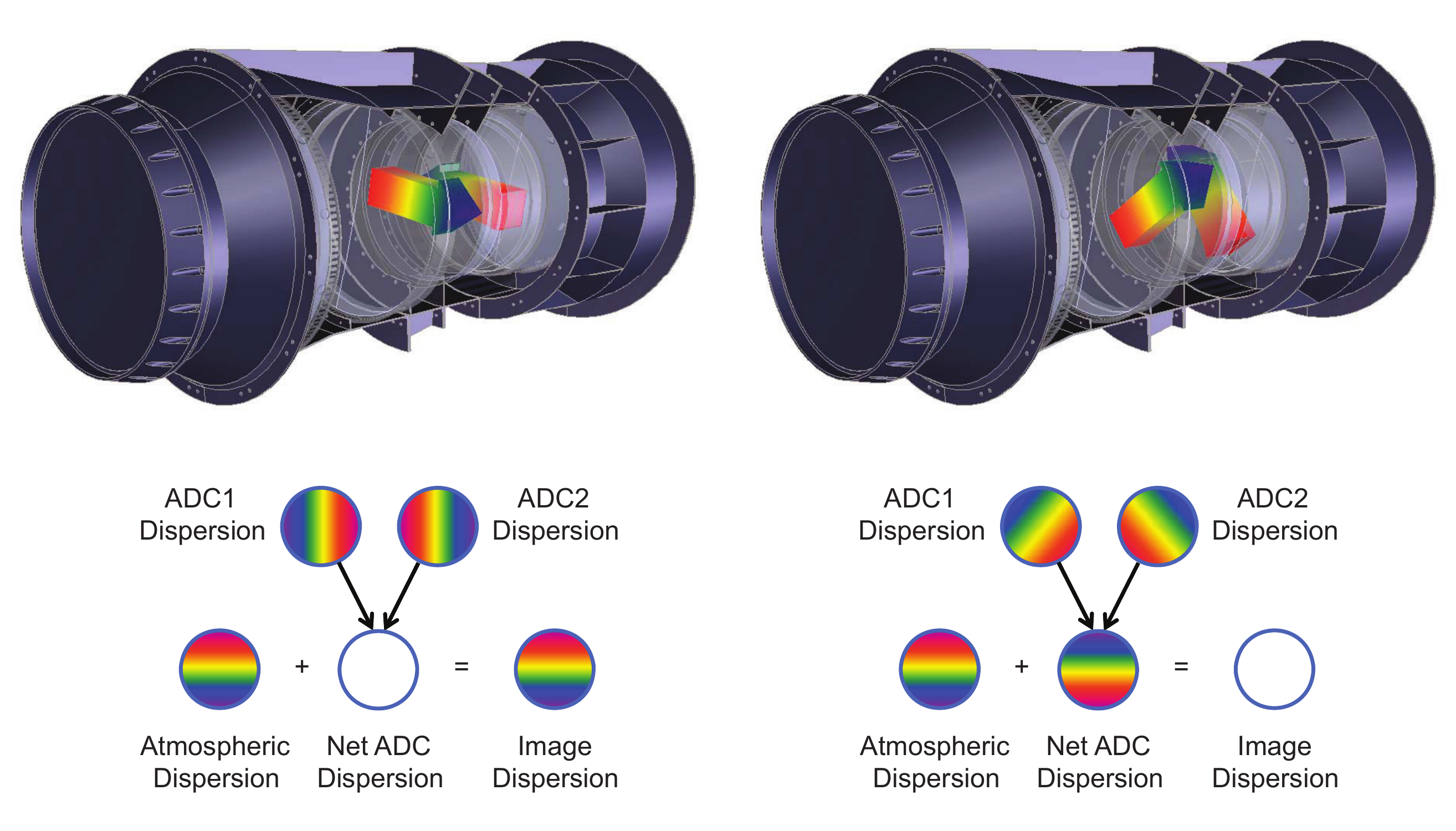}
\caption{Demonstration of the ADC concept. Left: the prisms produce leftward and rightward dispersions that cancel each other when oppositely oriented, leaving the input light unchanged. Right: the prisms can be rotated to other orientations to produce a net dispersion that negates atmospheric dispersion.} 
\label{fig:ADC}
\vspace{0.25in}
\centering
\includegraphics[height=4in]{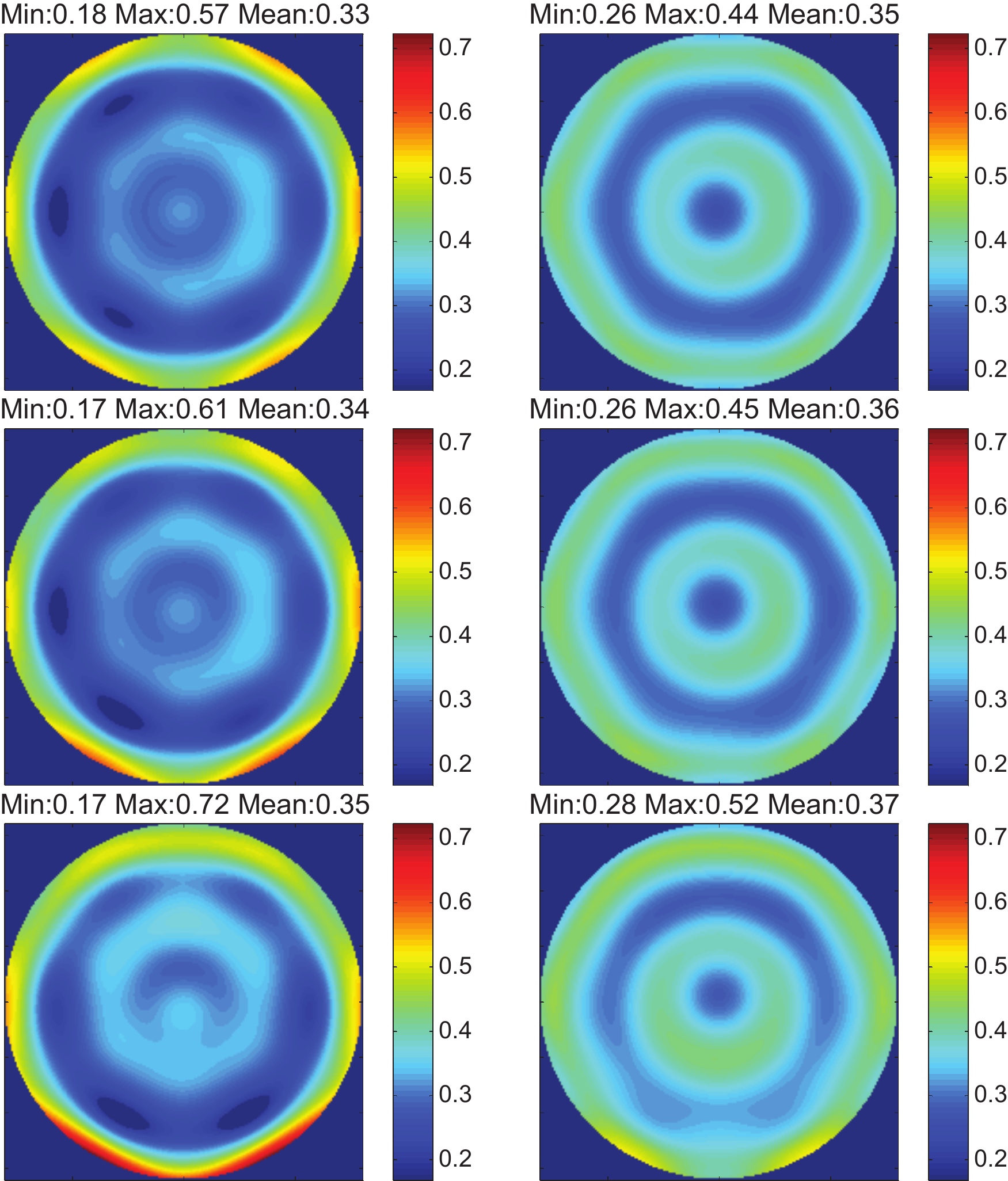}
\caption{Image quality of the corrector optical design. Blur is due to design aberrations, and is shown as FWHM blur in arcseconds. Each plot corresponds to the DESI $3.2^\circ$ circular field of view. Left column plots are calculated for a 360-450nm bandpass and right column plots for a 450-980nm bandpass.  Top, middle, bottom plots are for zenith angles of $0^\circ$, $30^\circ$, and $60^\circ$ respectively.}
\label{fig:blur}
\end{figure}

The ADC consists of two monolithic borosilicate lenses, each with two
spherical surfaces and an internal wedge angle.  Although designs exist which use fused silica for all lens elements, curved borosilicate ADC elements allow better control of lateral color on the blue end of the DESI band. The lens elements have a slight ($\sim0.25^\circ$) internal
wedge angle, which leads to predictable lateral color.  The net
dispersion magnitude and direction is set by rotating the ADC elements in opposite directions (Figure~\ref{fig:ADC}), around the corrector optical axis.  

Thin ring Kaydon\textsuperscript{\textregistered}~bearings are employed to support the ADC lens cells, and allow independent rotation of the ADC elements.
When oriented in opposite directions, the wedges introduce minimal
chromatic aberrations.  Opposite direction rotations introduce
increasing dispersion, while identical rotation of the elements
changes the direction of dispersion (required for an equatorial mount
telescope). 

Figure~\ref{fig:blur} shows the image quality of the corrector optical design after correction by the ADC.  The Figure shows the FWHM blur in arcseconds over two wavelength bands within the DESI full bandpass and for three zenith angles.  Since the corrector contains wedged ADC lenses that break the radial symmetry of the optical design, image quality can only be represented by a 2-D map over the focal surface, rather than a simple radial distribution.  The optical design image quality is excellent, achieving an average of 0.3\,arcseconds FWHM over the focal surface, and a worst case of 0.7\,arcseconds FWHM for some fields at $60^\circ$ from Zenith.  This meets the DESI requirements for image quality listed in Table~\ref{tab:corT2}.  Furthermore, the ADC design keeps the blur consistently small across all zenith angles.  Note that the Figure shows the theoretical performance of the initial optical design only, and does not include fabrication nor alignment errors; these are budgeted separately.


\subsection {Focal Surface}

\begin{figure}
\centering
\includegraphics[width=0.60\textwidth]{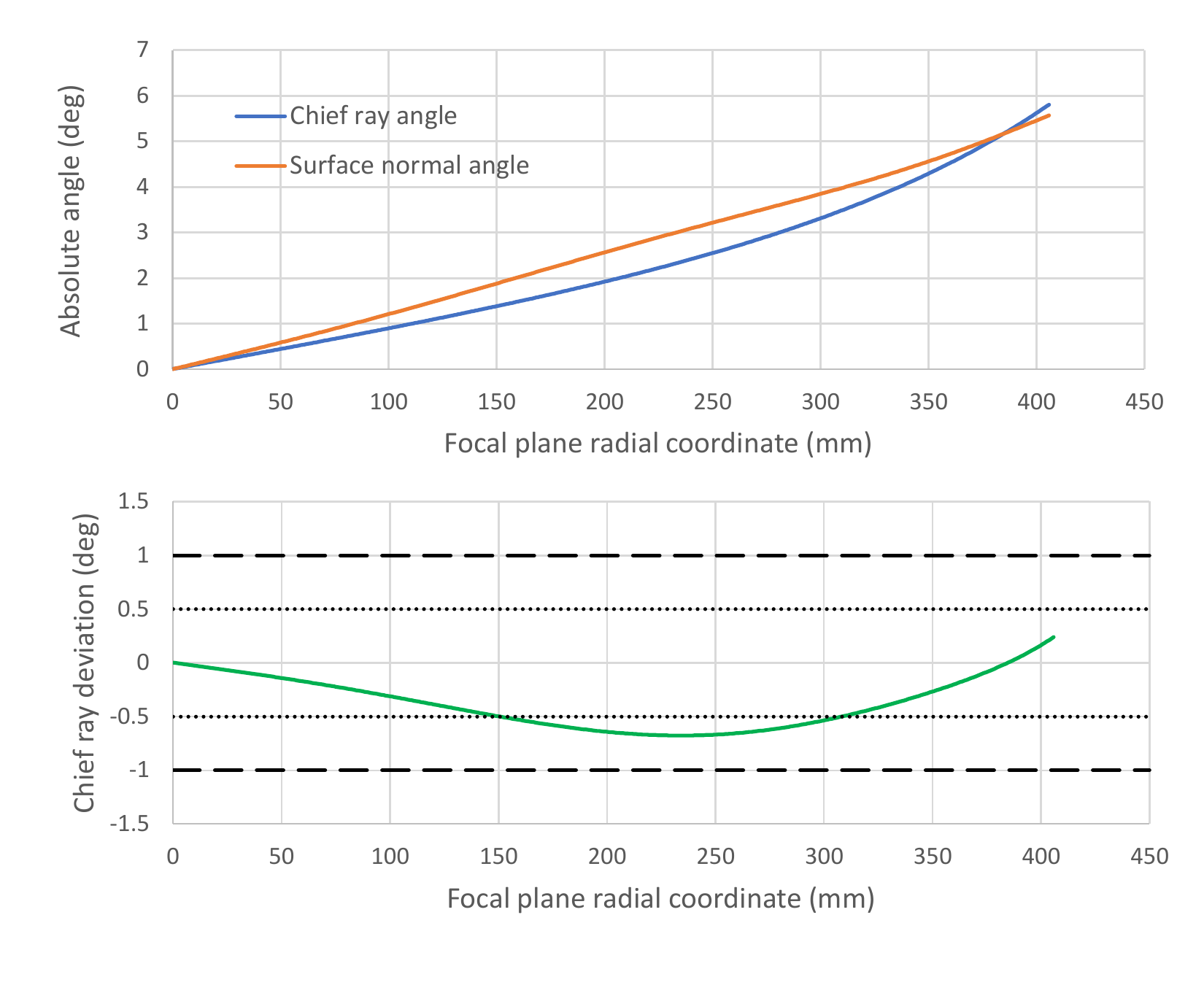}
\caption{Chief ray angle compared to the normal angle of the focal surface, as a function of radial position in the focal plane.  Top: the absolute angles of the chief ray at 500nm and of the local surface normal.  Angles are with respect to global z-axis.  Bottom: the chief ray deviation (i.e. the difference between the chief rays and surface normal angles) meets the requirements, shown as coarse and fine dashed lines for the maximum and average chief ray deviation allowed, respectively.} 
\label{fig:chiefray}
\vspace{0.25in}
\includegraphics[width=0.70\textwidth]{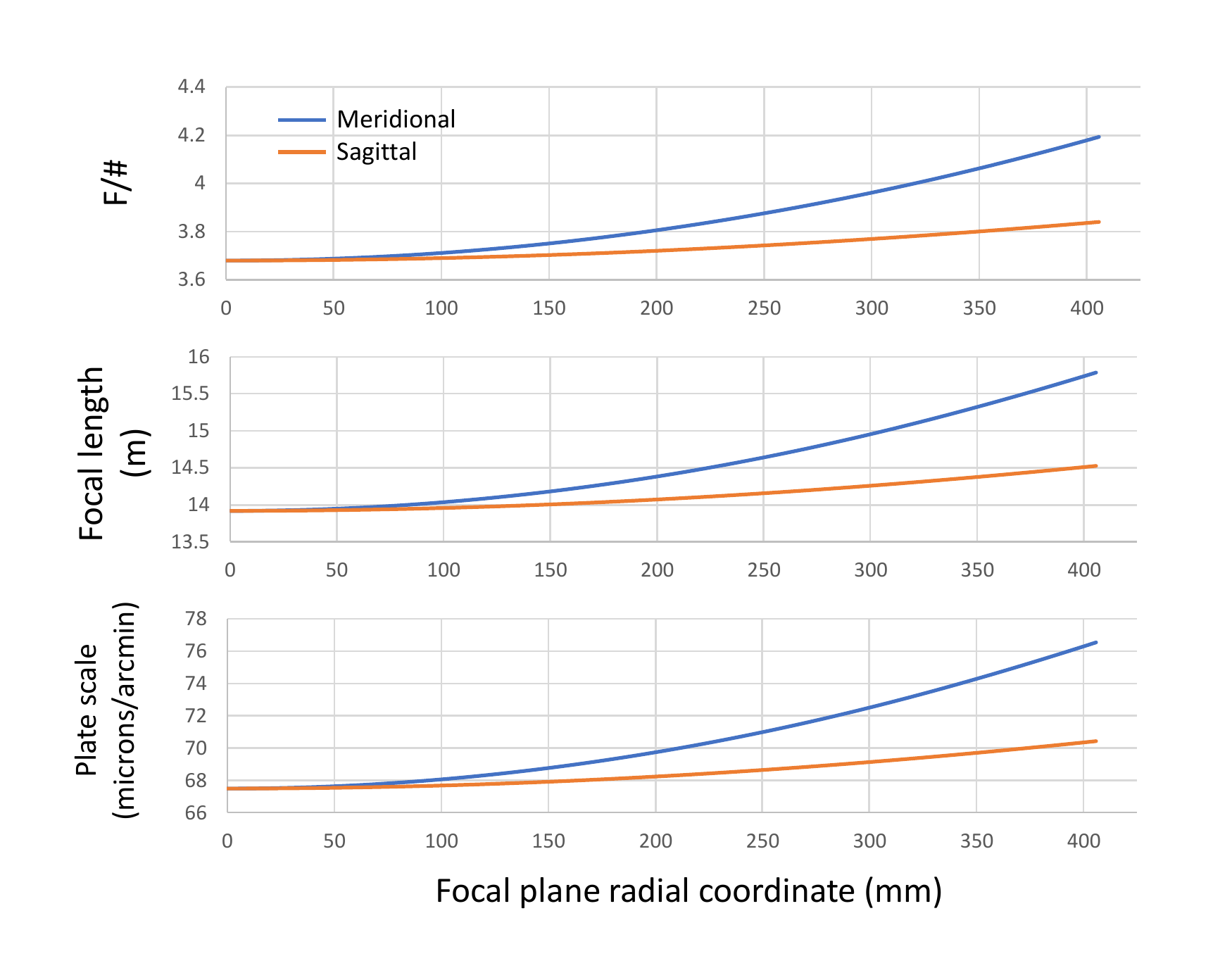}
\caption{Telescope parameters calculated from the optical design.  Parameters vary due to anamorphic pupil distortion as a function of field position and whether the meridional or sagittal directions are considered.  Top: F/\#.  Middle: focal length.  Bottom: plate scale.}
\label{fig:anamorphic}
\end{figure}

One constraint placed upon the optical design was that the angles of the chief rays for all fields were required to closely match the normal angle to the focal surface where they intersect.  The first reason for this is to minimize any defocus added when the fibers move within their 6\,mm patrol radii.  The second reason is to keep the cone of focused light incident on a fiber tip aligned to the fiber normal angle for maximum coupling efficiency.  Therefore the optical design must be optimized to keep the chief rays across the field of view normal to the curved focal surface.  Figure~\ref{fig:chiefray} shows that the chief ray angle in the optical design was kept close to the focal surface normal, meeting the requirement of $1^\circ$ maximum and $0.5^\circ$ average over the field of view (see Table~\ref{tab:corT2}).  It is worth noting that chromatic aberration in the corrector lenses causes the chief ray angle to vary slightly with wavelength.  Choosing the 500-nm chief ray for the design proved to best balance the chromatic error over the full DESI bandpass.  The DESI fibers are aligned so that their central axes are aligned to the prescribed chief ray angle as a function of location across the focal surface.


Figure~\ref{fig:anamorphic} shows the anamorphic (non-symmetric) distortion of the F/\# cone that is present as a function of position on the corrector focal surface in the optical design.  This corresponds to the distortion of the exit pupil as a function of field angle, and it means that the fixed fiber diameter maps to a sky area that varies in size and shape.  Note that F/\#, focal length, and plate scale change more quickly at locations farther from the optical axis.  Also note that the F/\# cone narrows, and the plate scale increases more in the meridional plane than the sagittal plane.  This means that the circular fiber subtends less solid angle to the sky in that direction and thus samples astronomical targets anamorphically.  This variation of the F/\# over the field of view is acceptable to DESI, and does not cause any significant performance loss.  The variation was analyzed and tracked in the design process, but not specifically minimized in the design optimization.

\subsection{Ghosts} \label{sec:ghostsection}

Ghost images and ghost pupils can be a potential problem in refractive optical designs such as the DESI corrector, when some of the transmitted light can reflect from various glass-air interfaces to form extra images on the detector.  Ghosts would be especially problematic for DESI, since observations involve a large dynamic range of bright and dim objects simultaneously: the ghost image from a bright star can couple to other fibers and disrupt the fainter objects. 

Ghost reflections are hypothetically possible between all optical surfaces of the corrector, including all lenses, the Mayall primary, and focal array sensors.  Avoiding potential ghosts between all possible surfaces while creating the optical design was too constraining, and could have led to worse overall performance, so optimization of the design did not specifically take ghosts into account.  Rather, we analyzed possible ghosts in the design in a separate exercise to confirm that their effects were acceptably small.

We analyzed ghosts early in the design process by first-order calculation of their relative brightness.  For all possible double-bounce ghosts, this analysis calculated their size at the focal surface and their relative transmission from reflection at all surfaces.  Conservatively, it assumed that all lens surfaces have a reflectance of 2\%, the primary mirror a reflectance of 90\%, the fibers at the focal plane a reflectance of 2\%, and the guider CCDs at the focal plane a reflectance of 10\%.  Table~\ref{tab:ghosttable} quantifies the brightness of the ghosts found in the corrector optical design by this analysis.  The brightest ghost results from a two-bounce reflection between the inner surfaces of the ADC lens pair, which are nearly concentric and therefore create a ghost at the focal surface with only a small amount of defocus.  This ghost has a diameter of 13\,mm, or about 0.024\,deg$^2$ on the sky, and a relative irradiance of $8.84 \times 10^{-9}$ compared to the source.  All other ghosts are dimmer than this, i.e. they introduce less unwanted light into the fibers.

\begin{table}
\centering
\caption{Relative irradiances of possible double-bounce ghosts in the DESI corrector}
\tiny
\begin{tabular}{c|c|c|c|c|c|c|c|c|c|c|c|c|c} \hline 
&	Primary	&	C1	&	C1	&	C2	&	C2	&	ADC1	&	ADC1	&	ADC2	&	ADC2	&	C3	&	C3	&	C4	&	C4	\\
&	Mirror	&	front	&	back	&	front	&	back	&	front	&	back	&	front	&	back	&	front	&	back	&	front	&	back	\\  \hline
C1 front	&	1.27E-12	&	...	&	...	&	...	&	...	&	...	&	...	&	...	&	...	&	...	&	...	&	...	&	...	\\	\hline
C1 back	&	7.64E-12	&	3.14E-13	&	...	&	...	&	...	&	...	&	...	&	...	&	...	&	...	&	...	&	...	&	...	\\	\hline
C2 front	&	3.02E-12	&	8.31E-13	&	2.15E-12	&	...	&	...	&	...	&	...	&	...	&	...	&	...	&	...	&	...	&	...	\\	\hline
C2 back	&	9.66E-13	&	2.11E-12	&	2.00E-13	&	4.04E-13	&	...	&	...	&	...	&	...	&	...	&	...	&	...	&	...	&	...	\\	\hline
ADC1 front	&	1.34E-09	&	1.83E-12	&	1.79E-11	&	5.92E-12	&	1.69E-12	&	...	&	...	&	...	&	...	&	...	&	...	&	...	&	...	\\	\hline
ADC1 back	&	4.58E-12	&	1.30E-12	&	6.43E-13	&	8.93E-13	&	3.07E-12	&	7.96E-12	&	...	&	...	&	...	&	...	&	...	&	...	&	...	\\	\hline
ADC2 front	&	4.68E-12	&	1.25E-12	&	6.47E-13	&	8.87E-13	&	2.87E-12	&	8.10E-12	&	8.84E-09	&	...	&	...	&	...	&	...	&	...	&	...	\\	\hline
ADC2 back	&	2.19E-12	&	9.76E-13	&	3.45E-13	&	5.28E-13	&	3.16E-12	&	3.90E-12	&	7.98E-11	&	6.81E-11	&	...	&	...	&	...	&	...	&	...	\\	\hline
C3 front	&	3.81E-12	&	7.04E-11	&	1.12E-12	&	3.80E-12	&	4.46E-12	&	7.71E-12	&	4.61E-12	&	4.85E-12	&	3.86E-12	&	...	&	...	&	...	&	...	\\	\hline
C3 back	&	1.56E-12	&	2.15E-11	&	3.75E-13	&	9.20E-13	&	9.07E-12	&	3.04E-12	&	3.24E-12	&	3.36E-12	&	3.14E-12	&	3.54E-11	&	...	&	...	&	...	\\	\hline
C4 front	&	8.73E-12	&	1.68E-12	&	1.09E-12	&	1.39E-12	&	3.28E-12	&	1.48E-11	&	2.92E-10	&	4.30E-10	&	4.27E-11	&	7.90E-12	&	5.47E-12	&	...	&	...	\\	\hline
C4 back	&	2.56E-12	&	1.36E-12	&	4.21E-13	&	6.67E-13	&	5.15E-12	&	4.62E-12	&	5.37E-11	&	4.86E-11	&	5.09E-10	&	5.70E-12	&	5.35E-12	&	4.78E-11	&	...	\\	\hline
Fiber tip	&	4.46E-12	&	1.02E-12	&	5.89E-13	&	7.80E-13	&	2.16E-12	&	7.65E-12	&	5.34E-10	&	1.06E-09	&	3.79E-11	&	4.72E-12	&	3.41E-12	&	1.91E-09	&	4.22E-11	\\	\hline
Guider CCD	&	2.23E-11	&	5.09E-12	&	2.95E-12	&	3.90E-12	&	1.08E-11	&	3.82E-11	&	2.67E-09	&	5.28E-09	&	1.90E-10	&	2.36E-11	&	1.70E-11	&	9.56E-09	&	2.11E-10	\\	\hline
\end{tabular}
\tablecomments{Irradiances are given as a fraction of the irradiance of a parent image within one square arcsec, and so are useful for calculating ghost brightness relative to astronomical target images on the fibers.  Each axis lists the possible surfaces for reflection, including lenses, the primary mirror, and the focal surface.  Lenses are designated C1 through C4, or A1/A2 for the ADCs, with a suffix 1 or 2 indicating the front or rear surface.  Focal surfaces are designated as either a fiber tip F or a guiding CCD Fccd.}
\label{tab:ghosttable}
\end{table}

\begin{figure}
\centering 
\includegraphics[width=\textwidth]{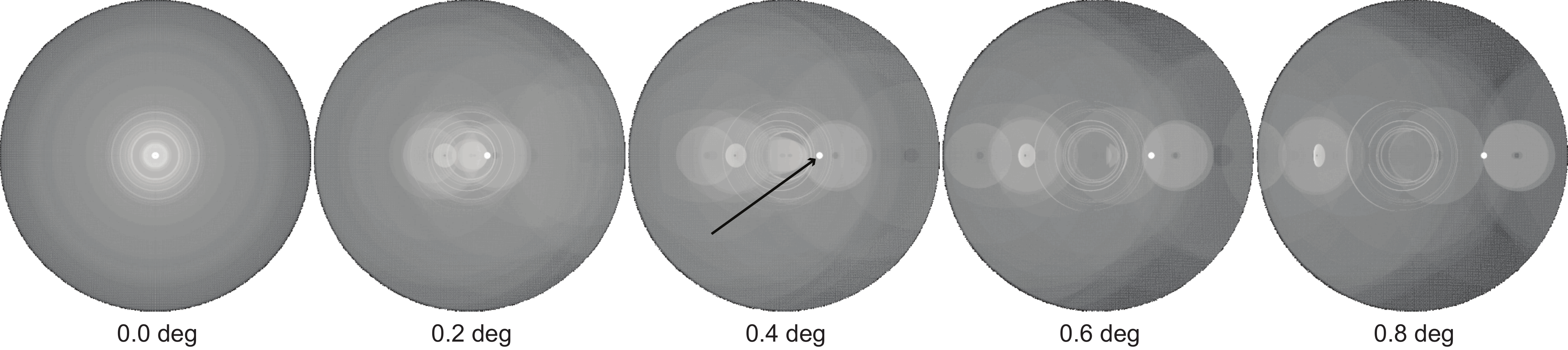} 
\caption{Focal ghosts present at the focal surface. Plots left-to-right show ghosts at several field locations across the focal surface. The arrow in the middle plot identifies the brightest focal ghost feature, a reflection between the two ADC lenses.}
\label{fig:corF3}
\end{figure}

The ghosts were calculated separately using non-sequential raytracing, and they results matched the first-order analysis.  Figure~\ref{fig:corF3} shows the raytraced images of the collected set of double-bounce ghosts, with a notional bright star at various field angles as the source.


DESI’s requirement for ghost contamination is that $<$ 1\% of the field of view be contaminated by ghosts that are brighter than 1\% of the night sky background.  An early calculation showed that the 1575 brightest stars in the Hipparcos2 I-band catalog (see \url{https://www.cosmos.esa.int/web/hipparcos/hipparcos-2}) will have a ghost reflection greater than 1\% of the NIR sky background.  The aggregate area of these ghosts is 4.8\,deg$^2$, or 0.036\% of the 14,000\,deg$^2$ survey, and therefore DESI’s ghost requirement is met.  Furthermore, the final lens coatings achieved a reflection of about 0.5\% per surface, and thus double-bounce ghost reflections will be 16 times fainter than the above early calculation suggests.

As a validation of the ghost modeling during the instrument commissioning phase, we steered the Mayall telescope to put a bright star at several angles within the instrument's field of view.  As expected, any ghosts were within the scattering wings of the star's bright PSF, and thus were not able to be distinguished.  In order to deliberately find ghosts, we steered the telescope to put a bright star on the focal plane's guider assembly, a configuration that reflects light strongly from its r-band filter and produces atypically bright ghosts at the focal plane.  Then we analyzed spectrograph data from all the fibers, looking for any contamination from the spectral signature of the star, and created a map of the bright star's light at the focal plane.  The map showed several ghosts that matched what was expected for this configuration, thus validating the ghost modeling.  Furthermore, we set a rule in the survey planning to avoid telescope pointings that place bright stars above 4th magnitude onto the guiders during operations, thus avoiding this ghost configuration; this is a minor constraint, and does not cause any loss of survey area.

Stray light considerations in the corrector beyond ghosts are discussed in section \ref{sec:straylightmain}.

\subsection{Lens Specification and Tolerancing}

The DESI corrector lenses were specified carefully for two significant reasons.  First, since throughput is the primary performance metric, the corrector must maximize the light from our science targets that enters each small fiber core.  Therefore excellent image quality was paramount to avoid losses, and it drove all specifications on the lenses.  Second, early in the design process, DESI made the decision not to verify the fully-integrated corrector performance in a test, since such a test would be cost-prohibitive and take many extra months.  Instead the corrector underwent full verification at the component level, i.e. the lenses, and the performance of the corrector at a higher level was verified by modeling and analysis of the as-built lenses.  This decision put great emphasis on fabricating the lenses correctly, and therefore also specifying them correctly, including requirements that the vendors perform rigorous verification testing.

Performance of the lenses was modeled by use of an image quality budget, aka a “blur budget”, which tracked all error contributors that degraded image quality.  We used this budget to create tolerances on all the lenses, including radii, thickness, material index variations (inhomogeneity), figure errors, surface roughness, centration and tilt. The estimate of blur included the design-residual blur described above in Figure~\ref{fig:blur}.  We allocated the various errors such that their values were all fairly similar \citep{miller18}.  This budget flowed down from the DESI fiber coupling systems engineering budget which gave an allocation to the lens design and assembly for a minimum throughput into the fiber cores \citep{besuner16}.

Instead of specifying tight tolerances on the lens thicknesses, radii of curvature, and glass indices, we used a compensation scheme that allowed the final lens spacings to be adjusted based on actual measurements of lens thickness and radii after they were fabricated, and on melt data from the glass vendors (measurements of the actual index and dispersion).  This respacing allowed looser manufacturing tolerances to be specified for these parameters.  This spacing correction was done after the individual lenses were fabricated and measured, resulting in a slightly different design prescription; the lens shifts were small, within the ranges expected \citep{miller18}.

Besides errors in the lenses themselves, blur also arose from lens misalignment errors.  The lenses are held in lens cells, which are in turn held by the barrel assembly.  Errors in the positions of the lenses result from a stackup of tolerances including lens mounts, cell mounts, and corrector barrel thermal and mechanical distortions.  Allowable limits for decenter and tilt in all these terms were established by Monte Carlo simulation, and the terms were controlled in the blur budget.  These alignment limits are described further in section \ref{sec:mechdesign}.

Table \ref{tab:reqspectable} summarizes the primary requirements for all lens characteristics.  (The table also shows the final measured values; these are discussed in the next section.)  As mentioned, these requirements were driven by an even allocation of tolerances in order to meet high-level performance, primarily throughput into any fiber core at the telescope focal plane.

Specification of the lens surface figure (height errors of the surface from its desired shape) is complicated because the effect of a particular figure error depends on its spatial frequency.  Therefore, the requirement is broken down into low, medium, and high frequency cases, plus roughness.  Low frequency errors are defined as those with about four cycles across the diameter or less, and correspond to what optical vendors traditionally consider ``figure error''.  (The specific frequency bands are defined differently for every surface, depending on their diameter.)  Middle frequency errors correspond to between about four and twenty cycles across the lens diameter.  This is a defined as an RMS slope error of the surface, a useful metric that both corresponds directly to a spot size error allocation and also is easily calculated by the vendor from surface measurements.  The effect of both low and middle frequency errors is a blurring of the PSF, but they are typically considered separately because they tend to be controlled separately by the vendor's choice of tooling for polishing.  High frequency errors are defined as those of over twenty cycles per diameter, i.e. with periods as small as 1-mm.  The effect of these errors is not so much a widening of the PSF but rather that light is scattered far outside of the PSF and counts as a throughput loss.  Roughness errors correspond to even higher frequencies, with periods less than 1-mm, and also have the effect of scatter loss.  They are differentiated from high frequency errors because they arise from different polishing mechanisms, and are measured differently as well.

A scratch-dig spec of 80/50 (not listed in Table \ref{tab:reqspectable}) was also specified for all lens front and back surfaces, although this was largely a cosmetic specification since the subsequent transmission loss is negligible.  

The lenses were also specified with a variety of mechanical specifications, applying to both the blanks and the final polished surfaces.  Diameter tolerances were $\pm$0.5mm.  All non-optical surfaces (edges, chamfers and flat faces) were fine ground to 2.5\,micron Ra or better.

Finally, all lenses were specified to maintain fiducial markings on their edges as a reference for all surface figure maps, wedge measurements, etc.  This ensured a known orientation of the lenses for when they were installed in the corrector, and supported accurate corrector models.

DESI consulted with several optics vendors to ensure that the challenging requirements in Table \ref{tab:reqspectable} were reasonably within the capabilities of most vendors, and to identify specific requirements where relaxation might be needed to ensure successful fabrication.  In particular, the surface height errors, broken into various frequency bands, were reviewed with vendors to make sure they were written in a way that could be accurately measured and achieved using their standard processes.  As an example of this, note that the slope error requirement is chosen to be higher for the front surfaces of C2 and C3 than it is for other surfaces.  This is because these are the two aspheric surfaces in the optical design, and meeting this requirement was anticipated to be more difficult than for the other surfaces that are spherical.  DESI modeled the effects of allocating a higher value for the two aspheres and found it acceptable, thus keeping the requirements within the vendor's capability.


\begin{table}
\footnotesize
\caption{Primary fabrication requirements and final measured values for each lens}
\label{tab:reqspectable}
\centering
\begin{tabular}{c|c>{\itshape}c|c>{\itshape}c|c>{\itshape}c|c>{\itshape}c}
\hline
\multirow{3}{4em}{\centering Lens} 	&	\multicolumn{2}{c}{\multirow{2}{11em}{\centering Substrate Inhomogeneity (parts per million)}}			&	\multicolumn{2}{c}{\multirow{2}{12em}{\centering Radius Tolerance \newline (\%)}}			&	\multicolumn{2}{c}{\multirow{2}{10em}{\centering Thickness Tolerance (mm)}}			&	\multicolumn{2}{c}{\multirow{2}{9em}{\centering Wedge Tolerance (microns of runout)}}			\\
 & & & & & & & & \\
	&	spec	&	value	&	spec	&	value	&	spec	&	value	&	spec	&	value	\\ [3pt]
\hline
C1	&	3	&	1	&	0.08 , 0.09	&	0.02 , 0.04	&	0.3	&	0.2	&	100	&	51	\\
C2	&	3	&	1.8	&	0.16 , 0.08	&	0.08 , 0.01	&	0.5	&	0.3	&	100	&	72	\\
ADC1	&	4	&	3.5	&	0.22 , 0.15	&	0.02 , 0.003	&	0.3	&	0.15	&	140	&	1	\\
ADC2	&	4	&	2.3	&	0.14 , 0.10	&	0.001, 0.01	&	0.3	&	0.15	&	140	&	113	\\
C3	&	10	&	1.6	&	0.15 , 0.10	&	0.05 , 0.07	&	0.15	&	0.09	&	100	&	58	\\
C4	&	5	&	3	&	0.11 , 0.19	&	0.01 , 0.03	&	0.15	&	0.07	&	100	&	76	\\
\hline \hline
\multirow{3}{4em}{\centering Lens} 	&	\multicolumn{2}{c}{\multirow{2}{11em}{\centering Overall Figure Error (waves peak-to-valley)}}			&	\multicolumn{2}{c}{\multirow{2}{12em}{\centering Surface Slope Error \newline (rms microradians)}}			&	\multicolumn{2}{c}{\multirow{2}{10em}{\centering High-frequency Figure Error (nm RMS)}}			&	\multicolumn{2}{c}{\multirow{2}{9em}{\centering Surface Roughness
(nm RMS)}}			\\
 & & & & & & & & \\
	&	spec	&	value	&	spec	&	value	&	spec	&	value	&	spec	&	value	\\ [3pt]
\hline
C1	&	0.5	&	0.09	&	1.5	&	1.25 , 0.6	&	6	&	3 , 5	&	2	&	0.5 , 0.6	\\
C2	&	2	&	1 , 0.2	&	5 , 2	&	6.3, 1.5	&	6	&	4 , 2	&	2	&	1.9 , 0.2	\\
ADC1	&	1	&	0.7 , 0.6	&	2	&	$<$2, $<$2	&	6	&	3 , 2	&	2	&	0.7 , 0.6	\\
ADC2	&	1	&	0.5 , 0.6	&	2	&	$<$2, $<$2	&	6	&	4 , 3	&	2	&	0.9 , 0.6	\\
C3	&	4	&	1.7 , 0.7	&	5 , 2	&	4.97 , 1.9	&	6	&	5 , 2	&	2	&	1.1 , 0.9	\\
C4	&	2	&	0.3 , 0.1	&	3	&	2.2 , 2.4	&	6	&	2 , 2	&	2	&	0.7 , 0.5	\\
\hline
\end{tabular}
\tablecomments{Table gives values for both front and back surfaces, if they differ. The polishing vendor for the ADC lenses provided structure function curves for the measured slope error instead of single values; the table shows the original single-value spec. The vendors met all requirements with the exception of the C2 surface slope which was analyzed and accepted by DESI. \citep[Previously published in][]{miller18}}
\end{table}

\subsection{Corrector Lens Acquisition}  \label{sec:2p7}

\begin{table}
\centering
\caption{Vendors and milestones for the corrector lenses}
\begin{tabular}{c|c|c|c|c|c|c} \hline 

Optic	&	Blank Vendor	&	Blank Delivered	&	Polishing Vendor	&	Lens Polished	&	Coating Vendor	&	Lens Coated	\\	\hline
C1	&	Corning Inc.	&	Dec-14	&	L3 Brashear	&	Jan-16	&	Viavi Solutions	&	Mar-17	\\	
C2	&	Ohara Corp.	&	Nov-14	&	Arizona Optical Systems	&	Oct-17	&	Viavi Solutions	&	Jan-18	\\	
C3	&	Ohara Corp.	&	Nov-14	&	Arizona Optical Systems	&	Jun-17	&	Viavi Solutions	&	Aug-17	\\	
C4	&	Corning Inc.	&	Dec-14	&	L3 Brashear	&	Jan-16	&	Viavi Solutions	&	Apr-17	\\	
ADC1	&	Ohara Corp.	&	Aug-15	&	Rayleigh Optical Corp.	&	Jan-17	&	Viavi Solutions	&	Nov-17	\\	
ADC2	&	Schott AG	&	Jul-15	&	Rayleigh Optical Corp.	&	Nov-16	&	Viavi Solutions	&	Jul-17	\\	\hline
\end{tabular}
\label{tab:vendors}
\end{table}

\begin{figure}
\centering 
\includegraphics[width=\textwidth]{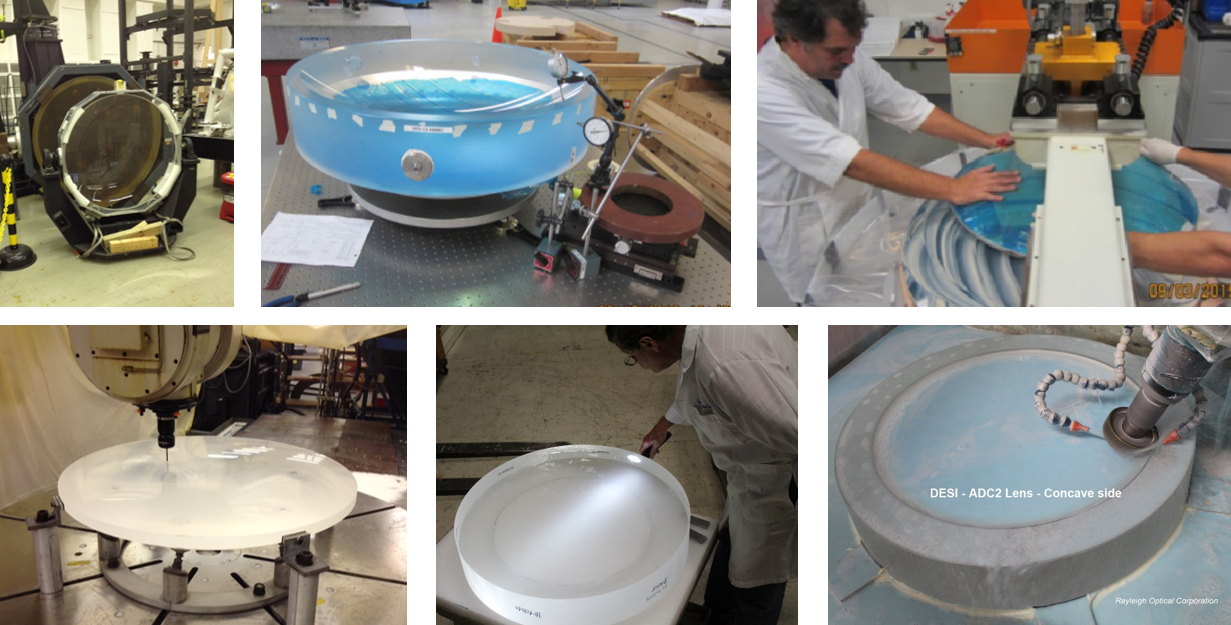} 
\caption{Corrector lenses in various phases of fabrication. From top left to bottom right: C1 in transmission test configuration, mechanical measurements of C2 lens, polishing of the C3 convex surface, dimensional probe of C4 after its generation, inspection of the ADC1 blank, and machining of ADC2. \citep[Images of C1, C2, C3, ADC1 previously published in][]{miller16_corrfab}}
\label{fig:corrlenses}
\end{figure}

Table~\ref{tab:vendors} lists the vendors and milestone dates for the glass blank procurement, lens polishing, and lens coating, for all six corrector lenses.  Care was taken to choose vendors with demonstrated ability to fabricate similar large lenses for other projects.  For each fabrication step, a pool of vendors was narrowed down by a competitive bid process weighing their relative capability and quoted cost and schedule, before finally contracting with the selected vendors.  Note that the dates in Table \ref{tab:vendors} are largely determined by the needs of the project schedule and the complexity of the various lenses and steps.  Figure~\ref{fig:corrlenses} shows the six lenses in various stages of work.

The various fabrication steps can take months or even years to complete, and so lens procurement was among the earliest operations started in the overall DESI build program.  The schedule for all lenses included several months of margin, at various stages, before the completed and coated lenses were due to be integrated into the corrector assembly.  This allowed for unforeseen delays at each vendor that might otherwise affect the higher-level DESI schedule.  The C2 lens in particular required more time than anticipated due to a challenging asphere on a convex surface that required slow polishing with a small tool; in this case the delayed lens still fit into the overall DESI schedule.

DESI also worked closely with all vendors throughout their contracts to assure they would meet their cost, schedule, and performance requirements, typically via weekly status meetings.  We required vendors to document compliance to all requirements and supply all test procedures and test data, and in many cases we used vendor measurements to confirm the corrector performance at a higher level.  DESI furthermore encouraged vendors to measure their critical parameters with multiple independent methods in order to be assured that requirements were truly met.  This was especially necessary because DESI chose not to verify the fully-integrated corrector performance in a test, as described above.

As shown in Table \ref{tab:reqspectable}, the delivered lenses met almost all their contractual specifications, typically by a significant margin.  The exception was the maximum slope error on the C2 asphere (see discussion below); in this case we analyzed the discrepancy to assure that any effect on performance was acceptable, and then granted a waiver.  Note that the polishing vendor for the ADC lenses provided structure function curves for the measured slope error instead of single values, though in the end the lenses met the equivalent spec; the table shows the original single-value spec.  Not shown in Table \ref{tab:reqspectable} are the diameters of the delivered lenses and the indices of refraction of the lens glasses, both of which were within specification.

The following sections describe highlights of the fabrication process at each vendor, including their verification testing to demonstrate that they met DESI's requirements.

\subsection{Corrector Lens Blank Procurement}

The excellent optical performance of the DESI lenses requires glass with excellent characteristics.  The first consideration is the availability of a suitable material glass in the required size with excellent internal quality, which immediately limits options to fused silica and borosilicate.  Fortunately these glasses work well in our optical designs, and these materials are available in sufficiently large blank sizes for all the lenses from several possible vendors with a history of successful delivery.

The main parameter of concern is glass homogeneity, the spatial variation of the refractive index in the raw material.  Homogeneity is measured as the minimum-to-maximum variation of the phase of the transmitted wavefront over the glass diameter divided by the thickness of glass; this is a unitless value, and is typically described in parts per million (ppm).  The index variation directly impacts the corrector image quality and causes wavefront distortion mostly in the low-order Zernike modes, \ie variations of up to about three cycles per lens diameter.  To determine its effect on corrector image quality, we modeled inhomogeneity as random Zernike modes within a limited band of spatial frequencies, and scaled the magnitude to keep the resulting image blur within an allocation in the blur budget.  The magnitudes were then flowed into the specifications for all of the lens blanks.  

\begin{figure}
\centering 
\includegraphics[width=.6\textwidth]{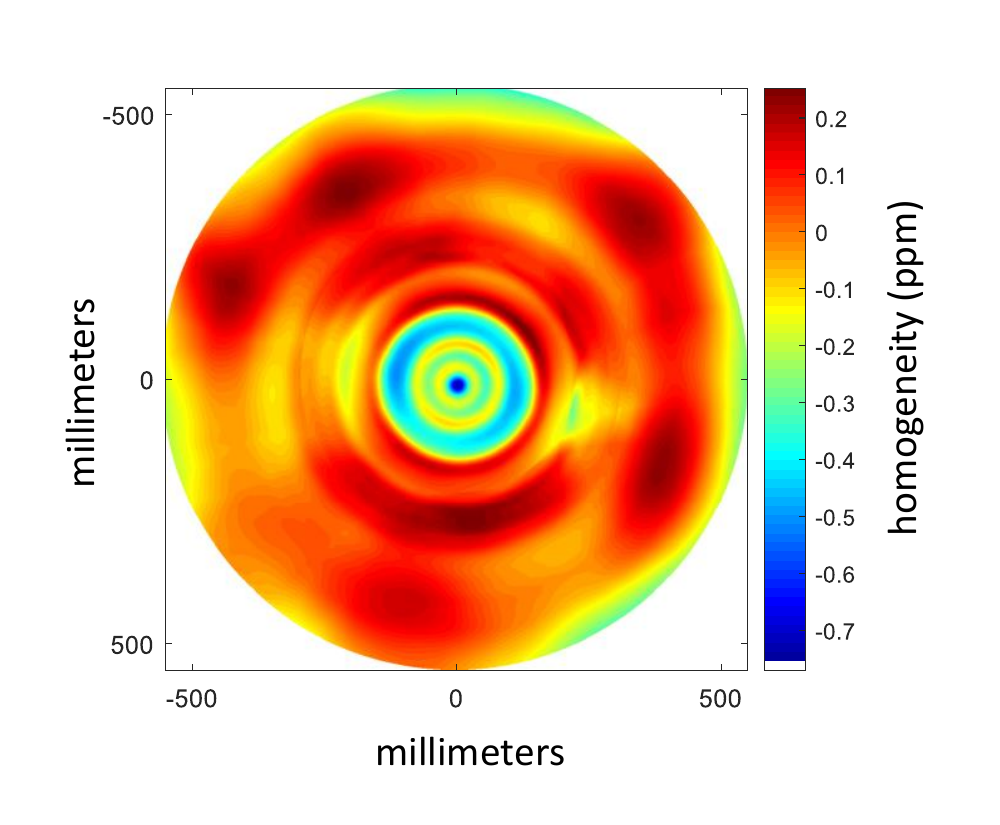} 
\caption{Homogeneity map of the C1 lens blank.  The measured range of index variation across the full diameter is 1.0 parts per million (ppm) which meets the 3\,ppm requirement.  The requirement allows an overall defocus term to be removed from the data.} 
\label{fig:homogmaps}
\end{figure}

The final homogeneity requirements are shown in Table~\ref{tab:reqspectable}.  All vendors delivered blanks that met their homogeneity requirements, mostly by a large margin.  Note that homogeneity is not easily captured in a single number, since it is generally a two-dimensional variation across the glass blank.  Therefore homogeneity maps of the blanks were requested along with delivery, and we modeled the maps using Zemax software to confirm that the inhomogeneity caused an acceptably small amount of blur.  Figure~\ref{fig:homogmaps} shows a map of the C1 blank homogeneity measured by the vendor as an example.  Its radially symmetric pattern is a typical consequence of the blank manufacturing process.  None of the glass vendors were able to measure homogeneity over the full diameter of their blanks because their interferometer test stations were not capable of measuring such large blanks.  Instead, vendors measured multiple subapertures of the full aperture instead, enough to cover the full aperture area, and stitched them together in software to compile a full aperture map.  The exception is the ADC2 lens, where DESI accepted verification testing by subapertures only by the vendor, after analyzing this case and determining that the performance would be acceptable.


Besides inhomogeneity, another requirement is tolerance on the absolute refractive index of the blank.  That is, the average index as a function of wavelength must be close to the nominal values expected for fused silica or borosilicate.  To verify this, the vendors measured the actual index of the produced blanks, known as melt data.  This was done by separating a small piece from the same glass production run, typically a piece outside the physical dimensions of the final blank, and measuring its index for a series of wavelengths over the DESI bandpass.  This is a standard procedure in glass manufacturing.  Melt data was only taken for the C1, C2, ADC1, and ADC2 blanks: analysis showed that the optical design is not sensitive to expected glass index variations in C3 and C4, and so the measurements were not taken.  In most cases the index data met the DESI specification for absolute index.  The glass used to make ADC1 had an absolute index slightly out of its tolerance, but our analysis showed that we could accept the glass with an insignificant effect on performance.

Inclusion class, a measure of the number and size of internal defects within the glass, was also a consideration, but it typically correlates with homogeneity.  Specifying low-inhomogeneity glasses led to a low inclusion class as well, with internal inclusions being less than 0.1\,mm$^2$ in cross section and less than 0.28\,mm in size (0.15\,mm for ADC1-ADC2).  These inclusions cause insignificant throughput loss, and are acceptable for the lenses.


\subsection{Corrector Lens Polishing}

The finished glass blanks were shipped to one of several optical shops for polishing.  Each went through the stages of machining, grinding and polishing, and final verification testing.  We next describe the fabrication of each pair of lenses at their respective vendors, focusing on how the vendors demonstrated that they met requirements.  In the interest of space, we will not describe every fabrication detail and test result, but we will show some highlights that represent typical results, along with some notable cases.

\subsubsection{The C1 and C4 lenses}

The C1 and C4 blanks arrived at L3 Brashear on different dates, and therefore each could progress through their fabrication steps on a staggered schedule.  As much as possible, work on both lenses was done in parallel using different work stations.  The following descriptions apply to both C1 and C4.

The C1 and C4 blanks were first placed on a large 6-axis Computer Numerical Control (CNC) machine, to quickly cut away the excess volume from the cylindrical blanks, resulting in a shape close to that of the final lens.  This is known as shaping, or generating.  The CNC machine cuts down the blank to the final outer diameter, adds bevels, and generates the desired curvature into the front and back surfaces.  This is done very precisely: the CNC machine can generate large optics with form errors less than 25\,microns and leave clean surfaces with little subsurface damage.  The machining step leaves some extra material thickness on the front and back surfaces which will be removed as part of the more careful polishing in the next steps.  The glass is machined using diamond-encrusted grinding cups of progressively finer grit.  The final grit used is 400, leaving a finish of better than 1\,micron RMS.  This is within the requirements for the edges and flats surfaces, and no more treatment is done on these surfaces; all future work focused on the curved front and back surfaces.

After the generation, the vendor checked all the dimensions of the blank using a large CMM, including overall wedge, and radii of the front and back surfaces, to confirm that all tolerances were met before proceeding.  The CMM probed a series of points on the various surfaces of the lens to establish absolute relations between them, accurate to a handful of microns.

Next, the front convex surface of the lens was polished using a large pitch lap.  The four laps were designed and fabricated in advance, before the blanks arrived at the vendor facility.  Since the surfaces are spheres, the laps can be as large as a meter in diameter - near the size of the lens itself - and progress in polishing was quick due to the large areas being polished.  The vendor polished with increasingly fine polishing compound, leading to a very smooth spherical surface.

As the surface approached its final shape and smoothness, the vendor periodically measured the surface to confirm the correct figure.  This was done using swingarm profilometry \citep{anderson95}.  A high-resolution profilometer is mounted on a precision rotation mount that “swings” the profilometer across the diameter of the lens, over a predetermined diametrical arc in space that closely matches the target radius of the surface.  The profilometer measures height errors over the length of the arc as differences from the ideal predetermined path.  The lens is turned on a rotation table underneath the swingarm, and a series of evenly-spaced arcs is collected.  Finally, the series of arcs, combined with several calibration measurements, is converted by software into a two-dimensional (2D) map showing the height errors of the entire surface.  This is a reliable, high-precision, and quick method for measuring the figure of a surface directly in situ while it is mounted face-up on the polishing station.  Note that the swingarm measures figure errors very accurately but is not sensitive to the overall radius of the surface; radius must be checked and verified by another method. 

When the surface was well polished and close to the final requirement for figure, there remained some residual low-spatial-frequency errors on the order of 1\,micron.  At this point, the lens was moved to a commercial magnetorheological finishing (MRF) machine for its final figuring.  The MRF machine is programmed to deterministically remove small figure errors from a surface, using a fine jet of slurry applied to the surface in a raster pattern.  The surface errors are removed precisely, and a very smooth surface is maintained.  The MRF is programmed with a “removal map”, a description of the surface figure error calculated from its last profilometry measurement.  After the MRF process, the C1 lens was remeasured; this process was repeated several times to bring the final figure within the requirement.

Once the front convex surface of the C1 lens was complete, the lens was turned over, and a similar process performed on the back concave surface.  This time however, after the surface was close to its final polish, the vendor measured the surface using interferometry.  Interferometry is a conventional technique for measuring an optical surface with high precision, and is preferred over swingarm profilometry, though profilometry is more convenient than interferometry for convex surfaces like the front of the C1 lens.  After polishing, the C1 lens was mounted on its edge, supported by two rollers that add repeatable mounting forces and gravity distortions of the lens during measurement that can be calibrated out.  A commercial interferometer was set up at the center of curvature of the concave surface to directly measure the surface; this is a traditional configuration for testing spherical concave surfaces \citep[e.g.,][]{malacara}.  This test showed small residual low-frequency errors, again on the order of 1\,micron.  The lens was then moved to the MRF station for its final figuring as before, this time using a removal map based on the interferometer data.

After their respective MRF operations, both C1 and C4 were considered complete and next underwent verification testing.  Every requirement was addressed by at least one measurement to verify it was met.  Some measurements were taken during the fabrication process, and some were after the lenses were complete.  In all cases, the finished lenses were within specification, as shown in Table \ref{tab:reqspectable}.

\begin{figure}
\centering 
\includegraphics[width=.8\textwidth]{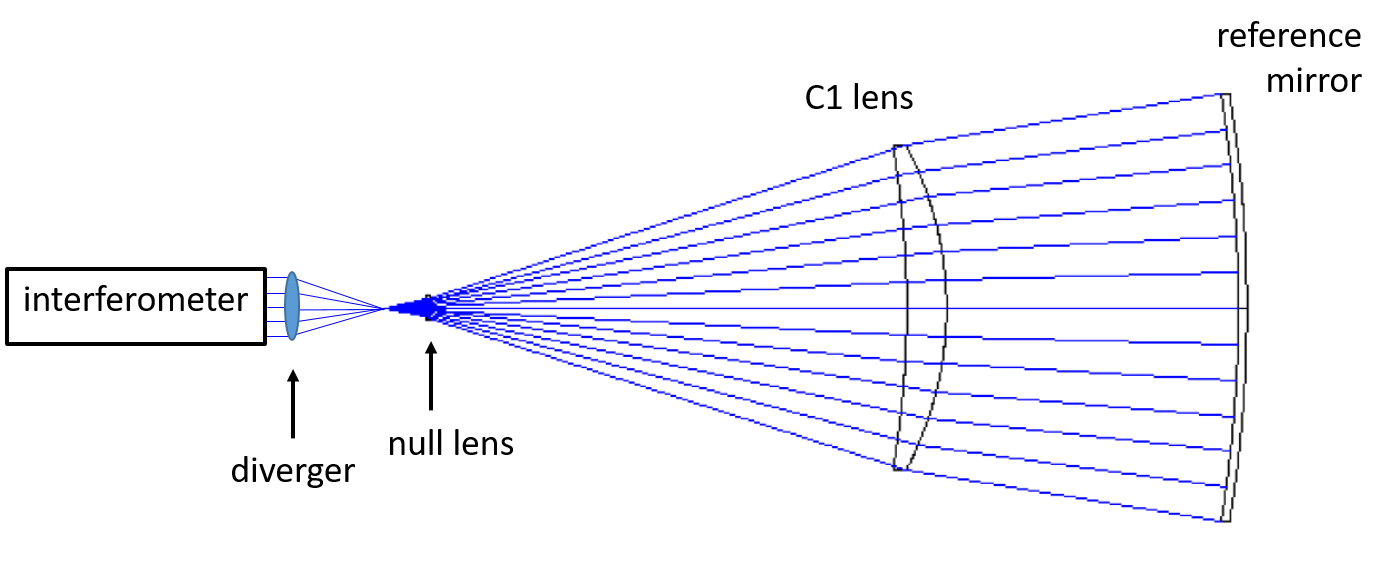}
\caption{Layout of the C1 transmission test.  Configuration includes a 1.5\,meter diameter spherical reference mirror and small null lens to remove test aberrations.  (Credit: L3 Brashear)}
\label{fig:C1translayout} 
\end{figure}

The low and middle frequency surface error requirements were verified simultaneously by interferometry for the C1 lens.  However, the front and back surfaces were not measured individually, as the requirement would suggest.  After some discussion with the vendor, DESI decided to instead test the lens performance with a single transmission test by interferometry, and updated the surface requirement to an equivalent total lens transmission requirement.  The first advantage to this was that it resulted in a simpler and faster test; measuring the surfaces separately would have been more onerous, especially the convex surface.  The second advantage was that it was a better representation of how the lens would be used in the corrector.  The third advantage was that it provided an extra verification of the homogeneity of the lens material, since its effect would appear in the test data.  As shown in Figure \ref{fig:C1translayout}, the transmission interferometry test was configured as a double-pass test, using a 1.5\,meter diameter concave reference mirror.  The reference mirror was first measured by itself, so that its own surface errors could be removed from the test data using software.  The test also included a small singlet lens near the interferometer focus to remove aberration from the total configuration, making it a null test.  During the test the C1 lens was measured with different orientations, in order to remove the effects of gravity in the data.

\begin{figure}
\centering 
\includegraphics[width=.6\textwidth]{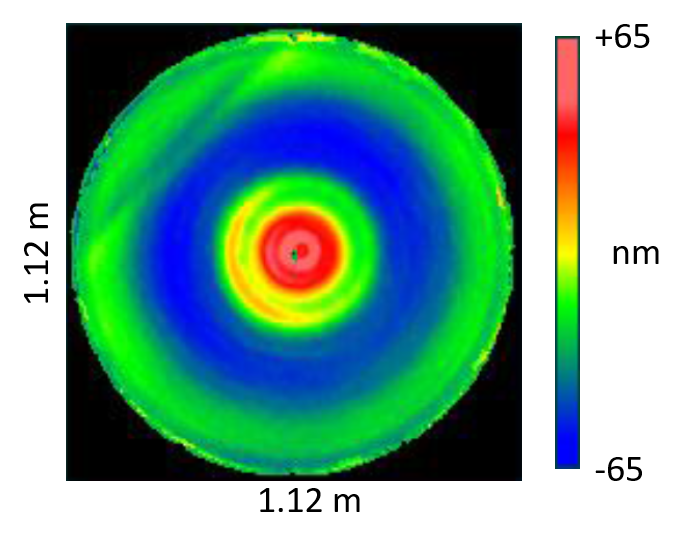}
\caption{Measured wavefront error of the C1 lens.  Plot shows the calculated single-pass transmission error in nm.  The central rings are due to the known inhomogeneity in the C1 glass.  The step feature in the upper left is due to a MRF process error during final polishing.  (Credit: L3 Brashear)}
\label{fig:C1transmdata} 
\end{figure}

The resulting interferogram was used to calculate the single-pass wavefront error due to the lens transmission.  Figure \ref{fig:C1transmdata} shows this measured single-pass error map.  One particular feature is the presence of ``rings'' in the center of the map.  These rings in the wavefront are not surprising: they correspond to the known inhomogeneity in the C1 glass, as seen in Figure \ref{fig:homogmaps}.  Another feature is a step-like error at the upper left in the Figure, caused by a known small process error in the last MRF run on the convex surface.  The vendor filtered the measured data by the defined bands for low and middle spatial frequencies using software, and the surface errors were shown to meet the requirements for both.

In the case of the C4 lens, the surfaces are both strongly convex, making interferometric testing impractical, and so the final figure requirements for low and middle frequency surface errors were verified by swingarm profilometry.  As before, each surface was measured with a dense array of radial arcs, which were assembled into a 2D surface error map.  The vendor then filtered the map by the defined bands for low and middle spatial frequencies, and the results compared favorably to the two spatial frequency requirements.

However, because swingarm profilometry stitches the 2D map together from 1D arcs, DESI and the vendor decided to measure the surface again in a different way to build confidence that the figure was well known.  The concern was that the radial arcs might not capture middle-frequency surface errors accurately in the azimuthal direction.  Therefore, the vendor measured several azimuthal data sets as well: the table was rotated continuously on its air bearing while the profilometer head was held at a certain radial position above the lens surface.  The data were then filtered as before for the appropriate spatial frequencies (10mm and above).  The vendor repeated this process at three different radial positions, and demonstrated that the surfaces indeed met their requirements.  The data points were not dense enough to assess the slope error between frequencies from 10mm to 20mm; in this case they confirmed that the requirement was met by filtering the data from the high spatial frequency verification test (see below).

\begin{figure}
\centering 
\includegraphics[width=.8\textwidth]{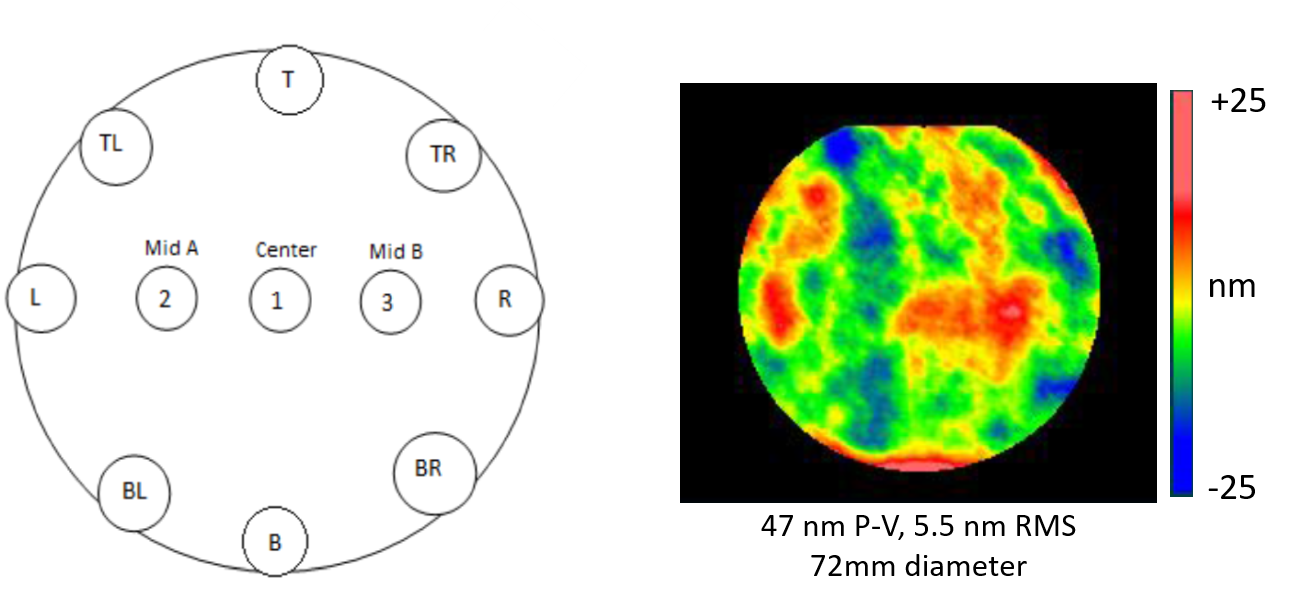}
\caption{Measurement of high frequency figure error.  Left: locations of measurements on C1 convex surface.  The large circle represents the surface clear aperture, and small circles represent the 11 sample locations.  Locations are noted with the labels used by the vendor.  Right: data for the "TR" sample; one edge of the sample is truncated since the sample is located near the lens edge.  (Credit: L3 Brashear)}
\label{fig:C1hifreqlayout} 
\end{figure}

To verify the high spatial frequency requirement for surface error, the vendor measured 11 locations evenly distributed over the clear aperture using a 4” interferometer and diverger attachments that appropriately matched the surface radius.  Figure \ref{fig:C1hifreqlayout} shows the 11 locations for the case of the concave surface of C1 and data for one sample.  Measurement of high frequencies was limited by the detector resolution of the interferometer mapped to the surface, so the areas to be measured were required to be much smaller than the full aperture – from 62mm to 220mm diameter, depending on the surface being measured.  These small areas corresponded to detector resolutions of well under 0.5mm, allowing verification of the high spatial frequency requirement at 1mm cycles and greater.  The lens rests face-up for this measurement, and the interferometer is mounted safely to the side of the lens, with its beam path directed to the lens using a securely mounted 45-degree fold mirror.

After collecting the data map, the vendor filtered it with software to isolate the specified frequencies as before and verified the requirement was met.  As an additional check on C1, the vendor inspected the interferometric data from both the transmission test and the back-surface reflection test to confirm that they were consistent with the measured high-frequency data.

Surface roughness was verified in a similar way.  The vendor measured nine locations over the lens diameter, this time inspecting a small area of view of approximately 1\,mm$^2$.  This small area was sufficient to measure the very fine spatial frequencies defined for roughness.  The vendor measured roughness by use of RTV replicas, which the vendor used to imprint the roughness pattern, and measure using a white light interferometer.  Applying RTV replicas is a standard method for measuring roughness, and the transfer introduces only a slight error that is well under the requirement.

\begin{figure}
\centering 
\includegraphics[width=.8\textwidth]{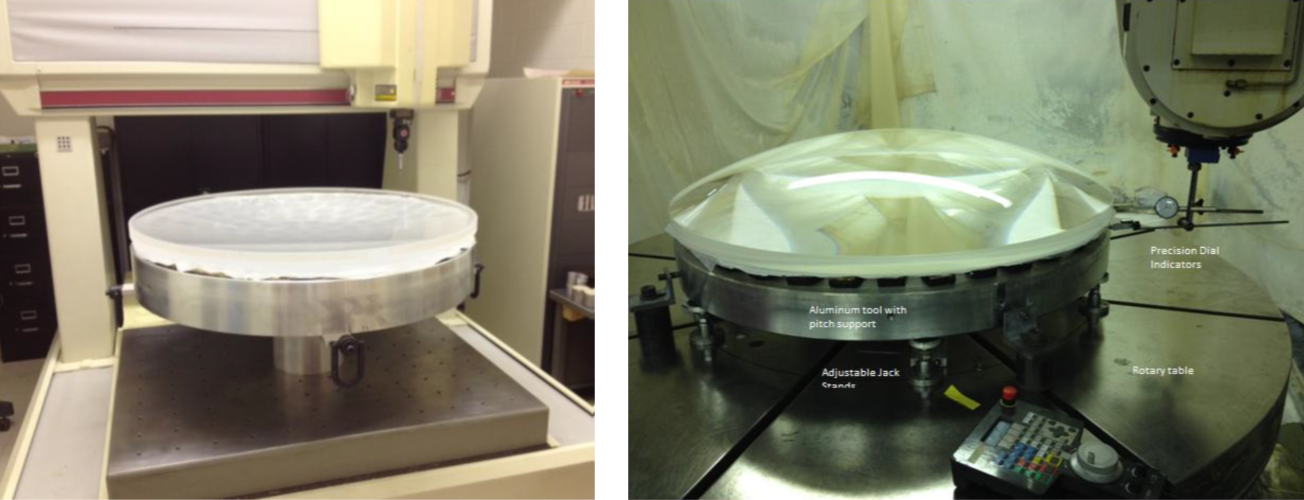}
\caption{Measurement of C1 lens dimensions.  Left: radii and thickness of the lens measured by a CMM.  Right:  lens wedge measured by dial indicators on a rotary table.  (Credit L3 Brashear)}
\label{fig:L3CMM} 
\end{figure}

The final verifications of surface radii and lens thicknesses were performed using a large CMM (Figure \ref{fig:L3CMM} left).  The vendor measured an array of points across two different diameters of each surface and fit them with curves in software to find the radii.  The CMM was not able to directly measure the center thickness of such large lenses, so separate measurements were taken and pieced together.  For C1, the CMM measured the locations of the center of the front and back surfaces with respect to the edge flat on the back side, and calculated the difference.  For C4, the lens was mounted on the CMM table front-surface-up, and a stack of shims and a gauge block were inserted below the back surface to just touch the center vertex; then the stack was moved to the side, and the CMM measured the difference between the stack height and the front vertex.

The lens wedge requirement was verified by turning the lens on a rotary table, and measuring the runout on the edges of the front and back surfaces with a dial indicator (Figure \ref{fig:L3CMM} right).   The indicator is accurate to 2.5\,microns, and so the error in the wedge measurement well below the requirement.

\subsubsection{The ADC1 and ADC2 lenses}

The overall approach to processing and verifying the ADC lenses at Rayleigh Optical Corporation was similar to that of C1 and C4, described above, and the steps in common will not be repeated here.

\begin{figure}
\centering 
\includegraphics[width=.8\textwidth]{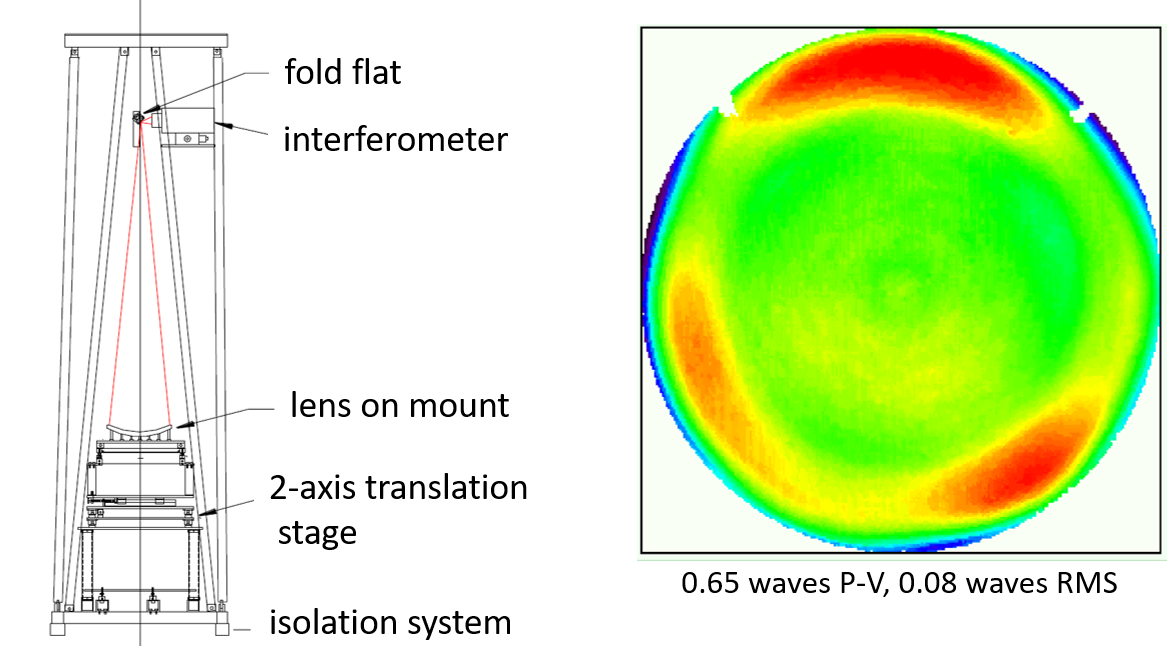}
\caption{Measurement of the concave lens surfaces.  Left: the test tower configuration.  Right: the measured concave surface of ADC1.  The small gaps at the top are fiducial markings used to orient the lens during measurement.  (Credit Rayleigh Optical Corporation)}
\label{fig:ROC_concavelayout} 
\end{figure}

As the lens surfaces approached the end of their polishing, the concave sides were measured in an interferometric test station to check when polishing should stop.  Figure \ref{fig:ROC_concavelayout} (left) shows the test station, referred to as the “test tower”.  The lens rests with its concave surface facing up on a rotary table in the test station, and a small 45-degree mirror above the lens folds the beam to a Fizeau interferometer mounted to the side.  The fold mirror and the interferometer's reference sphere are measured so that their own small figure errors can be subtracted from the test data.  Furthermore, the concave surface is measured multiple times with the lens rotated in 90-degree steps, so that systemic errors can be removed from the test data.  These calibrations are accounted for in software after the measurements are taken, and the result is a corrected map of the concave surface.  Figure \ref{fig:ROC_concavelayout} (right) shows the calibrated map of the ADC1 concave surface, as an example.  The surface figure is quite smooth with very low slope errors, and the errors are strictly low-order, as is to be expected since the surface was polished with a large tool.  The surfaces easily met their low and middle spatial frequency requirements.

\begin{figure}
\centering 
\includegraphics[width=.8\textwidth]{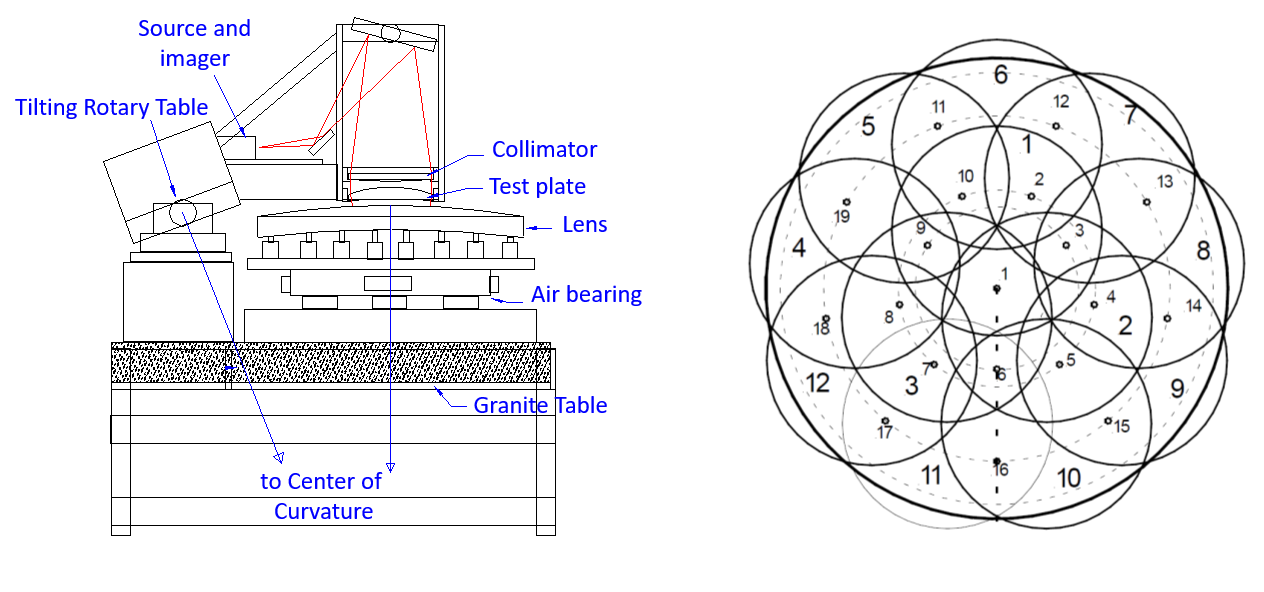}
\caption{Measurement of the convex lens surfaces.  Left: diagram of the Convex Surface Interferometric Test Stand.  The interferometer, fold mirrors, collimator, and test plate were moved as a unit by the tilting rotary table in order to measure multiple subapertures across the full surface aperture.  Right: map of the 12 subapertures covering the 800-mm diameter clear aperture of ADC1.  The 19 small circles show the locations of fiducials used for stitching together the subapertures into a full-aperture map.  (Credit Rayleigh Optical Corporation)}
\label{fig:ROC_CXTS} 
\end{figure}

In the case of the convex side, the vendor measured large subapertures across the full diameter using test plates.  Test plates are a traditional method for measuring large spherical surfaces in situ: they are fabricated ahead of time to match the nominal surface curvature, and when placed against the surface they act as a Fizeau interferometer \citep[e.g.,][]{malacara}.  The vendor fabricated 368-mm diameter test plates that matched the nominal convex surfaces of ADC1 and ADC2, and placed them at various locations across the ADC surfaces for measurement by an interferometer as shown in Figure \ref{fig:ROC_CXTS}.  The figure shows the tilting rotary table that moves the interferometer, fold mirrors, collimator and test plate as a unit, in order to measure multiple subapertures across the full surface aperture.  The axes of the rotary table were aligned to the center of curvature of the lens surface being measured, so that the test plate stayed aligned to the lens as it was moved to measure different subapertures.  Once the subaperture data were taken, the vendor stitched them together in software to produce a complete map of the full surface.

Unlike the C1 and C4 lenses, the prescription of the ADC lenses includes a wedge of roughly 0.25\,degrees.  More accurately, one of the lens surfaces has a 0.25\,degree tilt, and the other is normal to the center axis.  This wedge is large enough that it can simply be machined into the blank during the generation process.  First the CNC machine generated the unwedged surface of the lens, centered on the blank’s axis.  Then, because a tilted spherical surface is equivalent to a decenter of that surface, the blank was decentered on the CNC machine to generate the second curved surface.  The result was an overall wedge in the lens.  The vendor verified the wedge by rotating the lens on an optical table, and measuring the runout with a dial indicator against the outer edges of the front and rear surfaces.  The radii of curvature were checked with a precision spherometer.

\subsubsection{The C2 and C3 lenses}

The approach to polishing and verifying the C2 and C3 lenses at Arizona Optical Systems is similar to the other lenses, and we will not describe the common steps again.

However, the C2 and C3 lenses are significantly different from the other lenses in that they each feature an aspheric curved surface.  The surfaces are 10th-order polynomial aspheres with significant magnitude (see Table~\ref{tab:corT5}), and while they were designed to be within the vendor’s capabilities, aspheres require a different approach to tooling and polishing.

Simple spheres can be polished using large-area tools moved across their entire diameter and thus material is polished away quickly.  As described above, the tools used for the four spherical lenses were up to a meter in diameter, near the sizes of the lenses themselves.  But aspheres require small-area tools matched to the local curvature value, less than 25mm in size for some parts of the DESI lenses, and so material is removed at rates several orders of magnitude slower.  Furthermore, small tools polish aspheres to their correct shape effectively, but they also have the potential to leave residual polishing errors, particularly in the middle and high spatial frequencies.  Therefore polishing runs become shorter and more conservative, and are interrupted with more frequent measurements, which slows down the polishing process near its end.  Finally, the vendor must use a finer slurry for final figuring in order to achieve the required micro-roughness of the surface, which has a slower removal rate.  For all these reasons, the aspheres took much longer to fabricate than the other lenses: about three years for C2 and C3 together, compared to roughly 1 to 1.5 years for the other pairs of spherical lenses.  

The removal rate is asymptotic, i.e. progress towards the final figure is slowest at the end.  In the case of the C2 asphere, as the figure neared its final shape, DESI decided to model what performance degradation there might be if figuring were stopped, based on the current measured surface.  All lens requirements were otherwise met except for the mid-spatial slope requirement (as shown in Table \ref{tab:reqspectable}), and the analysis showed that the subsequent imaging error was acceptably small.  Thus we directed the vendor to stop the final figuring runs, and saved several months of schedule.

The C2 and C3 lenses received more attention due to their complex shape, and the vendor measured their parameters using multiple independent methods to be absolutely sure that the lenses were well understood and met all its requirements.  Methods included spherometry, swingarm profilometry, CMM profilometry, deflectometry, interferometry, and measurement by laser tracker.  Here we describe some verification tests that were unique to this vendor.

\begin{figure}
\centering 
\includegraphics[width=.8\textwidth]{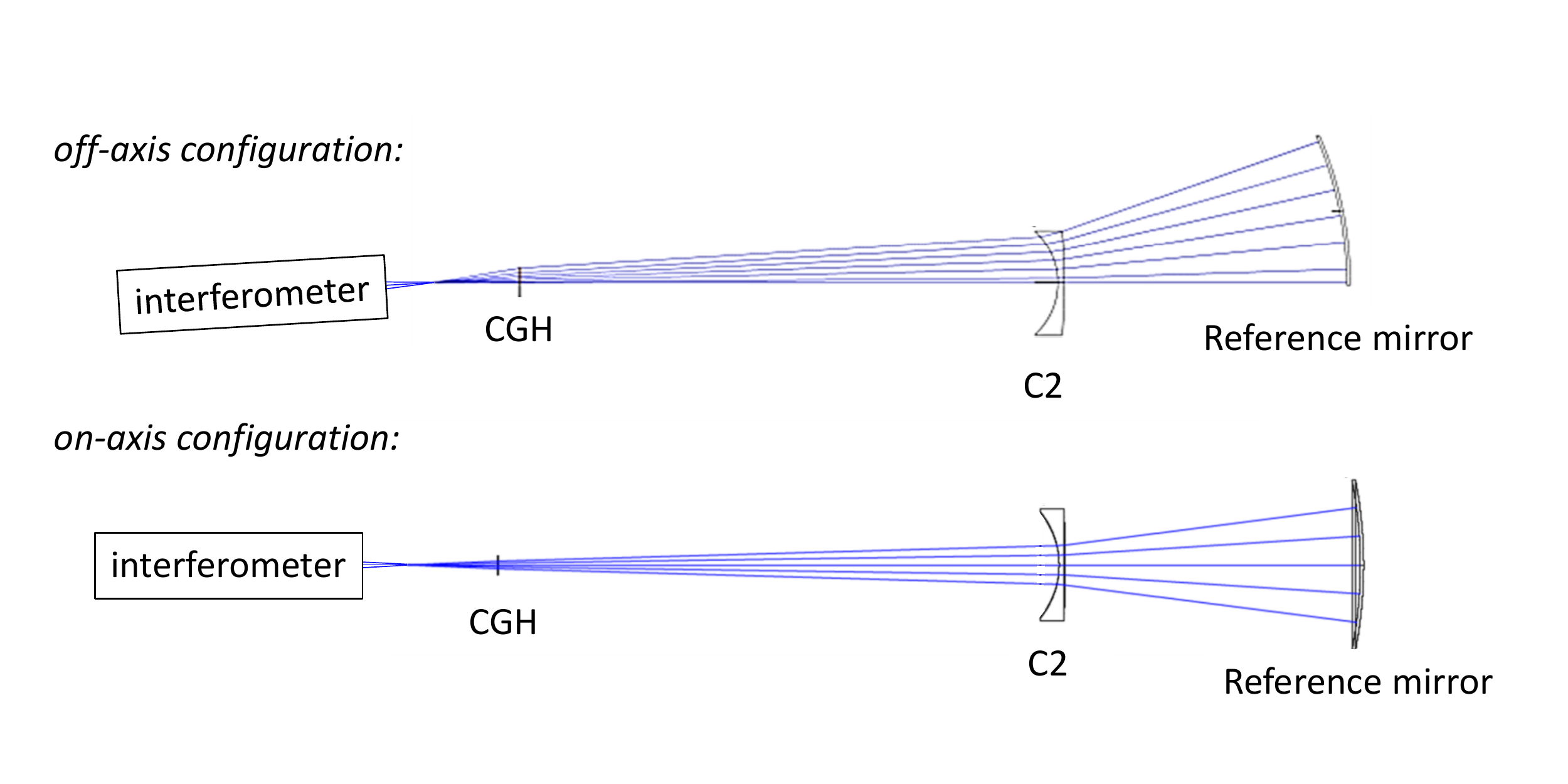}
\caption{Layout of the C2 transmission tests.  C2 is tested in a double pass configuration using an interferometer, a custom computer-generated hologram (CGH) to remove test aberrations, and a large spherical reference mirror.  Top: off-axis test.  The C2 lens is rotated to collect all off-axis subapertures.  Bottom: on-axis test.  (Credit Arizona Optical Systems)}
\label{fig:C2transmsetup}
\end{figure}

The middle and low frequency requirements were verified in a series of interferometric tests.  In the case of the convex C2 asphere, the lens was tested in transmission instead of measuring the surface directly, similar to the C1 lens.  Figure \ref{fig:C2transmsetup}, top, shows the test configuration.  The lens was measured in a traditional double-pass configuration using a 1.24-meter spherical reference mirror and a computer-generated hologram (CGH) which together allowed null testing of the asphere.  The test could only measure subapertures of the lens due to the limited size of the reference sphere, and so the test measured twelve subapertures of about 380mm diameter, clocked 30\,degrees apart, each of which spanned from the center of the lens to just outside the clear aperture.  The vendor then stitched the subaperture data into a full lens map using commercial software. 

The C2 lens was supported by a steel double rollerchain, a traditional method that minimizes mounting distortions in the glass, and the lens was protected with a thin Delrin strip wrapped inside a surrounding band of sheet metal to prevent metal-to-glass contact.  In order to measure the lens as accurately as possible, the vendor measured the reference mirror errors, the CGH errors, spacing errors in the setup, gravity effects, and calibrated out all of their contributions to the lens measurement.  

\begin{figure}
\centering 
\includegraphics[width=.6\textwidth]{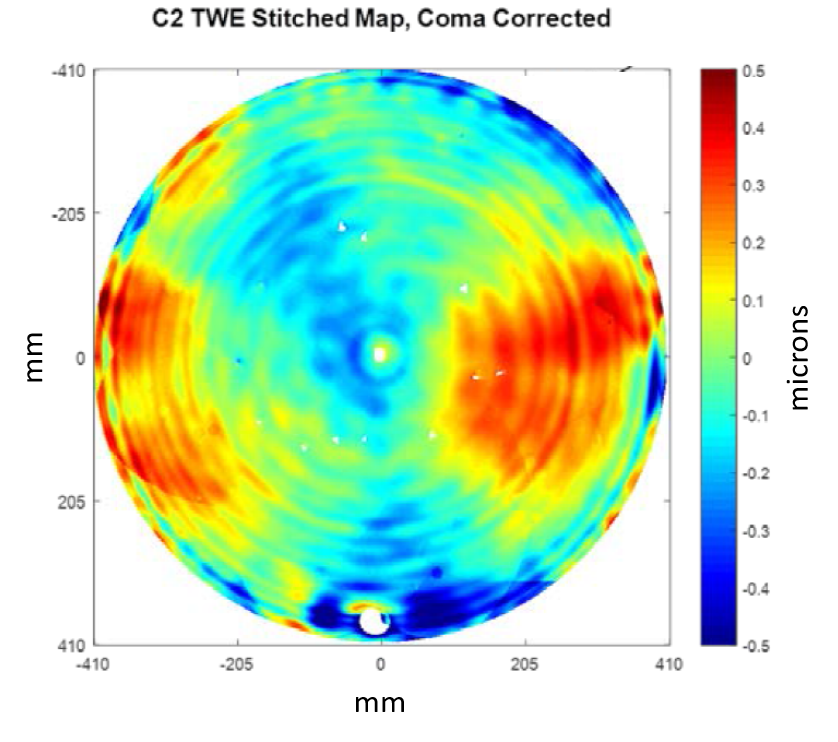}
\caption{Measured wavefront error of the C2 lens.  Plot shows the calculated single-pass transmission error in microns.  The map was created by stitching together 12 off-axis subaperture measurements and calibrating out test errors.  A radial pattern of middle-frequency polishing errors is visible, caused by the small tools used during polishing.  A series of small fiducial marks were placed on the lens to enable subapertures to be stitched together; these are blocked out in the data (white spots).  One particular low spot (at the bottom of the plot) is also blocked, being too steep for the interferometer to measure accurately; this was assessed separately.  This surface was analyzed and accepted by DESI.  (Credit Arizona Optical Systems)}
\label{fig:C2transmdata}
\end{figure}

Figure \ref{fig:C2transmdata} shows the final transmission measurement.  The residual polishing errors are visible in a radial pattern; these features are due to the small tools used to achieve the final figure.  There is also one particular low spot at the bottom of the plot that was too steep for the interferometer to measure accurately, and is blocked in the data; the vendor assessed this separately with the profilometer and combined its data with the interferometer map for an overall surface estimate.

To assess the low frequency requirement for the aspheric surface using the lens transmission data, the vendor made the assumption that the asphere was responsible for the entire transmission error as a conservative upper limit.  Given this assumption, the surface met its requirement with a large margin.  To assess the middle frequency requirement, the vendor subtracted the surface error of the spherical side, measured in a separate interferometric test, and the remaining surface error was compared to the middle frequency requirement.  As noted above, DESI assessed this error at the time when the figuring neared completion, and accepted a middle frequency figure slightly above requirement, to save several months of schedule.

\begin{figure}
\centering 
\includegraphics[width=.9\textwidth]{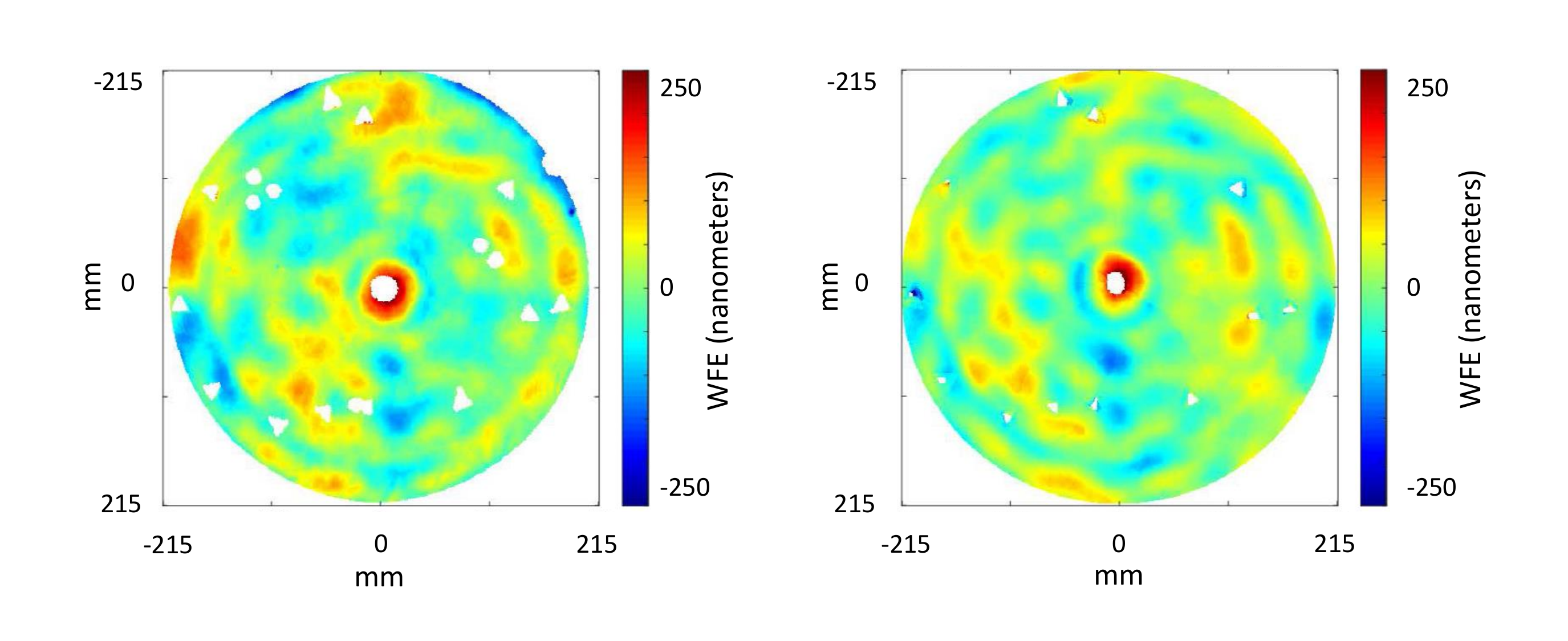}
\caption{Independent measurements of C2 transmitted wavefront error.  Left: on-axis test data.  Right: central portion of the off-axis test data.  The two data sets are very similar, thus validating each measurement process.  (Credit Arizona Optical Systems)}
\label{fig:C2comparemethods}
\end{figure}

The vendor also measured the on-axis transmission of the C2 lens using the test configuration shown in Figure \ref{fig:C2transmsetup}, bottom.  This was done as a validation of the off-axis test rather than a contractual verification, providing an independent check of the lens polishing errors.  Again due to the limited size of the reference sphere, the test only measured the central 430mm diameter of the lens.  A new CGH was also required for this test.  Figure \ref{fig:C2comparemethods} compares the on-axis data with the equivalent central portion of the stitched map from the off-axis test, giving confidence in the validity of the off-axis data.  Some low-order terms (Zernike terms 1-11) are filtered from the two data sets to remove residual alignment error contributions in each.

\begin{figure}
\centering 
\includegraphics[width=.8\textwidth]{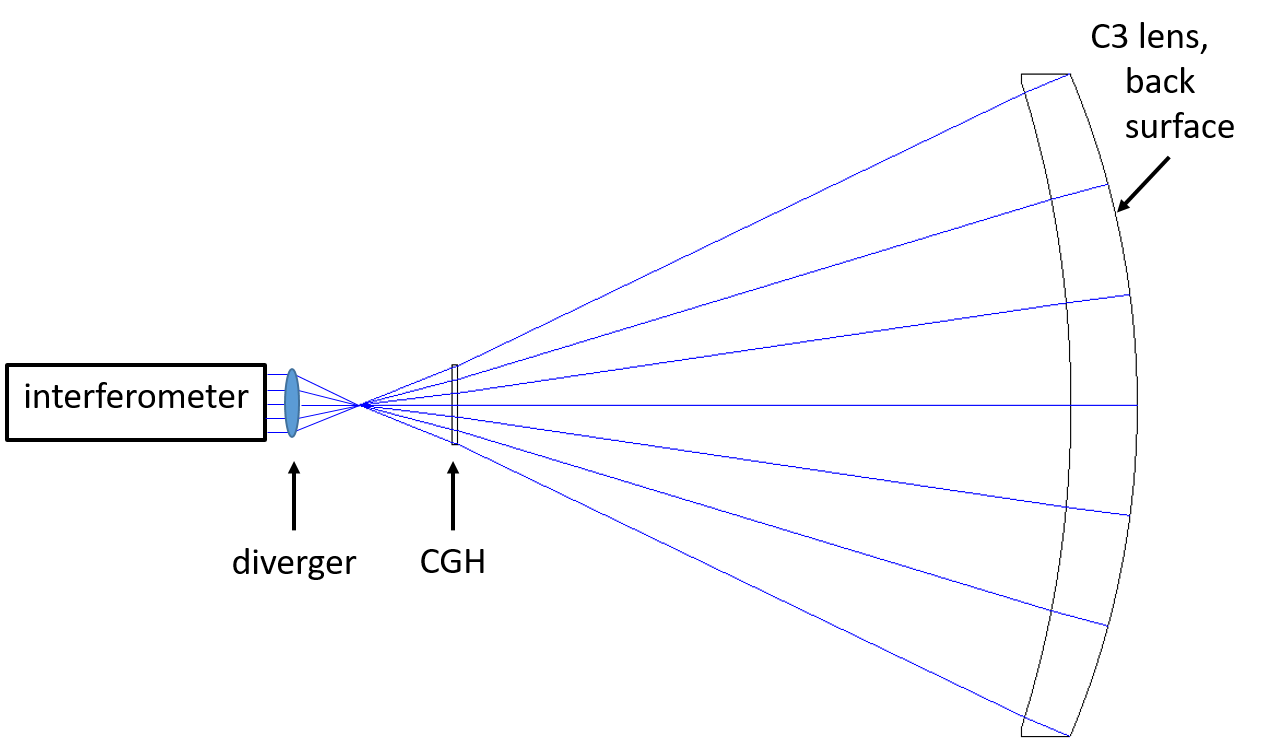}
\caption{Layout of the C3 convex surface measurement.  The ''Mangin'' configuration measures the total wavefront error of the double-pass transmission and back surface reflection of the lens.  The  vendor calculated the convex surface profile from this by subtracting a separate measurement of the concave surface.  The configuration includes a custom computer-generated hologram (CGH) to remove test aberrations.  (Credit Arizona Optical Systems)}
\label{fig:C3Manginlayout}
\end{figure}

The spherical convex surface of the C3 lens was measured interferometrically in transmission to verify its low and middle frequency surface requirements.  Figure \ref{fig:C3Manginlayout} shows the test layout, known as the “Mangin” configuration: the interferometer beam passes through the C3 aspheric concave surface, through the lens material, and reflects from the C3 spherical convex surface back through the lens.  In this case the geometry of the lens allows for a single full-aperture measurement of the convex surface.  Again this requires a custom CGH to allow for a null test.  As before, the test aberrations, CGH error, gravity distortions, and the contribution of the asphere surface were all calibrated out of the measured data, resulting in an accurate measurement of the convex surface.  The surface met its requirements.  The vendor did not remove the error due to the transmission through the glass itself, because its contribution to the total was acceptably small.  The spherical surface was also measured periodically during its fabrication to confirm that the radius and figure were on track to meet the final requirements.  The vendor did this using 180-mm test plates in a Fizeau configuration.

Both of the concave surfaces of C2 and C3 were measured directly by interferometry, using a center-of-curvature configuration similar to that used by the ADCs.  These tests measured the full diameter of the surfaces, so no subaperture measurements were needed.  The C3 test required a CGH to correct for the lens surface asphericity, whereas the C2 test did not require a CGH since the surface is spherical.  Both surfaces met their low and middle spatial frequency requirements.  The C3 aspheric surface, similar to the C2 asphere, had a particular low area that was too steep to be measured by the interferometer, and thus was blocked out in the surface map generated by this test.  However, because the vendor used swingarm profilometry to monitor the surface figures during fabrication (as was done by the vendors of the other DESI lenses), the swingarm data contained a measurement of that low area.  The vendor isolated the measurement of the low area from the last swingarm test and combined it with the full interferometric map to verify that the surface met its requirements.


In addition to the C3 lens surface measurements previously described, the vendor performed a transmitted wavefront error test of C3 as an independent check of the lens performance. The test layout used a 1.4\,m reference sphere and a CGH to measure the lens in double pass, covering the central 720\,mm of the lens.  The resulting data map was consistent with the other interferometric measurements of C3, providing confidence that the lens was well-understood and within its requirements.

To verify the radius of curvature, the vendor measured the lens surfaces using spherometers, both as a final measurement and also to check progress during the polishing progression.  The spherometers were calibrated using the vendor’s CMM, and the radius calculations were carefully reviewed by the vendor as well as by the DESI team.  Some radii were also measured using additional, independent methods.  For example, the C2 and C3 spheres were also measured using a laser tracker; the tracker collected an array of data points across the surface, and software fit the best spherical surface to them.  These tests are complementary: the spherometer is a fast, convenient test but takes a single data point that is indirectly converted to a radius, whereas the laser tracker is less convenient but captures more data points directly from the lens surface.  

The C3 concave surface radius of curvature is difficult to measure by spherometer or laser tracker because of its aspheric shape, so the vendor performed a further verification of the C3 radius during its final interferometric testing.  After the setup was well aligned, the separations between the test components (interferometer focus, CGH, lens) were measured accurately by laser tracker, the test was modeled in Zemax software, and the radius determined from the model.  This independent check again confirmed the C3 radius.

Finally, the C2 lens surfaces and radii were measured independently early in their fabrication by NASA's Goddard Space Flight Center using a large CMM at their facility.  The surfaces were monitored during the fabrication by swingarm profilometry directly at the polishing station, but this method is insensitive to lower-order errors such as radius.  Spherometer measurements filled in the gap, but these were indirect measurements prone to error, especially with aspheric surfaces.  Therefore, DESI and the vendor agreed to pause the polishing and ship the lens to the Goddard facility.  Both lens surfaces were measured there by directly probing a 2D grid of points over the full diameter, and generating accurate surface maps.  This check caused a three-week delay, but it successfully validated the vendors process for tracking low-order errors, and reduced the risk of uncovering an incorrect radius or figure later.

\begin{figure}
\centering 
\includegraphics[width=.3\textwidth]{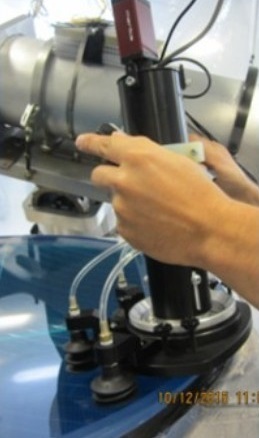}
\caption{Measurement of the C3 concave aspheric surfaces using the MSF-FORM instrument.  (Credit Arizona Optical Systems)}
\label{fig:C3MSFFORM}
\end{figure}

To verify the high frequency surface requirements, the vendor used a deflectometry method.  In deflectometry, the user illuminates an optical surface using a device that projects a particular test pattern, observes the image of the reflected pattern with a camera, and calculates the surface error based on the measured distortions of the pattern \citep[e.g.,][]{su2010,Maldonado2014}.  The vendor had developed its own portable deflectometry method for testing optics, Flexible Optical Ray Metrology (FORM).  The vendor's FORM instrument was designed specifically for measuring middle and high spatial frequencies (MSF-FORM) and was capable of measuring slope errors with better than 100\,nanoradian accuracy, with a range of several milliradians and 100-micrometer spatial resolution.  This instrument was ideal for verifying the DESI lens requirements.  Since the lens surfaces were too large to measure all at once with sufficient spatial resolution, the vendor measured a series of 10-20 samples distributed across the lens surfaces, enough to capture possible variations with location.  Each measurement covered a 40-mm diameter sample, enough area to calculate the high frequency surface error.  The vendor then calculated a weighted average of the samples, and showed that the surfaces met the requirements.  Figure \ref{fig:C3MSFFORM} shows the vendor using their MSF-FORM instrument to measure the high frequency surface error at a location near the edge of C3.

The vendor used a commercial interferometric microscope to verify the roughness requirement, the MicroFinish Topographer (MFT) from Optical Perspectives Group.  The portable and lightweight MFT measures roughness directly on an optic in situ, measuring a small area slightly less than a square millimeter.  The vendor measured 13 locations throughout the lens surfaces with the MFT, with arrays across two diameters, and showed that the average roughness met its requirement, with almost all samples meeting the requirement individually.

\subsection{Lens Coatings} 

The corrector’s need for high throughput means that all lens surfaces must be treated with an antireflection (AR) coating.  Without that coating, the expected 4\% Fresnel loss from each of the twelve surfaces would be unacceptable, and ghost reflections would be significantly stronger.

The coating requirements are shown in Table~\ref{tab:coatreqs}, the most important of which is that the surface transmission must be above 98.5\% average over the DESI bandpass.  This requirement is based on meeting a system-level throughput allocation for the overall corrector; this is a challenging requirement over the wide DESI bandpass, but it was chosen to be consistent with vendor capability.  This transmission is specified for unpolarized light.  The coating is required to survive the environments of the Mayall dome at Kitt Peak, as well as potential international shipping environments, and also needs to be durable to handle the risk of handling during testing and integration.  


\begin{table}
\centering
\caption{Coating requirements for each lens surface.}
\begin{tabular}{c|c} \hline 

Bandpass	&	360 - 980 nm	\\	
Average Transmission over bandpass	&	98.5\%	\\	
Minimum Transmission over bandpass	&	98.0\%	\\	
Coating Aperture	&	Match surface clear aperture	\\	
Coating Quality	&	Consistent with 80-50	\\	
Durability: humidity	&	MIL-C-48497A, humidity	\\	
Durability: adhesion	&	MIL-C-48497A, quick tape	\\	
Durability: abrasion	&	MIL-C-48497A, moderate abrasion	\\	
Durability: Solubility and Cleanability	&	MIL-C-48497A, acetone and alcohol	\\	
Survivable temperature range	&	-20$^\circ$ to +60$^\circ$C	\\	\hline
\end{tabular}
\label{tab:coatreqs}
\end{table}

Besides these coating requirements, the lenses present other significant challenges to the coating vendor.  First and foremost, the lenses are extremely large, requiring an unusually large coating chamber.  Second, the lenses are heavy, requiring custom handling and transfer fixtures to move them safely.  Third, the polished lenses are unique and expensive, and must be coated correctly only once; a new substrate cannot be supplied easily if the coating fails for some reason.  This uniqueness therefore requires a careful test and demonstration program.

\begin{figure}
\centering 
\includegraphics[width=.8\textwidth]{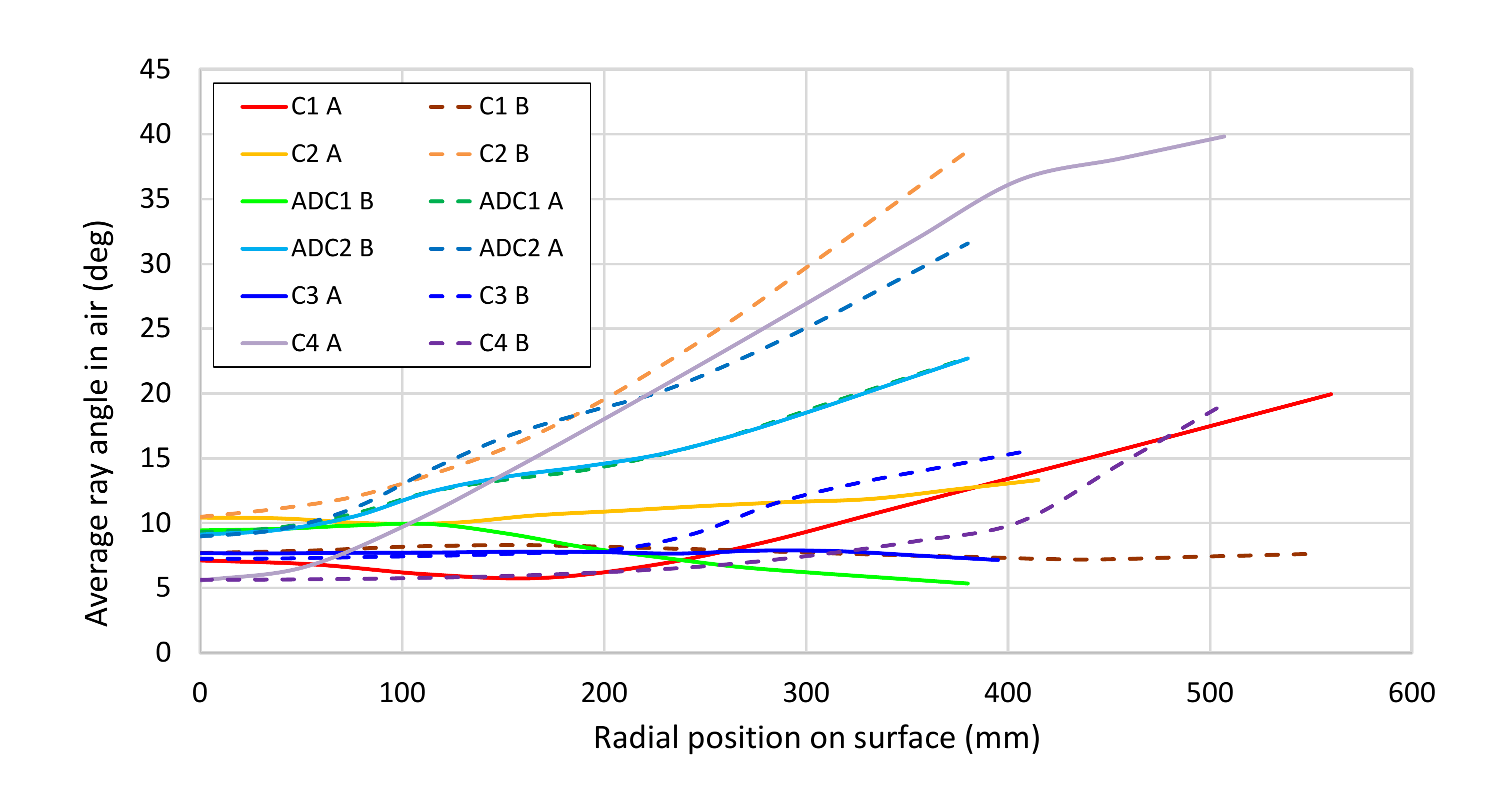}
\caption{Average ray angle vs. position for each coated lens surface.  This information calculated from the optical design was used by the vendor to correctly design the AR coating to work best at these AOIs across the diameters of the lenses. \citep[Image previously published in][]{kennemore18}}
\label{fig:coatingAOIs} 
\end{figure}

Another significant coating challenge is the strong curvature of the corrector lenses, both positive and negative.  Uniformity is not a specific requirement on the lenses, but the coating must have high transmission across their full apertures.  What this curvature means is that the ray angle of incidence (AOI) across each lens can vary considerably.  Figure~\ref{fig:coatingAOIs} shows the AOI as a function of radial position on the lens surface, averaged for all incident field rays there, for all lens surfaces.  Three of the surfaces have angle ranges over 30$^\circ$ with the highest angle approaching 40$^\circ$.  Since coating performance directly depends on AOI, this position variation presents a significant difficulty to the coating design.


The vendor successfully designed an AR coating that meets the demanding requirements above \citep{kennemore18}.  While the specific design details are proprietary to the vendor, the coating is based on a traditional multilayer design.  The coating process uses ion-assisted deposition (IAD), resulting in “hard” coatings that meet all durability requirements.

Solving the coating AOI difficulty means depositing a coating whose layer thickness varies slightly with radial position.  A common technique to achieve this is to use a rotating mask in the chamber that controls how much time is spent exposing different parts of the lens to the evaporated material.  This is a difficult technique that requires careful calibration as well as moving parts in the vacuum chamber.  The vendor took a simpler approach with the DESI lenses instead: the coating chamber geometry allowed multiple radial positions of the deposition sources as well as substrate height variability.  Using these two variables allowed the vendor to vary the coating thickness and avoid using any masking. 

The coating design is robust in that it performs over a bandwidth that is wider than required.  This compensated for the small unavoidable fabrication errors - specifically layer non-uniformity and angle variation - that cause the coating performance to “shift” up or down slightly in wavelength, therefore assuring that a coating deposited on a lens would be acceptable.  This larger bandwidth also allowed a single fundamental coating design to be used for all common substrate surfaces.  (Fused silica and borosilicate coatings used slightly different coating layer thicknesses.)

After the coating was designed, the coating process was verified in a test phase prior to coating a one-of-a-kind optic.  The vendor built a ``surrogate'' assembly to do this, a structure that matched the size and shape of each lens and held arrays of 25\,mm diameter witness samples.  The surrogate supported each witness at the appropriate angle to form a radial cross-section of the lens surface.  The small witnesses could be measured quickly and cheaply with a spectrophotometer, and the results used to adjust the location of the deposition sources in the chamber.  After several iterations, the source locations were optimized, and multiple measurements verified that the coating uniformity was correct and repeatable.  The optimal geometry was found for each of the twelve lens surfaces during the test phase.  When the time came to coat each DESI lens surface, its surrogate assembly test was repeated immediately beforehand several times as a validation of the process.

Having sensitive thin layers in the design meant that controlling layer thickness precisely was critical. The vendor controlled the manufacturing variability to within an acceptable level through the use of strategically-placed quartz crystal monitors in the chamber along with stable deposition plumes. Low deposition rates and high rotation speeds also allowed for better control.

The high value, uniqueness, and large size and mass of each lens necessitated great care in handling.  Each lens went through many steps at the vendor facility - removal from the shipping crate, flipping, cleaning, transport in and out of the coating chamber, heated to deposition temperatures and cooled back down - all while inside a Class 100 clean room.  To meet these needs of safe handling, the vendor designed cleaning rings and supports that allowed the lenses to be easily positioned in any orientation for cleaning, inspection, or transfer to other tooling. The cleaning tooling was also resistant to the cleaning solvents used in cleaning each lens surface. To minimize the thermal mismatch between the lens and tooling during coating operations, the coating tooling was constructed of Invar and length-tuned to allow both the low-expansion fused silica and higher-expansion borosilicate lenses to be safely held during thermal cycles, while also maintaining the required clear aperture. Two rings were used as the common load support for all the lenses, one for cleaning and one for coating, thus simplifying the interface with lifting equipment and other mounts. Each lens required simple adapters to be installed on the common rings to handle the different diameters and surface shapes. The tooling was analyzed using FEA to assure loads and stresses were well within safe limits. Finally, a load test was performed on each tooling at 1.5 times maximum weight before use with any lens.

Care was likewise taken to assure that the overall process was safe and efficient.  The vendor wrote a detailed procedure to step through each operation, and updated it after each lens coating to improve the process for the following lenses.  The procedure was performed each time by the same dedicated and skilled team.  Early versions of the procedure were walked through using dummy masses in place of the lenses in order to identify problems.  One concern was that the lenses might be heated or cooled too quickly in the coating chamber; thermal analysis determined a maximum heating/cooling rate that was put into the procedure.  Overall, all six lenses were processed successfully with no handling errors.

\begin{figure}
\centering 
\includegraphics[width=.8\textwidth]{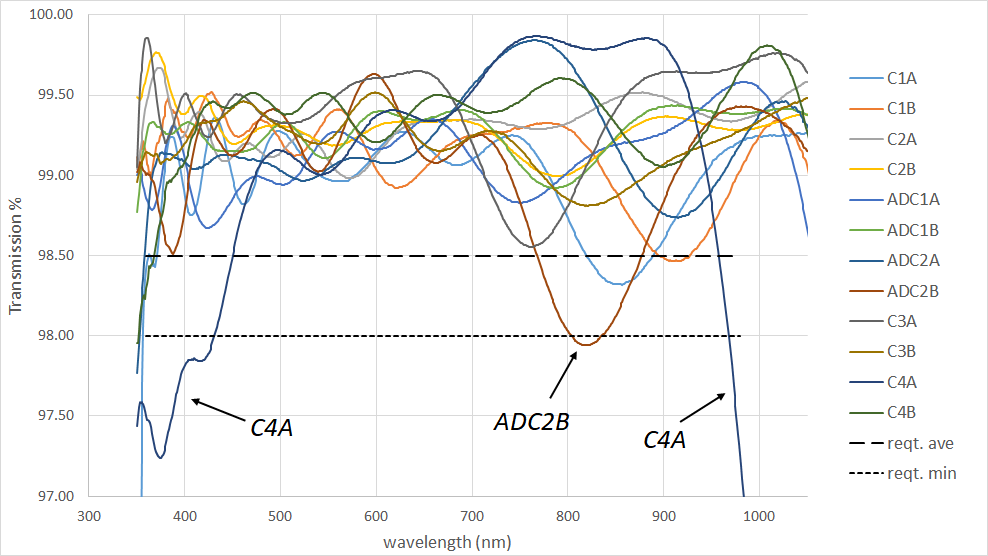}
\caption{Corrector AR coating performance. Data represents the averages of measurements across the lens surface at the specified AOI.}
\label{fig:coatperf} 
\end{figure}

Figure~\ref{fig:coatperf} shows the actual coating performances for all twelve lens surfaces.  These are based on data representing the averages of final measurements across the lens surface at the specified AOI.  Several lenses do not quite meet the exact DESI requirements.  C4A, which was the first surface to be coated, suffered a degradation due to incorrect design of the surrogate assembly that did not represent the actual lens correctly.  This error was tracked down through extensive testing and analysis, and the surrogate assembly fixed; when it was demonstrated to work correctly, DESI agreed to continue the coatings progress, and the problem did not occur again on any more coatings.  DESI also accepted C4A as it was instead of stripping and recoating it.  Besides C4A, the ADC2B coating dips below the minimum requirement at around 830nm, due to coating material index variation, but this coating was also accepted since it meets the requirement for average transmission.


All twelve coatings exceed the requirement for average transmission over the DESI bandpass.  The minimum transmission is not met in some cases.  However, when the total transmission is considered for all twelve surfaces together, the coatings that exceed performance more than make up for those coatings that are slightly below requirement.  Also, vendor data shows that the coating performance is quite uniform across the lens diameter.  This demonstrates that the coating thickness varies across the diameter as planned, successfully leading to high transmission at the appropriate AOI.

\begin{figure}
\centering 
\includegraphics[width=.8\textwidth]{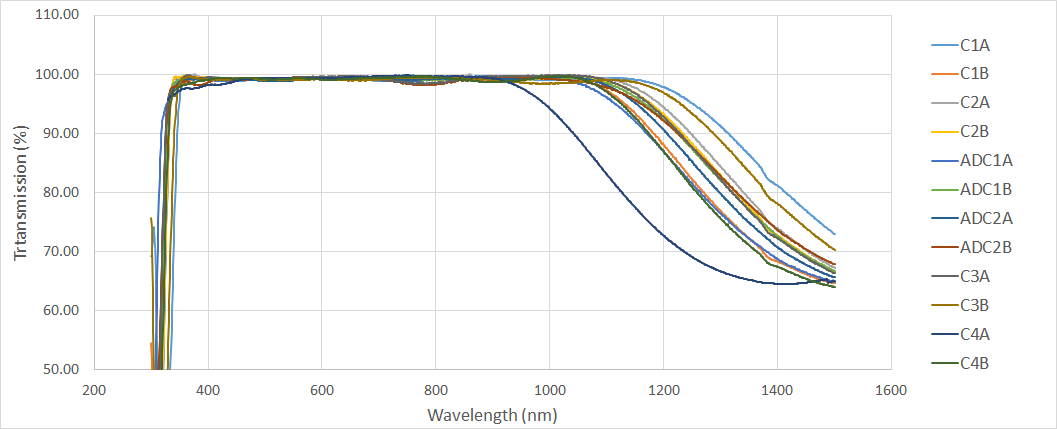}
\caption{Corrector AR coating performance measured over 300-1500nm, beyond the 360-980nm DESI bandpass.}
\label{fig:coatperfwide} 
\end{figure}

Figure~\ref{fig:coatperfwide} shows the measured transmission of the lens coatings beyond the DESI bandpass.  While the coatings are optimized for the DESI bandpass only, this suggests that potential future DESI upgrades might take advantage of their high transmission out into the infrared.  For example, the total transmission of the twelve coatings at 1200nm is 30\%, possibly leaving the door open for a near-IR spectrograph.


\subsection{Primary Mirror Coating and Maintenance} 

The 4-meter Mayall primary is coated with a bare aluminum coating.  The coating reflectivity degrades over time, mostly due to contamination (dust and humidity) from its exposure to the dome environment, and also somewhat due to aluminum oxidation.  Therefore, the mirror is cleaned on a regular maintenance schedule.  First, the mirror is ``snow cleaned'' with CO$_2$ once a week to remove loose dust.  Second, the mirror is carefully washed every six months to remove adhering dust, using a contactless pressure wash method to avoid the risk of scratching the mirror.  Finally, the mirror is realuminized with a new coating every three years using a devoted coating chamber located at the observatory.  The mirror reflectance and scatter are measured at least monthly, and more frequent cleaning and aluminizing are done if reflectance drops below a certain threshold value. 

\begin{figure}
\centering 
\includegraphics[width=.8\textwidth]{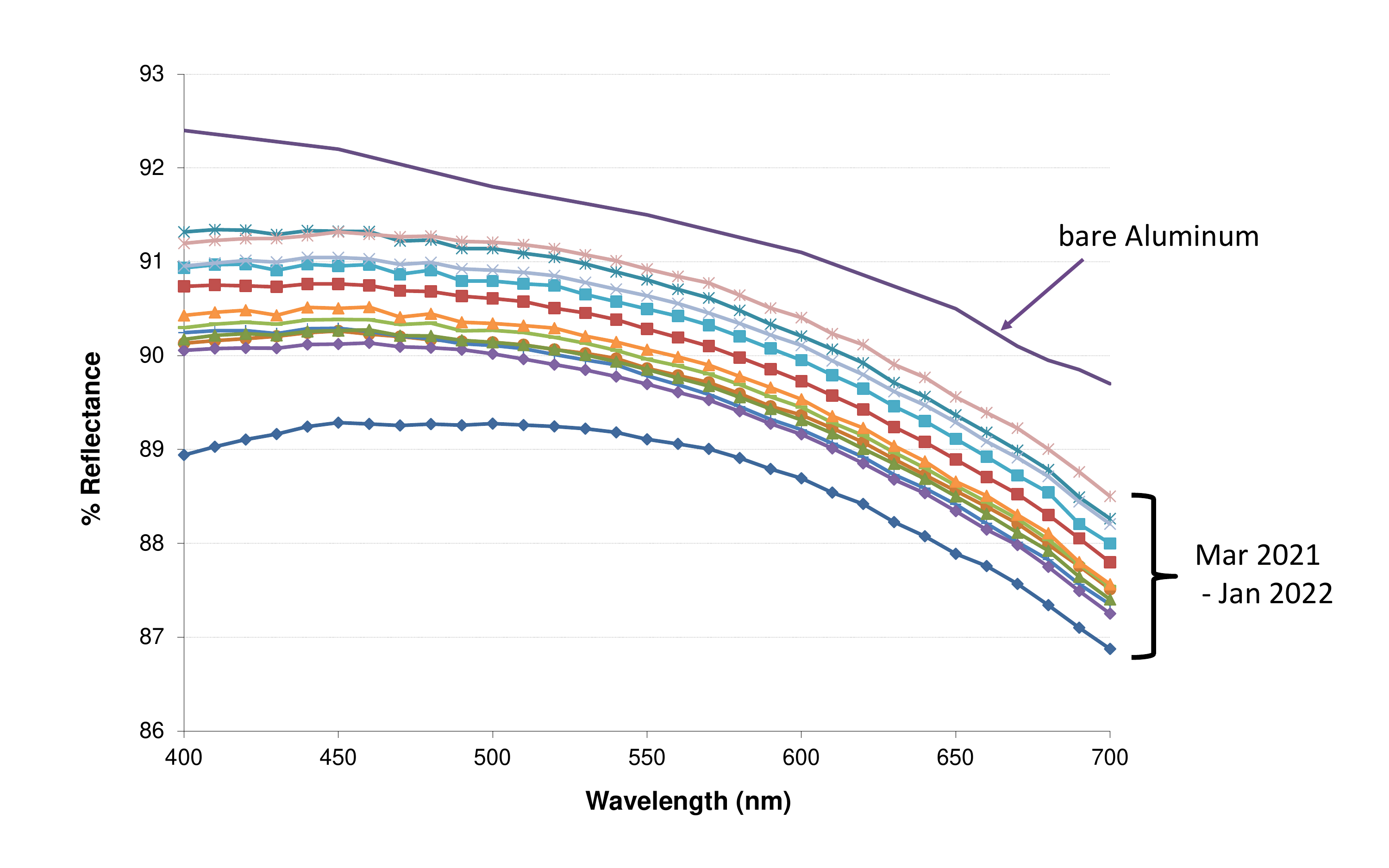}
\caption{Measurements of the primary mirror reflectivity.  The top curve shows the theoretical reflectance of bare aluminum.  The lower curves show a range of measurements taken before and after various cleanings over a ten-month period.  The mirror maintenance plan will keep the reflectivity consistently within the lower curves.}
\label{fig:M1_refl} 
\end{figure}

Figure~\ref{fig:M1_refl} shows several measurements of the primary mirror reflectivity over time; reflectivity varies depending on the mirror's cleaning schedule, but generally it will be between 89 and 92\,percent at 400nm, and 87 to 90 percent at 700nm.


\section{Corrector Mechanical Design and Fabrication} \label{sec:mechdesign}

The corrector mechanical design must support the heavy optical components and maintain precise alignments over a range of conditions in order to meet the performance required by DESI.  This section describes the requirements flowed to the mechanical design and explains how the cells and barrel were designed to meet those requirements.  We also describe the process for the integration and precision alignment of the lenses, cells, and barrel segments together at UCL and Fermilab.  We then describe the design of the ADC rotator, hexapod, and the top end assembly.

\subsection{Optical Component Alignment Requirements} 

The mechanical design needs to achieve and maintain precise alignment of its optical lenses in order to meet the corrector's high throughput requirements.  Alignment must be held during assembly, after motions of the ADC, after changing gravity orientations during science observations, and during temperature excursions in the demanding Mayall dome environment.  Maintaining these alignments drove the mechanical requirements for positioning and stability of all corrector mechanical elements.



\begin{table}  
\centering
\caption{Total alignment and deflection requirements for the corrector lenses
and focal plane.}
\label{tab:lens_place}
\addtolength{\tabcolsep}{4pt}
\begin{tabular}{lcccccc}
\hline
 & \multicolumn{2}{c}{Total Tolerances} & \multicolumn{2}{c}{Static Tolerances} 
& \multicolumn{2}{c}{Dynamic Tolerances} \\
\hline
 Element & Lateral & Tilt & Lateral & Tilt & Lateral & Tilt  \\
         & ($\pm\micron$) & ($\pm\mu$rad) & ($\pm\micron$) & ($\pm\mu$rad) 
& ($\pm\micron$) & ($\pm\mu$rad) \\
\hline
 C1         & 200 & 123 &  80 &  49 &  58 &  59 \\
 C2         &  75 & 176 &  55 &  83 &  20 &  60 \\
 ADC1       & 200 & 250 &  50 &  88 &  81 & 123 \\
 ADC2       & 200 & 175 &  50 &  87 &  81 &  79 \\
 C3         & 100 & 180 &  45 &  84 &  51 &  68 \\
 C4         & 200 & 105 &  70 &  66 &  62 &  38 \\
Focal Plane & 150 &  92 &  20 &  25 &  40 &  25 \\
\hline 
Element spacing & \multicolumn{6}{c}{$\pm 50\,\micron$} \\
\hline \hline
Corrector Decenter & \multicolumn{6}{c}{$\pm 300\,\micron$} \\
Corrector Tilt & \multicolumn{6}{c}{$\pm 48\,\mu$rad} \\
Corrector Defocus & \multicolumn{6}{c}{$\pm 30\,\micron$} \\
\hline \hline
\end{tabular}
\end{table}

The effects of misalignment on the lens elements were calculated from detailed Monte Carlo simulations of corrector optical performance.  This led to a lens positioning tolerance budget that met the requirements of the overall corrector throughput budget. These tolerances are shown in Table \ref{tab:lens_place}.
The total decentering and tip/tilt tolerance budgets for the lenses and focal plane are outlined in the second and third columns of the Table.
The total tolerances were then divided into static and dynamic tolerance budgets; this division was made based on the DESI team's previous experience with the similar DECam corrector \citep{flaugher15}.
The static tolerance, columns 4 and 5 in the Table, refer to the maximum allowable lens offsets introduced in the assembly of the corrector. 
The dynamic tolerances, columns 6 and 7 in the Table, refer to offsets due to flexure and ADC lens decenter error due to any misalignment of the ADC rotation axis to the optical axis of the system.
We included margin when dividing the total tolerances into smaller allocations; we were confident in doing so because we expected to meet those allocations easily based on estimates and experience from DECam.
The static and dynamic tolerances in the table were further subdivided into allocations for the various mechanical components of the corrector; these are discussed below.  These include the lens-to-cell alignment (see Table \ref{tab:deflec_cell}), the cell-to-barrel alignment (Table \ref{tab:deflec}), and the barrel segment-to-segment alignment (also Table \ref{tab:deflec}).  Again, we included margin when defining these allocations since we expected to meet them with reasonable effort.
The performance requirements in Table \ref{tab:lens_place} apply over a range of zenith angles from 0 to 60$^\circ$ and over the -10$^\circ$ to +30$^\circ$C operating temperature range expected in the Mayall dome environment.  The requirements do not need to be met over the corrector survival temperature range, defined as -20$^\circ$ to +60$^\circ$C, which comes from the extremes expected during shipping.

Table \ref{tab:lens_place} also shows the tolerances for absolute lens placement (spacing) along the optical axis.  Finally, Table \ref{tab:lens_place} shows the tolerances for the alignment of the complete corrector to the primary mirror of the telescope. These ranges fed into the design requirements for the hexapod and the Guide, Focus and Alignment (GFA) system.

\subsection{Lens Cells}\label{sec:lenscells}

The lens cells provide the interface between each lens and the barrel
of the wide-field corrector. DESI uses Momentive RTV560 at the lens-to-cell interface, an RTV (room temperature vulcanized) silicone elastometer with good heritage in aerospace applications. RTV silicones have much higher
coefficient of thermal expansion (CTE) values than typical glasses and
metals, with common CTEs for silicone ranging from 200 to 300\,ppm degree$^{-1}$ C.
RTV silicones (referred to here as simply RTV) are essentially incompressible, and exhibit varying effective stiffness and CTE, depending on boundary conditions: constrained and free surfaces determine where the RTV can expand under load.  Elastomeric mounts can be made to be essentially athermal
with a suitable choice of elastomer thickness, \ie, temperature changes cause little
induced stress in the components.

Discrete pads of RTV are used for both radial and axial support to mount the lenses in their cells. This method, which was adopted by both the MMT \citep{fata93} and DECam wide field correctors \citep{flaugher15}, has the advantage that the pads can be manufactured to high dimensional accuracy and consistency before mounting.  
We glued the pads to inserts made of the same material as the cells, then inserted them through holes in the side of the lens cell and attached with screws. The radial pads were glued to the lens with a thin layer of RTV, and the axial pads simply pressed against the lens.  The radial or axial pads exhibit different mechanical properties due to their configuration, and this was taken into account in the design. 

The cell material was an important choice in the design, since the relative CTE of the lens and cell materials affects the thickness of the radial pads needed to maintain an athermal design.  The radial pad thickness should not be too great, which would be needed for a large CTE mismatch, as this means the lens will not be held stiffly under a changing gravity vector. However, the pad thickness also should not be too small (less than $\sim$1\,mm), which would correspond to a close CTE match, as this poses problems for accurate attachment of the pads to the inserts.  This led to a material choice of nickel-iron alloy cells (CTE of $\sim$3\,ppm C$^{-1}$) for the fused silica lenses, and steel cells (CTE of $\sim$12\,ppm C$^{-1}$) for the borosilicate lenses.  These choices allow for an acceptable thickness of the radial pads.

\begin{figure}[h]
\centering
\hspace{0.25in}
\begin{minipage}[b]{0.45\linewidth}
\includegraphics[width=2.5in]{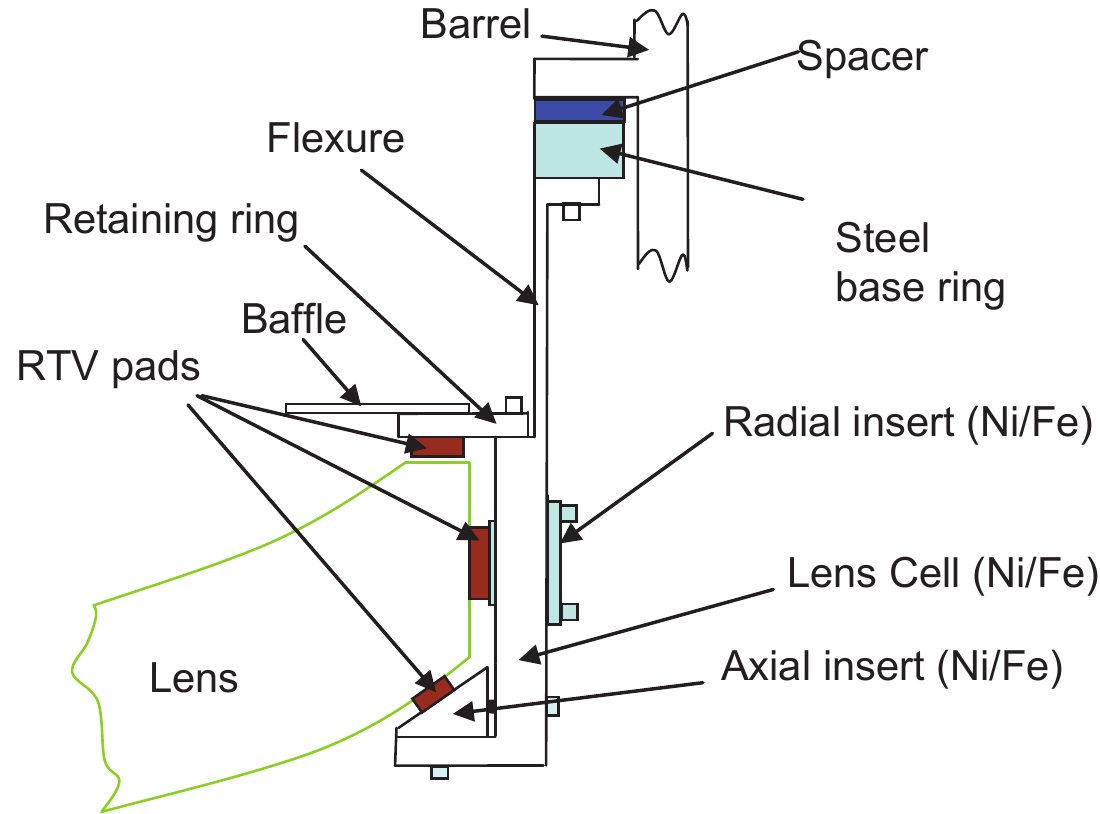}
\end{minipage}
\hspace{0.25in}
\begin{minipage}[b]{0.45\linewidth}
\includegraphics[width=2.5in]{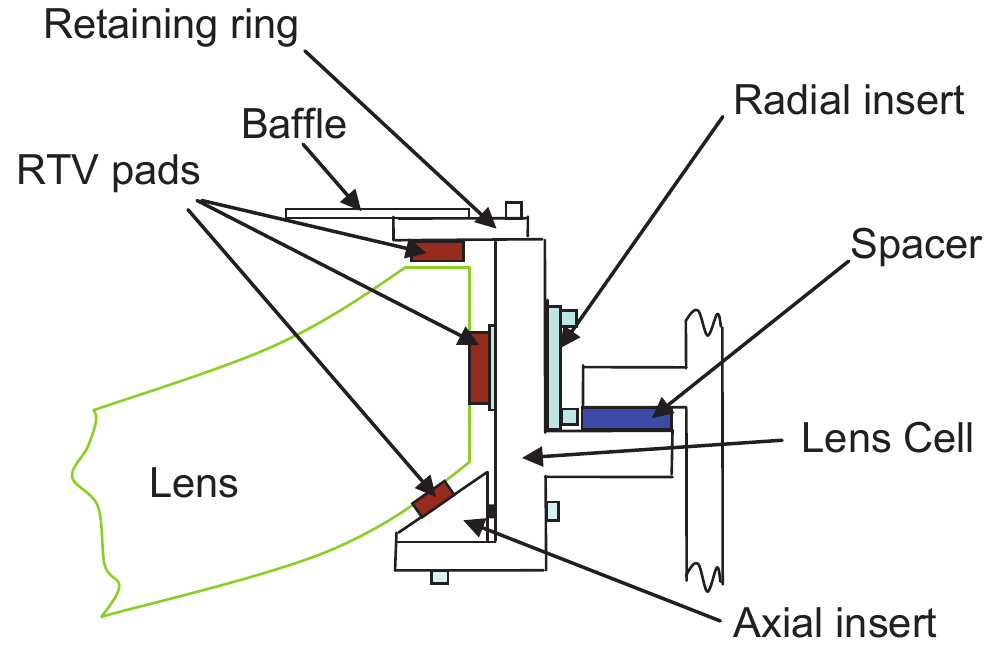}
\end{minipage}
\caption{Schematic of the lens cell designs.  Left: the cell for the fused silica lenses, made of a Ni/Fe alloy.  Right: the cell for the borosilicate lenses, made of steel. \citep[Image previously published in][]{doel16}}
\label{fig:UCL34}
\end{figure}

Schematic designs of the cells for the C1-C4 fused silica lenses and the ADC borosilicate lenses are shown
in Figure~\ref{fig:UCL34}. The lens cell design is similar to the one used for DECam.  Both the axial pads and radial pads are mounted on inserts. These inserts were precision-machined and accurately positioned to meet the tight tolerances on the lens mating surfaces derived from the optical system alignment
tolerances. The axial pads were glued to the axial inserts but not
to the lens, while the radial pads were glued both to the radial inserts and the lenses. 

The cells for the fused silica lenses have flexures that connect from the main body of the cell to a steel base ring. The base ring is attached via spacers to the barrel section flanges.  
The gaps and flexures are the same size in the C2 and C3 cell, whereas the C1 and C4 cells have narrow slots (5mm width) between the flexures; the thinner gaps for C1 and C4 allow for a stronger support for these heavier lenses, as well as easier sealing of the cells to restrict airflow and dust ingress.  The ADC cells do not require flexures between the cell and barrel structure, since both are steel and thus the CTEs are the same.

The cell designs include a retaining ring and baffles along with spacers between the cell mounting flange and the barrel flange. The retaining ring has pads attached but these do not touch the lens so as not to over-constrain it.  The static assembly tolerances for the alignment of the lenses into the cells, along with the dynamic tolerances, are given in Table~\ref{tab:deflec_cell}. 

\begin{table}[hbt]
\caption{Lens-to-cell assembly tolerances.}
\begin{center}
\label{tab:deflec_cell}
\begin{tabular}{lcccc}
\hline
 & \multicolumn{2}{c}{Static Tolerances} & \multicolumn{2}{c}{Dynamic Tolerances} \\
\hline
 Element & Lateral & Tilt & Lateral & Tilt  \\
         & ($\pm$ $\mu$m) & ($\pm$ $\mu$rad) & ($\pm$ $\mu$m) & ($\pm$ $\mu$rad) \\
\hline
 C1         &  50 &  44 &  8 &  14 \\
 C2         &  25 &  59 &  5 &  7 \\
 ADC1       &  25 &  63 &  6 & 10 \\
 ADC2       &  25 &  62 &  6 & 16 \\
 C3         &  25 &  60 &  6 & 13  \\
 C4         &  25 &  47 &  12 &  24 \\
\hline
\end{tabular}
\end{center}
\end{table}

Stress-induced birefringence limits the allowable stresses in the lens to $<3$~MPa. The target for the maximum cell stress was a factor of 2 below the yield strength (228\,MPa). For C1, due to its position above the primary mirror, we attached wire straps between the cell and barrel as an additional safety measure.

The axial pads have two functions.  First, they take up any irregularity
of the lens pad surfaces, and hence the lens cell surface, in order to ensure an
even spread of the loading on the lens.  Second, they hold the lens stiffly so that a change in gravity vector does not produce too much tilt or decentering of the lens.  The radial pads also provide a measure of resistance to tilt and decenter.
The optimal thickness of the radial pads that achieves an athermal behaviour can be calculated from \citet{doyle02}.  We derived the thickness for each case of lens and cell; results are shown in the last column of Table~\ref{tab:UCL2}. 


\begin{table}[h]
\centering
\caption{Lens cell pads.}
\begin{tabular}{c|c|c|c|c} 
\hline 
Lens & Mass & Axial Pad Size & No. of Axial  & Radial Pad Size\\ 
     &  (kg)  & (mm) & and Radial Pads &  (mm) \\ 
\hline 
C1    & 201   & 10, 10, 1.0    & 48  &  32, 32, 2.17 \\ 
C2    & 152   & 10, 10, 1.0    & 24  &  80, 80 ,1.62 \\ 
C3    & 84    & 10, 10, 1.5  & 24 &  45, 45, 1.60 \\
C4    & 237   & 10, 10, 1.0    & 48  &  25, 25, 2.00 \\ 
ADC1  & 102   & 10, 10, 1.0    & 24  &  60, 60, 3.50 \\ 
ADC2  & 89    & 10, 10, 1.5  & 24 &  50, 50, 3.55 \\ 
\hline
\end{tabular}
\tablecomments{Pad sizes are given as width, length, thickness.}
\label{tab:UCL2}
\end{table}

\begin{figure}
\centering 
\includegraphics[width=.95\textwidth]{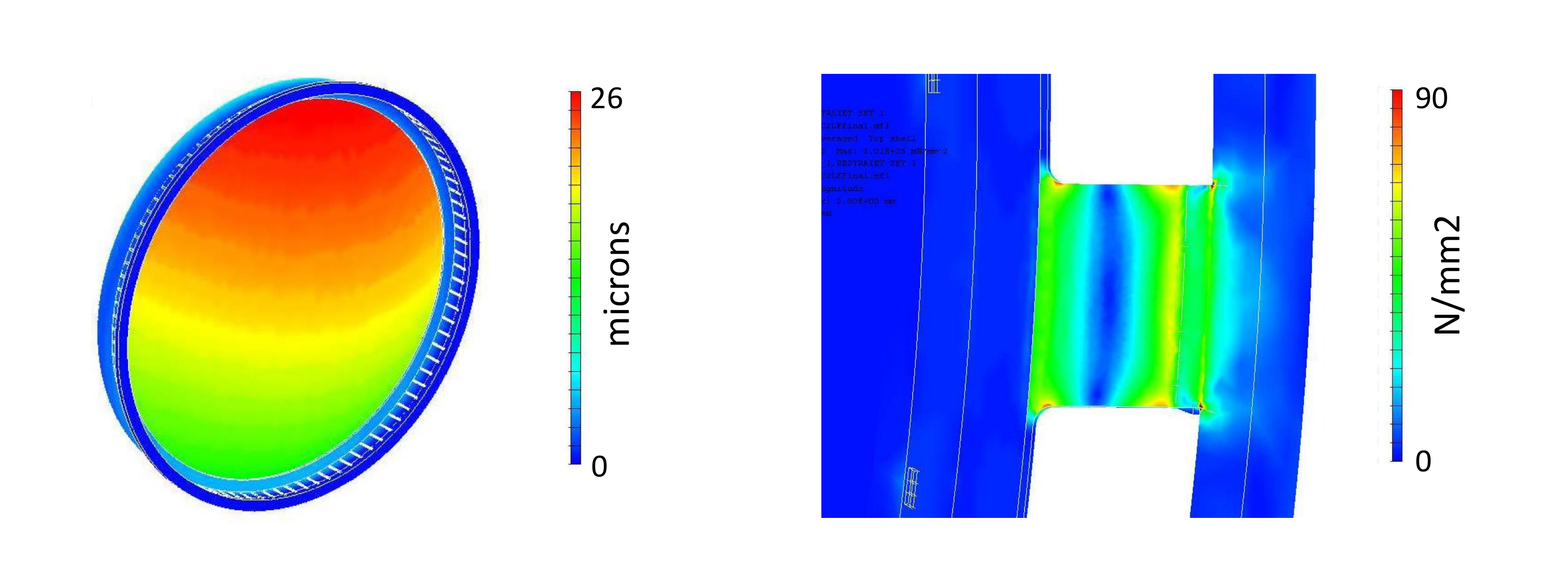}
\caption{Examples of the lens cell FEA.  Left: FEA model showing deformation due to tilt of the C4 lens and cell at a 60$^\circ$ Zenith angle.  Right: FEA model showing stresses in a C2 flexure due to a temperature change of -40$^\circ$C.}
\label{fig:UCL_FEAs} 
\end{figure}

The combined mass of the cells, base rings, and spacers are required to be less than 560\,kg.  Furthermore, for survey operations the lens and cells need to meet the alignment requirements over zenith angles of 0-60$^\circ$.  Furthermore, they must survive a temperature range defined for the corrector of -20$^\circ$C to +60$^\circ$C.  The DESI team at University College London (UCL) performed a finite element analysis (FEA) of the lens/cell designs to confirm that they met the stress and dynamic tolerance requirements over the defined ranges of temperatures and zenith angles.  In all designs the cell wall thicknesses were 20\,mm, the flexure
lengths for the Ni/Fe cells were 42~mm, and the flexure thicknesses were 2~mm.  The FEA models used slightly simplified designs for ease of modeling.  The radial pads were modeled as being attached to a continuous cell wall rather than as inserts in cell wall openings.  The axial pads were modeled as being directly attached to both the lens and the cell rather than just the cell.  These simplifications affected the value of the Young's modulus, and was accounted for by adjusting the modeled RTV material properties so that the RTV emulated the behavior of a linearized narrow regime of a force-displacement curve measured in pad studies at UCL.
Table~\ref{tab:UCL5} summarizes the predicted stresses for a maximum change of temperature and orientation change, along with the tilts and decenters at the most extreme Zenith angle.  The tilts and decenters are acceptably small contributions to the total tolerance budget shown in Table~\ref{tab:lens_place}.  Figure \ref{fig:UCL_FEAs} shows examples of the FEA modeling.

\begin{table}
\centering
\caption{FEA prediction of lens cell stresses and motions.} 
\begin{tabular}{c|cc|cc}
\hline 
Lens & Max Stress for & Max Stress for & Lens Tilt for & Lens and Cell Decenter \\
 &  -40$^\circ$ Change & 90$^\circ$ Zenith Angle & 60$^\circ$ Zenith Angle & for 60$^\circ$ Zenith Angle \\
 & (MPa) & (MPa) &   ($\mu $m) & ($\mu $m)  \\ 

\hline 
  C1       & 91 & 5.8  &12.7 & 6.0 \\
  C2       & 88 & 8.3  & 0.3 &2.8 \\
  C3       & 76 & 4.1  & 5.2 & 2.3 \\
  C4       & 80 & 5.0  &15.6 &10.5\\
  ADC1   & $<$1.0 & 0.43 & 2.1 &1.0 \\
  ADC2   & $<$1.0 & 0.21 & 6.9 &2.0 \\
\hline 
\end{tabular}
\tablecomments{Lens tilt given as microns of difference across the lens diameter.}
\label{tab:UCL5}
\end{table}

\subsection{Barrel}\label{sec:barrel}

The challenges of the barrel design stem from the DESI decision to leave the total mass atop the Mayall telescope unchanged (and thus maintain the moment of inertia about the telescope Declination axis) and preserve the previous primary mirror obscuration.  This is especially challenging given that the new corrector lenses are significantly larger than the lenses in the Jacoby corrector, not to mention the larger focal plane assembly and additional fiber cables.  This decision led to a substantial effort to optimize the corrector mass and to minimize the space around the lenses occupied by the barrel, hexapod and the ADC rotator mechanisms.  In this section we discuss some of the optimization efforts and describe the specifications and final design of the barrel.

Table \ref{tab:lens_place} shows the overall alignment and deflection requirements for the six corrector lenses and the focal plane.  These were broken down into requirements for the placement of the lenses in the lens cells (or cells for short) and the placement of the cells in the barrel.  These latter requirements are shown in Table \ref{tab:deflec}.  The static tolerance, shown in columns 2 and 3 in the table, refer to the maximum errors introduced in the process of aligning the cells in the barrel. The dynamic tolerances, shown in columns 4 and 5, refer to the barrel deflections due to telescope motion plus the errors introduced by the ADC rotation mechanism.  These performance requirements apply over a range of Zenith angles from 0 to 60\,degrees.

\begin{table}[hbt]
\caption{Cell-to-barrel tolerances, including focal plane.}
\begin{center}
\label{tab:deflec}
\begin{tabular}{lcccc}
\hline
 & \multicolumn{2}{c}{Static Tolerances} & \multicolumn{2}{c}{Dynamic Tolerances} \\
\hline
 Element & Lateral & Tilt & Lateral & Tilt  \\
         & ($\pm$ $\mu$m) & ($\pm$ $\mu$rad) & ($\pm$ $\mu$m) & ($\pm$ $\mu$rad) \\
\hline
 C1         &  20 &  18 &  50 &  45 \\
 C2         &  20 &  22 &  15 &  53 \\
 ADC1       &  20 &  25 &  75 & 113 \\
 ADC2       &  20 &  25 &  75 & 112 \\
 C3         &  20 &  24 &  45 &  54 \\
 C4         &  20 &  19 &  50 &  14 \\
Focal Plane &  20 &  25 &  40 &  25 \\
\hline \hline
Static axial tolerance of all barrel flanges & \multicolumn{4}{c}{$\pm$ 100 $\mu$m} \\
Flange distortion (circularity) & \multicolumn{4}{c}{50 $\mu$m peak-to-peak} \\
Air and light tightness & \multicolumn{4}{c}{yes} \\
Barrel material & \multicolumn{4}{c}{Carbon Steel} \\
Barrel assembly/disassembly repeatability & \multicolumn{4}{c}{$\pm$ 5 $\mu$m per flange} \\
\hline \hline
Barrel flange flatness & \multicolumn{4}{c}{ 15 $\mu$m} \\
Barrel flange parallelism & \multicolumn{4}{c}{ 15 $\mu$m} \\
\hline
\end{tabular}
\end{center}
\end{table}

The specifications of the cell alignment with respect to the barrel are $\pm 20$ $\mu$m for the lateral displacement and about $20$ $\mu$m at the edge of the lenses for the tip/tilt angles.  These requirements are similar to the tolerances achieved for DECam.  The second part of Table \ref{tab:deflec} contains the barrel flanges' axial tolerances, barrel out-of-roundness, air and light tightness, barrel material, weight limits for the barrel and the cells, and the assembly/disassembly repeatability requirements for the barrel sections.  The last part of Table \ref{tab:deflec} specifies the flatness and parallelism of the flanges.  These last two specifications simplified the machining of the spacers and the alignment process, and made it possible to achieve the assembly/disassembly repeatability tolerances.  The barrel met all the alignment requirements of Table \ref{tab:lens_place}.

Three main concepts were followed in the design of the barrel:

\begin{itemize}
\item  Use a single shell (or monocoque) design as a way of using mass effectively and to minimize primary mirror obscuration.  Ribs and gussets were only used for local reinforcement of connecting flanges.

\item Minimize the number of barrel sections to reduce the number of connecting flanges, and design these flanges such that they add very little to the barrel deflections.

\item Connect the hexapod through a single flange to the barrel.  This reduces primary mirror obscuration and allows for reduction in cost, schedule and risk by using essentially the same hexapod that was used in DECam.
\end{itemize}

\subsubsection{Rationale for a Single Shell Barrel Design} \label{sec:barrel_monocoque}

The studies summarized in this section showed that, given the dimensions of the DESI barrel, the most effective way to add mass in order to reduce barrel deflections is to increase the thickness of the cylindrical shell supporting the lenses.  Increasing the shell thickness is more effective than adding any other kind of reinforcements such as ribs.  The stress-strain relation for shells is given by:

\begin{equation}
\label{eq:hooke}
\left( \begin{array}{c}
 N_x \\ N_y \\ N_{xy} \\ M_x \\ M_y \\ M_{xy} \end{array} \right) =
\bar{K} \;\;
\left( \begin{array}{c} 
 \epsilon^0_x \\ \epsilon^0_y \\ \gamma^0_{xy} \\ \kappa_x  \\ \kappa_y  
\\ \tau  \end{array} \right)
\end{equation}

\noindent
where $\epsilon_x^0$, $\epsilon_y^0$ and $\gamma_{xy}^0$ are respectively the strains related to stretching the shell along the axis of the cylinder, stretching the shell along the circumference of the cylinder and shearing the shell along the axial and circumferential directions.  The bending strains $\kappa_x$, $\kappa_y$ and $\tau$ are related respectively to bending an axial strip of shell, bending a circumferential strip of shell and bending a strip of shell like a helix.  The $N_i$ and $M_i$ are essentially the forces and the moment per unit length.  For a uniform and isotropic shell the matrix $\bar{K}$ is

\begin{equation}
\label{eq:kbar}
\bar{K}=
\left( \begin{array}{cccccc}
  C & \nu C & 0 & 0 & 0 & 0 \\
 \nu C & C & 0 & 0 & 0 & 0 \\
  0 & 0 &  \frac{1-\nu}{2} C & 0 & 0 & 0 \\
  0 & 0 & 0 & D & \nu D & 0 \\
  0 & 0 & 0 & \nu D & D & 0 \\
  0 & 0 & 0 & 0 & 0 & \frac{1-\nu}{2} D  \end{array} \right)
  \;\;\;,\;\;\; \mbox{with } \;\;\; C=\frac{E \, t}{1-\nu^2} \;\;\; \mbox{and} \;\;\; D=\frac{E \, t^3}{12 \, (1-\nu^2)}
\end{equation}

\noindent
where $C$, $D$, $E$, $\nu$ and $t$ are respectively the extensional rigidity, bending rigidity, Young's modulus, Poisson's ratio and the thickness of the shell.  Since $D$ grows very fast with the shell thickness it will reach a point where the shell will be rigid enough that for a ``beam'' type of load the only important deformations of the shell will be the extensional ones, or in other words deflections will be dominated by $N_x$, $N_y$ and $N_{xy}$.  Deflection studies using FEA analysis of cylindrical shells of the same length-to-diameter ratio and the same shell mass-to-load ratio as the ones in the DESI barrel showed that deflections are inversely proportional to the shell thickness $t$.  This clearly indicates that deflections in the case of the DESI barrel are dominated by the extensional properties of the shell.  For example $C$ in Equation \ref{eq:kbar} has a linear dependence with the shell thickness $t$, which in turn means that for a given stress the strains will be inversely proportional to $t$ which will produce deflections that are also inversely proportional to $t$.

\begin{figure}[htb]
\begin{center}
  \includegraphics[width=.9\textwidth]{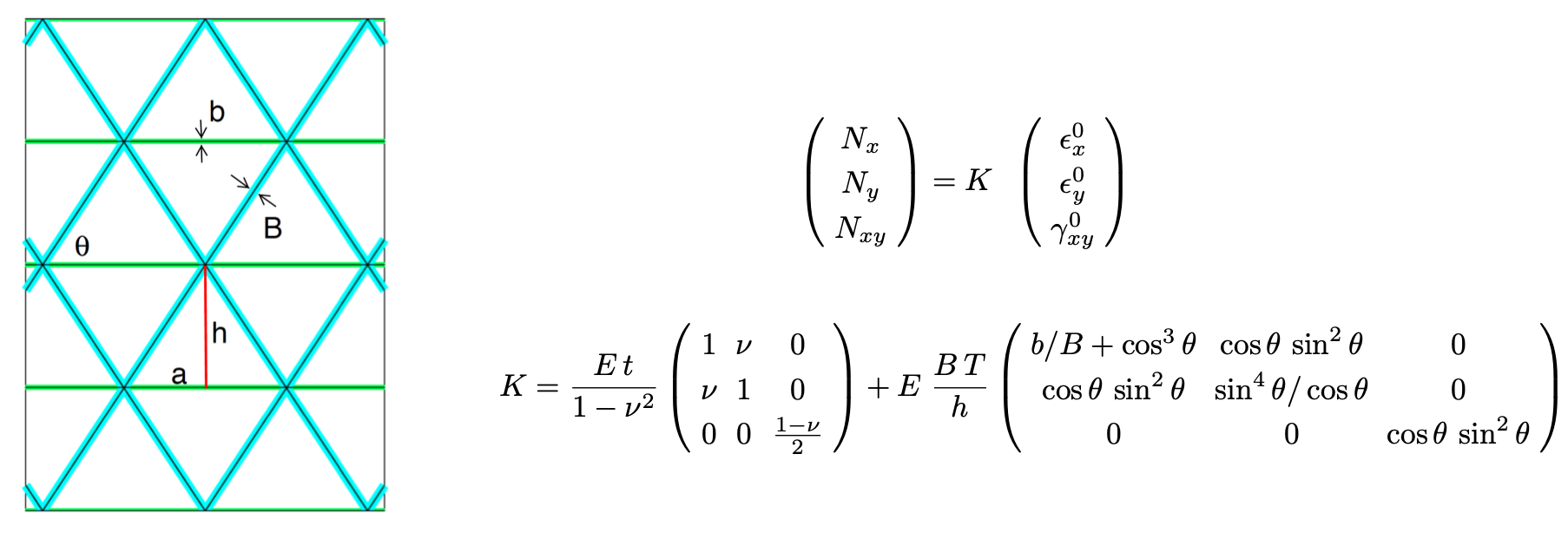}
  \caption{Isogrid pattern.  The base and height of the triangles are 
labeled $a$ and $h$ respectively.  The thickness of the slanted blue bars 
is labeled $B$, and that of the straight green bars is labeled $b$.  The 
angle at the base of the isosceles triangles is labeled $\theta$.  The stress-strain relation for the isogrid are given by the expressions on the right, where $t$, $T$, $E$ and $\nu$ are the thickness of the skin and the grid and the Young's modulus and Poisson's ratio respectively.}
\label{fig:isogrid}
\end{center}
\end{figure}

To understand if reinforcing the shell with ribs would be advantageous in terms of reducing barrel deflections we studied an isogrid design like the one shown in Figure \ref{fig:isogrid}.  This isogrid is a combination of a skin attached to a grid characterized by the parameters $b$, $B$, $\theta$, $a$ and $h$.  By selecting the values of $b$, $B$ and $\theta$ one can select axial ribs ($B=0$, $\theta=0$) or circumferential ribs ($b=0$, $\theta=90$) or any option in between.  The number or density of ribs is determined by the parameters $a$ and $h$.    
The stress-strain relation for the isogrid is given by the expressions to the right of the isogrid figure.  The first part of $K$ corresponds to the skin and the second part to the grid, with $t$ and $T$ being the thicknesses of the skin and the grid respectively.  The bending part of the stress-strain relation was ignored because FEA studies showed that the barrel deflections are dominated by the extensional properties of the shell.  If $\rho$ is the density of the barrel material then the average surface density $\delta$ of the shell can be written as $\delta = \rho \, \bar{t}$, with 

\begin{equation}
    \bar{t} = t + \frac{B \, T}{h} \left( \frac{b}{B} + \frac{1}{\cos\theta} \right) \left[ 1 - \frac{B}{4 h} \left( \frac{b}{B} + \frac{1}{\cos\theta} \right) \right]
\end{equation}

\noindent The quantity in square brackets is a small correction for the bar's overlap.  The larger the value of the matrix $K$ the smaller the strains and therefore the smaller the barrel deflections.  To determine whether it is advantageous to add ribs we need to maximize $K$ keeping $\bar{t}$, and therefore the mass of the barrel, constant.  This maximization can be done analytically and shows that all the elements of the matrix $K$ are maximized when $T=0$ regardless of the values of all the other isogrid parameters.  FEA studies using u-channels instead of regular ribs reached the same conclusion, once the shell is stiff enough in local bending any additional bending rigidity will not help to reduce the total deflections of the cylinder.  In this regime the most effective way to add mass to the cylinder to reduce total deflections is to add the mass to the single shell.  A monocoque design such as this one also allows the hexapod to get closer to the barrel, which reduces the diameter of the cage, which in turn reduces the obscuration of the mirror.  Given all the above considerations the decision was made to use a single shell design for the DESI barrel.

\subsubsection{Main Barrel Components}\label{sec:prel_barrel_design}

\begin{figure}
\begin{center}
 \includegraphics[width=.75\textwidth]{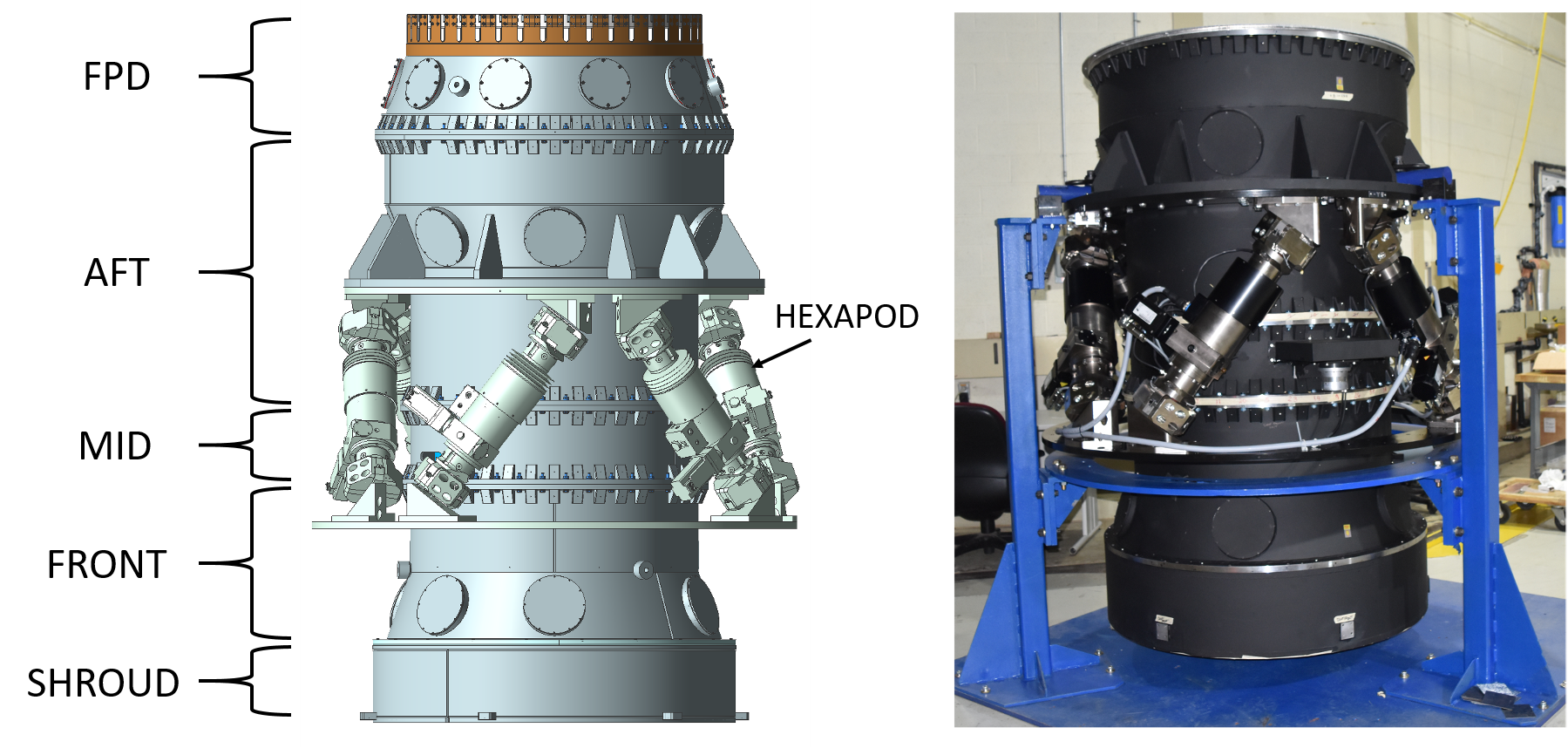}
  \caption{DESI corrector barrel assembly.  Left: barrel components with the hexapod and FPD.  Right: the barrel with hexapod mounted on a stand after final assembly at Kitt Peak National Observatory. \citep[Left image previously published in][]{gutierrez18}}
\label{fig:hex_oneside}
\end{center}

\begin{center}
  \includegraphics[width=0.9\textwidth]{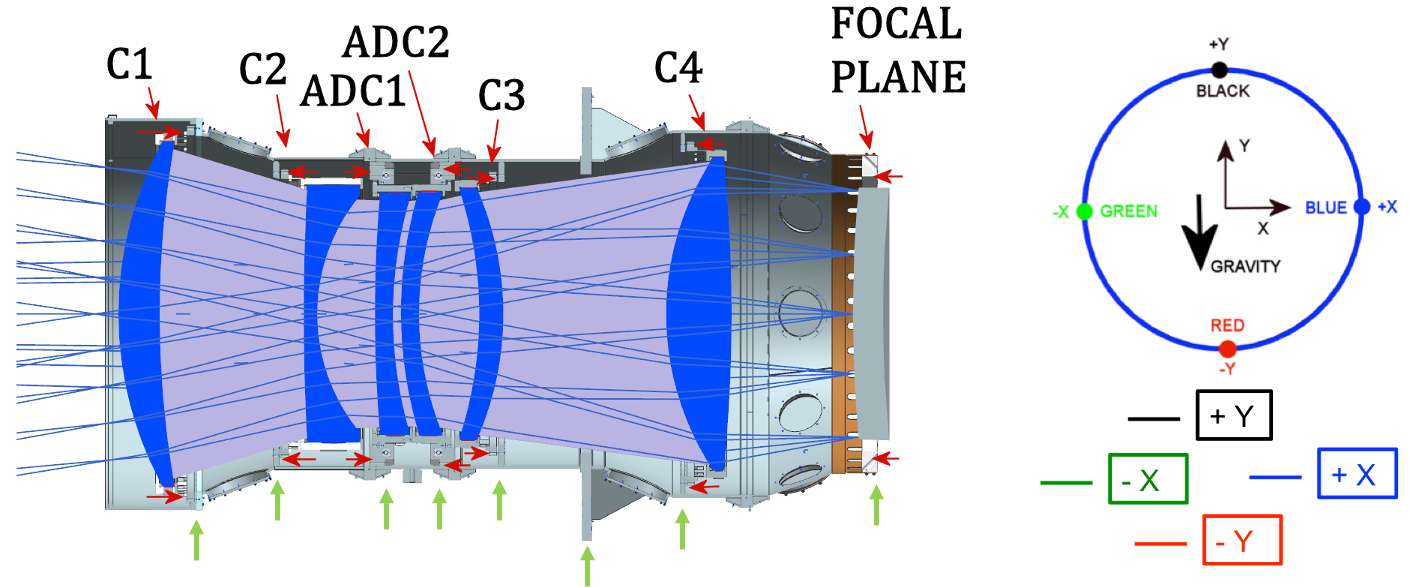}
  \caption{Layout of the barrel assembly.  Left: section view showing lens positioning inside the barrel.  The short horizontal red arrows show the direction in which the lenses attach to the barrel.  The light purple region shows the keepout volume for the transmitted light rays.  Green arrows show the reference locations used to track deflections in the FEA.  Right: four points were tracked in each xy-plane in the FEA as shown.  The points are color coded as black $+Y$, blue $+X$, red $-Y$, and green $-X$.  This color code is used in the FEA deflection plots in this Section. \citep[Left image previously published in][]{gutierrez18}}
\label{fig:FEA_points}
\end{center}
\end{figure}

The design of the corrector barrel meets all the mechanical requirements listed in Table \ref{tab:deflec}, provides a minimum number of flanges and sections, a homogeneous shell thickness, and the interfaces for the hexapod movable flange, the lens cells, the ADC rotation mechanism and the focal plane assembly.
Figures \ref{fig:hex_oneside} and \ref{fig:FEA_points} show two pictures of the barrel and a photograph of the final barrel assembly at Kitt Peak National Observatory.  The barrel consists of four main sections.  Following the light path (bottom to top in Figure \ref{fig:hex_oneside} and left to right in Figure \ref{fig:FEA_points}) the sections are: 1) the Shroud section which protects the C1 lens and allows the entire structure to rest vertically on the floor, 2) the Front section which houses the C1 and C2 lenses, 3) the Middle (Mid) section which supports the ADC rotation mechanism and the ADC1 and ADC2 lenses, and 4) the After (Aft) section which houses the C3 and C4 lenses and provides the flange that attaches the entire structure to the hexapod.  In the structural analysis that is described below, the Focal Plane Adapter (FPD) is also included; this section acts as the mechanical interface to the focal plane assembly.

The Shroud section extends 38\,mm beyond the C1 lens, in order to protect the C1 lens, and to allow the entire barrel structure to rest vertically on the floor. The front side of the Shroud includes tapped holes used to install a protective cover, and the other side is attached to the wide side of the Front section cone through a bolted joint.  The Shroud also supports six precision mounted retro-reflectors to allow referencing the position of the barrel to the primary mirror using a laser tracker.  Using a CMM the positions of these retro-reflectors were measured precisely with respect to the coordinate system of the barrel.  For position reproducibility the Shroud was pinned to the Front section using expandable pins. 

The Front section consists of a cone and a cylindrical piece welded together.  An inner flange near the weld supports the C2 lens cell and spacer.  A wide flange is welded to the end of the cone aperture to support the C1 lens cell and spacer and the Shroud section.  At the downstream end a gusset reinforced flange attaches the Front to the Middle section.  All three flanges were precision machined for flatness and parallelism.  Four moglice-filled precision pins were installed in the flanges connecting the Front and Middle sections to ensure the $\pm 5 \, \mu$m assembly/disassembly reproducibility requirement.  Six ports with an inner bore of 180~mm were provided to measure the lens spacing and to allow for lens cleaning.  

The Middle section is the shortest but tightest-packed section of the barrel, providing support for the ADC rotation mechanism and the ADC1 and ADC2 lenses.  Of the four flanges in this section two are internal and are attached to the two custom bearings that rotate the lenses.  The other two gusset reinforced flanges attach the Middle section to the Front and After sections.  All four flanges were precision machined for flatness and parallelism.  Each of the driving mechanisms for the ADC rotator consists of a harmonic drive gearmotor and two gears. The harmonic drive and the smaller of the two gears are located outside the Middle section, with the smaller gear attached to the harmonic drive.  The larger gear is inside attached to the bearings.  The small and large gears connect through two rectangular slits cutout in the wall of the Middle section.  The spacers and lens cells that support the ADC1 and ADC2 lenses are attached to the custom bearings.

The After section houses the C3 and C4 lenses.  The upstream and downstream pieces of the After section are welded to the Hexapod Mounting Plate (HMP) which connects the entire barrel structure to the hexapod.  Two inner flanges close to both ends support the spacers and the cells for the C3 and C4 lenses.  Two gusset reinforced flanges connect the After section to the Middle section and the FPD.  All four of these flanges were precision machined for flatness and parallelism.  Four moglice-filled precision pins were installed in the flanges that connect the After section to the Middle section to ensure the $\pm 5 \, \mu$m assembly/disassembly reproducibility requirement is met.  After aligning the entire structure the FPD was pinned to the After section using expandable pins.  The shell also provides six 180~mm ports; these provide access to measure the lens spacing and to allow for cleaning of the lenses.

The Focal Plane Adapter (FPD), while not considered part of the barrel, is an important contributor to the structural design of the corrector and so is described briefly here.  The FPD is the structural interface between the corrector barrel and the focal plane assembly; since those are managed by FNAL and LBNL respectively, the two institutions managed the FPD collaboratively.  LBNL designed the FPD, FNAL procured and aligned it, and then LBNL aligned and integrated it to the focal plane assembly.  The FPD consists of two main components: a steel cone with flexures, and an aluminum ring permanently bolted to the cone's flexures.  The FPD thus mitigates the thermal expansion mismatch between the steel barrel and the aluminum focal plane assembly mount.  The FPD has two bolted interfaces - one to mate to the barrel and the other to mate to the focal plane assembly - which were match-drilled when the FPA was colocated with the barrel and focal plane assembly.  Match-drilling was necessary in order to place the focal plane and lenses at the correct and precise locations required by Table \ref{tab:deflec}.  These locations were established by measurements at FNAL of the optical axis and image location with reference to tooling balls on the FPD (see Section \ref{sec:alignment}).  The FPD has ten 150~mm ports that are used to provide air cooling to the focal plane.  More description of the FPD is in \cite{silber2022}.


The barrel shape was designed to accommodate the light flow through the barrel and at the same time minimize the barrel diameter so that the barrel and the hexapod fit inside the required 1.8-m maximum diameter for the cage.  The light purple volume in Figure \ref{fig:FEA_points} shows the keepout volume for the light; this volume is created by joining the clear apertures for each of the six lenses inside the barrel.

The bolted joints between the barrel sections were designed to minimize barrel deflections.  Each flange is bolted with an equal spacing between bolts, and since the flanges may have different diameters, they may have different numbers of bolts as indicated in Table \ref{tab:bolts}.  These flanges were reinforced with gussets in order to minimize deflections and to make them stiff enough for the high precision machining required to meet the $15 \,\mu$m flatness and parallelism specifications.  All the inner flanges have a corresponding set of tapped holes to bolt the cell spacers and the cell base rings to the barrel.  After final alignment of the cell inside the barrel and the installation of the lenses inside the cells the cell spacers were machined to achieve the final axial tolerance of $\pm 50 \,\mu$m between lenses (see Table \ref{tab:lens_place}).  The weights, material and shell thicknesses for each section of the barrel are reported in Table \ref{tab:section_weights}. 

\begin{table}
\begin{center}
\caption{Bolted Joint Details.} 
\label{tab:bolts}
\begin{tabular}{l c c c}
\hline
 & Bolt Type & No. of Bolts & Joint Thickness \\
 &  &  & (mm) \\
\hline
ADC1 Joint & Class 10.9, M10 x 1.5 & 40 & 37  \\
ADC2 Joint & Class 10.9, M10 x 1.5 & 40 & 37  \\
C4 Joint   & Class 10.9, M10 x 1.5 & 50 & 38  \\
\hline
\end{tabular}
\end{center}
\end{table}

\begin{table}
\begin{center}
\caption{Barrel Sections.} 
\label{tab:section_weights}
\begin{tabular}{clccc}
\hline
 & Section & Wall Thickness & Weight & Material \\
 & & (mm) & (kg) & \\
\hline
1 & FPD ring & & 25 & Aluminum 6061-T6 \\
2 &  FPD & 3 & 82 & AISI 4130 \\
3 & AFT section & 10 & 698 & ASTM-A36 \\
4 & MID section & 10 & 145 & ASTM-A36 \\
5 & FRONT section & 10 & 240 & ASTM-A36 \\
6 & SHROUD section & 10 & 104 & ASTM-A36 \\
7 & Bolts, nuts, etc &  & 11 & Class 10.9 \\
 & Total Mass & & 1305 \\
\hline
\end{tabular}
\tablecomments{Weights do not include lenses or cells.}
\end{center}
\end{table}

The hexapod is attached to the barrel through the Hexapod Mounting Plate (HMP).  This mounting plate is located about 150\,mm downstream of the center-of-gravity (CG) of the barrel.  This way, the CG of the barrel was placed in between the hexapod moving and fixed flanges to reduce the barrel rotation induced by the hexapod actuators' finite stiffness, and to increase the space allowed for the ADC rotator mechanisms.  All the main components were included in the calculation of the center of mass, such as the barrel sections, the cells, the lens, the focal plane assembly, etc. 


\subsubsection{Finite Element Analysis of Barrel Deflections}

\begin{figure}
\begin{center}
  \includegraphics[height=3.5in]{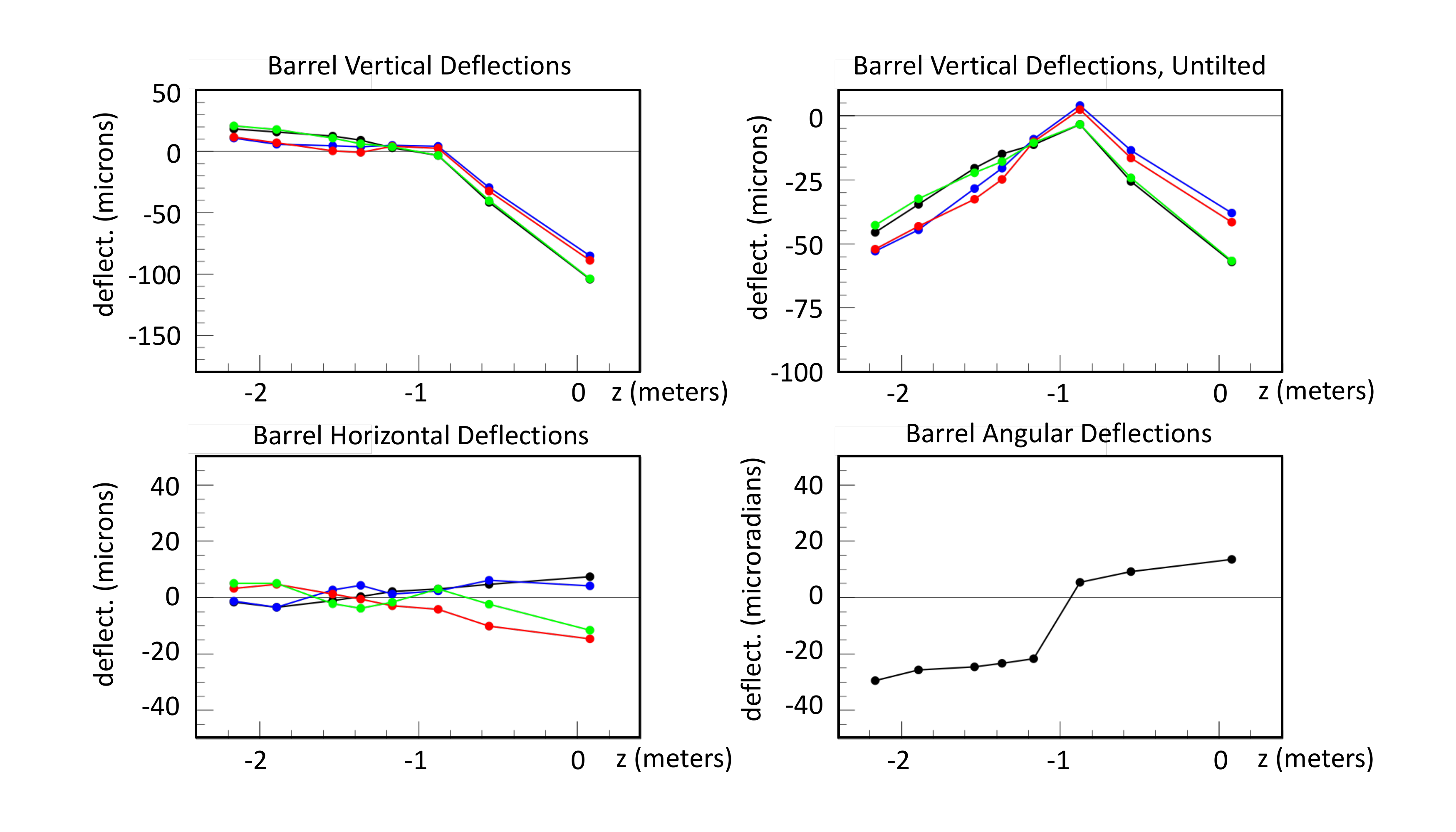}
  \caption{Barrel deflections predicted by FEA.  Telescope is in the horizontal position and oriented at a 0$^\circ$ rotation around the barrel axis.  From left to right the points correspond to C1, C2, ADC1, ADC2, C3, HMP, C4 and the focal plane.  The color code is given in Figure \ref{fig:FEA_points}.  The plots show the barrel vertical deflections (top left), the same deflections after an overall rotation of the barrel to make the deflections at both ends equal (top right), the horizontal deformations (bottom left), and the angular deflections after the barrel has been rotated to make deflections at both ends equal (bottom right).}
\label{fig:full_joe_18}
\end{center}
\end{figure}

We used Finite Element Analysis (FEA) software to track deflections in eight different barrel planes as shown by the green arrows in Figure \ref{fig:FEA_points} and at four points in each plane as shown in the right hand plot in the same figure.  The points are color coded as black $+Y$, blue $+X$, red $-Y$ and green $-X$.  Plane tilts were calculated as the axial displacement of the top point minus the lower one divided by the diameter.  Figure \ref{fig:full_joe_18} shows the results of the FEA when the telescope is in the horizontal position and there is no rotation around the axis of the hexapod.  The top left plot in this figure shows the vertical deflections.  The same deflections, after an overall rotation of the barrel to make the average deflections at both ends equal, are shown in the top right plot and the horizontal deformations are shown in the bottom left plot of the same figure. Finally the angular deflections, after the barrel has been rotated to make deflections at both ends equal, are shown in the bottom right plot.  Two other cases were also studied in which the barrel was rotated 40 and 80\,degrees around the axis of the hexapod.  Due to the hexapod symmetry results will repeat after 120\,degrees, so we studied three points between 0 and 120\,degrees. 

\begin{figure}
\begin{center}
  \includegraphics[width=0.8\textwidth]{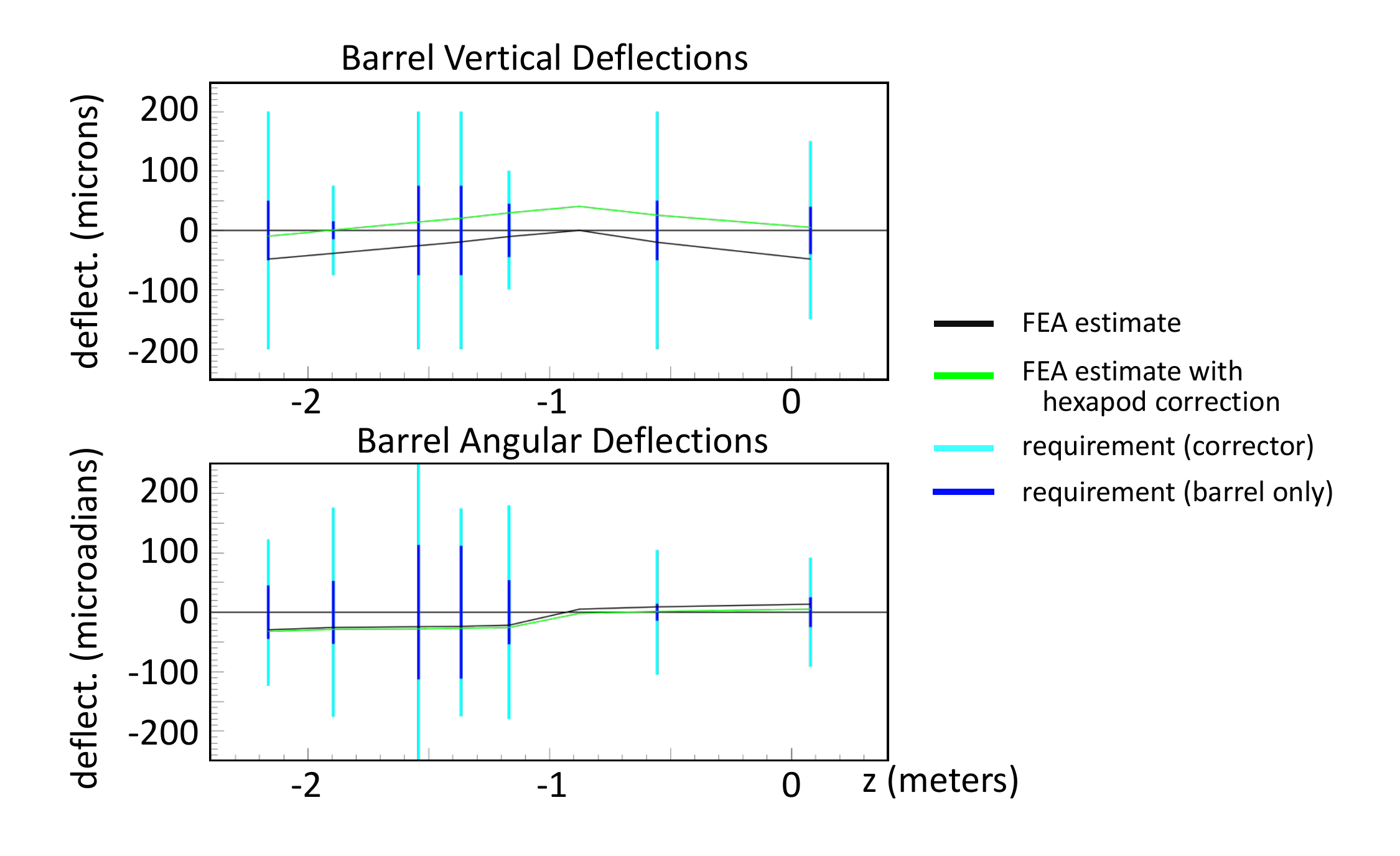}
  \caption{Barrel deflections predicted by FEA.  
Black lines show the deflections at the locations of the lenses and focal plane along the barrel axis.
Green lines show deflections after an overall rotation of the barrel to correct average deflections, as is performed by the hexapod. 
Light blue bars show the lateral and tilt requirements for the entire corrector.
Dark blue bars show the lateral and tilt requirements due to barrel deflections alone. 
The FEA prediction including the hexapod correction (green) is within requirements (dark blue).
The barrel is in the horizontal position and at a 0$^\circ$ rotation around the barrel axis.
Top: lateral deflections. Bottom: angular deflections.}
\label{fig:comp_21}
\end{center}
\end{figure}

Figure \ref{fig:comp_21} shows the FEA results for the barrel deflections compared to requirements from Table \ref{tab:lens_place} and Table \ref{tab:deflec}.  Deflections are found from the average of the deformations shown in the right plots of Figure \ref{fig:full_joe_18}.  Figure \ref{fig:comp_21} shows that some predicted deflections are larger than the requirements.  This is not a concern since image quality is primarily degraded by deflections of the lenses relative to each other and to the focal plane, and the full barrel can be rotated using the hexapod to correct for an overall barrel deflection.  Given this, the corrected deflections (green lines) satisfy all requirements.  Table \ref{tab:comp_deflec} summarizes the FEA results.  The corrected deflections are in column three, and columns four and five of the same Table show the FEA results for the case in which the barrel has been rotated 40 and 80\,degrees around its axis.

\begin{table}[htb]
\caption{Comparison of positioning requirements and the FEA results.}
\begin{center}
\label{tab:comp_deflec}
\begin{tabular}{ccccc}
\hline
\hline
 Element & Decenter Reqt & \multicolumn{3}{c}{FEA result for cases of barrel rotation around its axis} \\
\cline{3-5}
 & $\pm\mu$m &  0$^\circ$ &  40$^\circ$ &  80$^\circ$ \\
\hline
 C1    &  50 &    -9.9 &    -9.8 &    -9.7 \\
 C2    &  15 &     0.3 &     0.2 &     0.4 \\
 ADC1  &  75 &    13.7 &    13.5 &    13.6 \\
 ADC2  &  75 &    20.4 &    20.2 &    20.4 \\
 C3    &  45 &    29.7 &    29.8 &    29.6 \\
 C4    &  50 &    25.5 &    25.5 &    25.7 \\
 FP    &  40 &     5.3 &     6.4 &     5.5 \\
\hline
\hline
 Element & Tilt Reqt & \multicolumn{3}{c}{FEA result for cases of barrel rotation around its axis} \\
\cline{3-5}
 & $\pm\mu$rad &  0$^\circ$ &  40$^\circ$ &  80$^\circ$ \\
\hline
 C1    &  45 &   -32.3 &   -31.8 &   -31.9 \\
 C2    &  53 &   -28.9 &   -28.4 &   -28.5 \\
 ADC1  & 113 &   -28.1 &   -27.7 &   -27.8 \\
 ADC2  & 112 &   -27.0 &   -26.6 &   -26.7 \\
 C3    &  54 &   -25.5 &   -24.7 &   -25.1 \\
 C4    &  14 &     1.3 &     0.8 &     0.9 \\
 FP    &  25 &     5.0 &     4.6 &     0.0 \\
\hline
\hline
\multicolumn{2}{l}{Hexapod trans. in $\mu$m (\% range)} &480.8 (6.0\%)&474.2 (5.9\%)&464.8 (5.8\%)\\
\multicolumn{2}{l}{Hexapod rotat. in arcsec (\% range)} & 11.3 (4.5\%)& 10.5 (4.2\%)& 9.5 (3.8\%)\\
\hline
\end{tabular}
\tablecomments{The FEA was performed with the barrel in the horizontal position.  The bottom two rows give the virtual hexapod translations and rotations that eliminate the overall translation and rotation of the barrel as a unit.}
\end{center}
\end{table}

\begin{figure}[htb]
\begin{center}
  \includegraphics[scale=0.5,viewport=  15 412 273 650,clip]{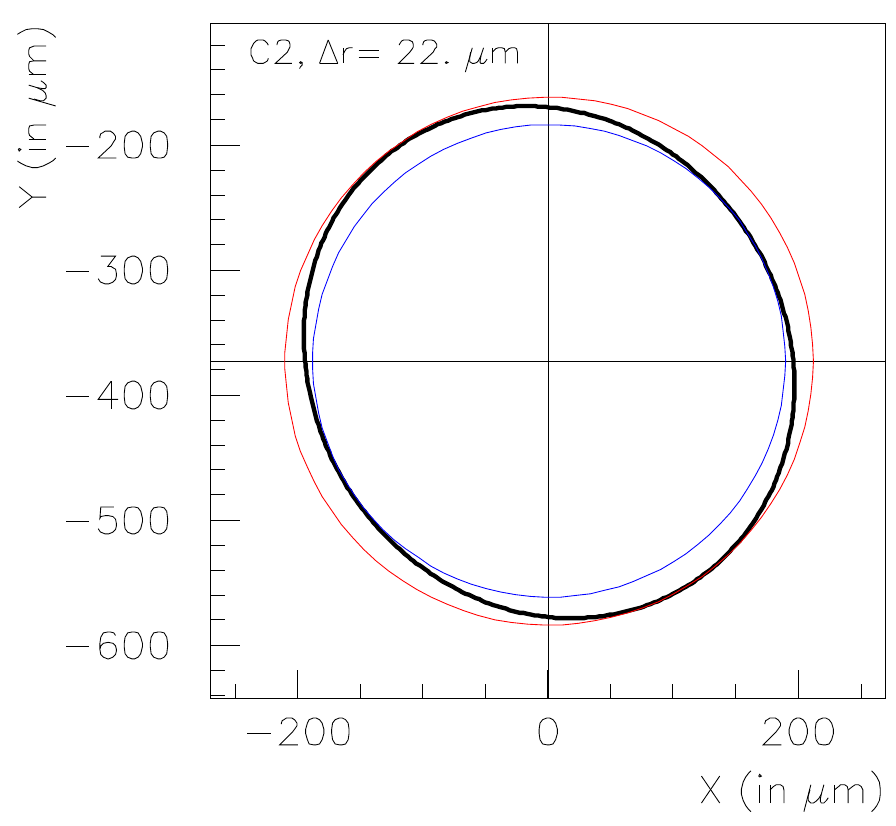}
  \includegraphics[scale=0.5,viewport= 273 412 531 650,clip]{Case1r_ofr.pdf}
  \includegraphics[scale=0.5,viewport= 273 163 531 398,clip]{Case1r_ofr.pdf}
  \includegraphics[scale=0.5,viewport=  15 412 273 650,clip]{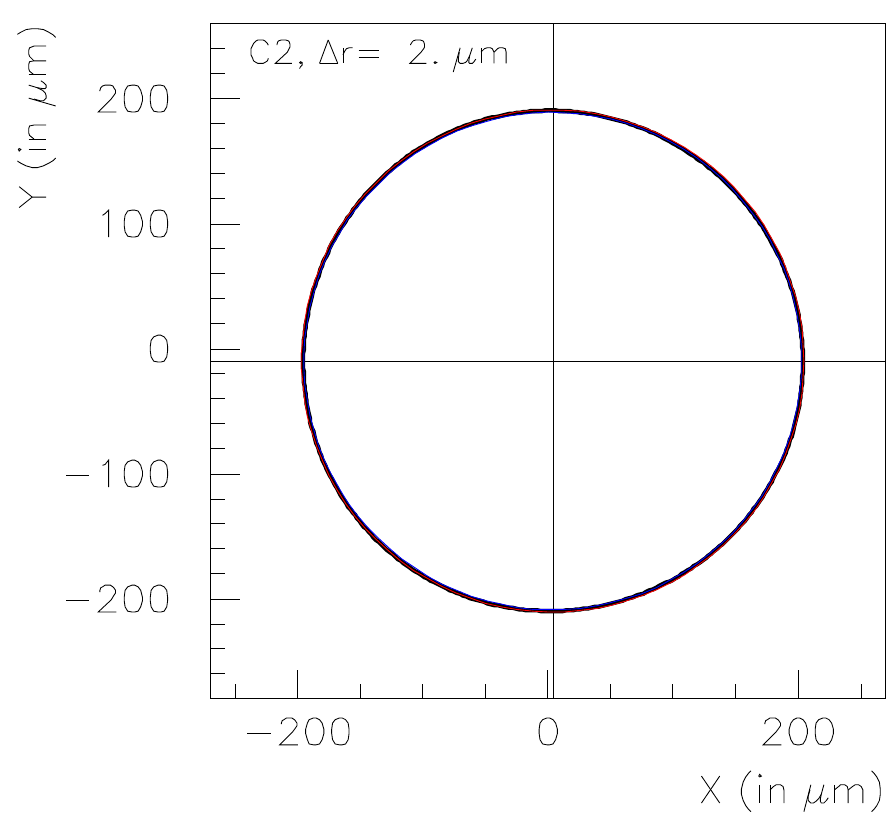}
  \includegraphics[scale=0.5,viewport= 273 412 531 650,clip]{Casev_ofr.pdf}
  \includegraphics[scale=0.5,viewport= 273 163 531 398,clip]{Casev_ofr.pdf}
  \caption{Barrel out-of-roundness plots.  The black curve shows the FEA results and the blue and red circles correspond to the minimum and maximum deformations.  The distance between the two circles is indicated by $\Delta r$. Top row: telescope is in the horizontal position and the barrel is rotated 0$^\circ$ around the barrel axis.  Bottom row: the telescope is in the vertical position.  Left to right: C2 flange, C4 flange, focal plane.}
\label{fig:ofr1}
\end{center}
\end{figure}

\subsubsection{FEA Analysis of Barrel Out-of-Roundness}

The barrel out-of-roundness for the telescope in the horizontal and vertical positions are shown in Figure \ref{fig:ofr1}.  The plots show the results for the C2 and C4 flanges and the focal plane because these are the worst cases.  The C1 flange is further away from the HMP than the C2 flange, but since the C1 flange is wider, because the Shroud also attaches to it, the out-of-roundness of C1 is smaller than that of C2.  The out-of-round deformations are produced by the axial forces exerted by the hexapod, through the HMP, on the barrel.  These axial forces produce deformation in the barrel, or out-of-roundness, that is opposite on both sides of the HMP.  The hexapod exerts forces on the HMP at essentially three points, so for example when the telescope is in the horizontal position, the forces produced by the HMP on the barrel shell are not perfectly axial, they also produce a moment.  The rotation produced by this moment will push the barrel shell on one side of the HMP and pull from the shell on the other side.  This will produce an out-of-round pattern that is 90\,degrees out of phase on either side of the HMP, and that is what is observed in the top row plots in Figure \ref{fig:ofr1}.  Also, since the HMP moments when the barrel is horizontal will produce a two-fold pattern on the same side of the HMP, e.g. top and bottom, one would expect a two-fold pattern on the out-of-roundness deformation and again this is what is observed in the FEA results.  When the telescope is in the vertical position the forces on the actuators of the hexapod are all the same, so there is a three-fold pattern of axial forces on the HMP which will produce a three-fold pattern of out-of-roundness in the barrel, and this is what is observed in the bottom row of plots in Figure \ref{fig:ofr1}.  The blue and red circles in Figure \ref{fig:ofr1} correspond to the minimum and maximum deformations.  The distance between the two circles is indicated by $\Delta r$.  The maximum deformation in this case is $45\,\mu$m.  We also performed FEA studies with the barrel rotated by 40 and 80\,degrees around its central axis.  In these cases the barrel out-of-roundness shape stays approximately the same but rotates.  The maximum out-of-roundness, $50\,\mu$m, occurs for the focal plane when the barrel is rotated around its axis by 80\,degrees.  Thus in all cases that were studied, $\Delta r$ does not exceed the 50\,$\mu$m specification of the DESI requirements (see Table \ref{tab:deflec}).

\subsubsection{FEA Analysis of Barrel Stresses}

We also performed FEA studies to understand the peak stresses in the corrector barrel structure.  We examined three cases.  The first two cases considered the fully-assembled barrel in the horizontal and vertical positions respectively.  The third case considered the barrel to be in the horizontal position but without the focal plane assembly (FPA).  This last case simulates the FPA installation when the telescope is at the maintenance platform, fully assembled but without the FPA.  In all cases the stresses are well below the requirement of one fourth of both the Yield and the Micro-Yield Stress.

\subsection{Alignment}\label{sec:alignment}

\begin{figure}
\begin{center}
  \includegraphics[width=0.45\textwidth]{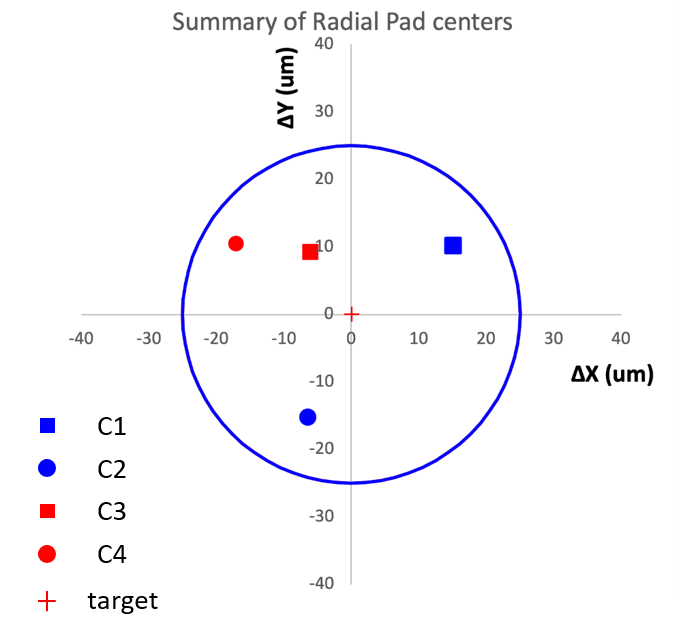}
  \includegraphics[width=0.45\textwidth]{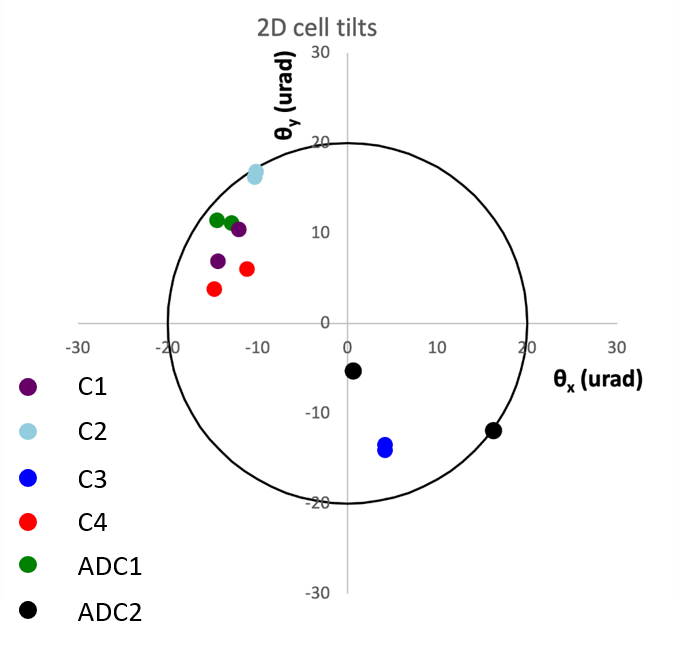}
  \caption{Lens cell centers and tilt measurements after the last alignment iteration at FNAL.  The left plot shows the centers of the C1, C2, C3 and C4 cells in the barrel coordinate system.  The center of this coordinate system was defined as the best fit to the center of rotations of the ADC1 and ADC2 cells.  The plot on the right shows the tip/tilt measurements of all the cells, the double sets of points in this plot correspond to two different measurements, giving an idea of the measurement repeatability errors.  The large circles show the alignment requirements. \citep[Left image previously published in][]{gutierrez18}}
\label{fig:cell_alignment}
\end{center}
\end{figure}

The assembly and alignment of the corrector was performed in stages at UCL and Fermi National Accelerator Laboratory (Fermilab or FNAL).
The process started at FNAL where the flatness and positions of the barrel-barrel section flanges and the barrel-cell flanges were measured using a large-range CMM, the Hexagon Delta Slant 30-51-25.
The lens cells were then integrated into the barrel sections and the barrel sections assembled.
Using the CMM the cells were set at their optimal position for centering the lenses and measurements were made of various external and internal reference surfaces which were subsequently used during lens installation at UCL.
The cells and barrel were drilled and doweled to facilitate their removal and subsequent replacement after the lenses are mounted in them. 
The accuracy of centering and realignment on the dowels is of order $\pm10~\micron$ per pin.
The doweled cells and barrel sections were then shipped to UCL where alignment and assembly of the full corrector system was performed.

Figure \ref{fig:cell_alignment} shows the measurement of the centers and the tilt of the cells after the last alignment iteration at FNAL.  All cells were measured at the same time using the large CMM at FNAL.  The barrel's $z$-axis runs very close to the center of the barrel and the $xy$-plane is parallel to all the barrel flanges that support the cells.  The center of the $x-y$ coordinate system was defined as the average of the center of rotations of the ADC1 and ADC2 cells.  The left plot in Figure \ref{fig:cell_alignment} shows the measurements of the C1, C2, C3 and C4 cell centers.  The right plot shows the tilt angles for all the cells, including ADC1 and ADC2.  The big circles in both figures show the alignment requirements.  In the case of the centers the requirement included the $\pm 20\,\mu$m for the cell centers and the $\pm 5 \mu$m for the assembly/disassembly repeatability requirements.  

After checking that the requirements were satisfied, the barrel was fully disassembled and the cells were pinned.  After pinning, the cells were removed from the barrel and reinstalled, the expandable pins were inserted back in the cell base rings, and the barrel was fully assembled and remeasured.  In this measurement of repeatability, all cell centers and tilt angles were found to be within requirement with the exception of ADC2 (see Figure \ref{fig:cell_alignment}, right).  The ADC2 cell tilt was at the limit of the requirement and was deemed acceptable.  Given the size and weight of the barrel we consider the repeatability of the cells to be quite remarkable.

At UCL the axial RTV pads were first installed in the lens cells. Each cell was placed in turn on a rotary table and adjusted using dial gauge measurements so that the previously measured centering and tilt reference surfaces on the cell were aligned such that the center of the x-y coordinate system defined by FNAL was centered on the rotary table. The axial pads were then glued to the axial inserts (the exception was the C3 axial pads that were glued directly to the cell) and these were then positioned in the cell using the feedback from a Micro-Epsilon laser sensor so that the RTV pad surfaces ran true to the cell center to an accuracy of better than $\pm$20\,$\mu$m. To attach the RTV pads to the inserts a thin layer of RTV560 was used. The metal insert surfaces (and the glass lens surfaces for the radial pads) were first treated with a primer, SS4004P from Momentive, before gluing.  

The positioning of the axial pads for the ADC2 lens was slightly more involved. As the wedge for the lens is expressed on the axial pad side, the axial pads were aligned by first offsetting the lens cell by an amount corresponding to the lens surface offsets and then aligning the pads to a common height. When the cell is re-centered this produced the correct pad heights for the tilted surface.  
Once the axial inserts were installed in the cells, the interior cell surfaces were painted black using Lord Corporation Aeroglaze 9947 primer and Z306 top coat. 

\begin{figure}
\centering
\begin{minipage}[b]{0.45\linewidth}
\includegraphics[width=\textwidth]{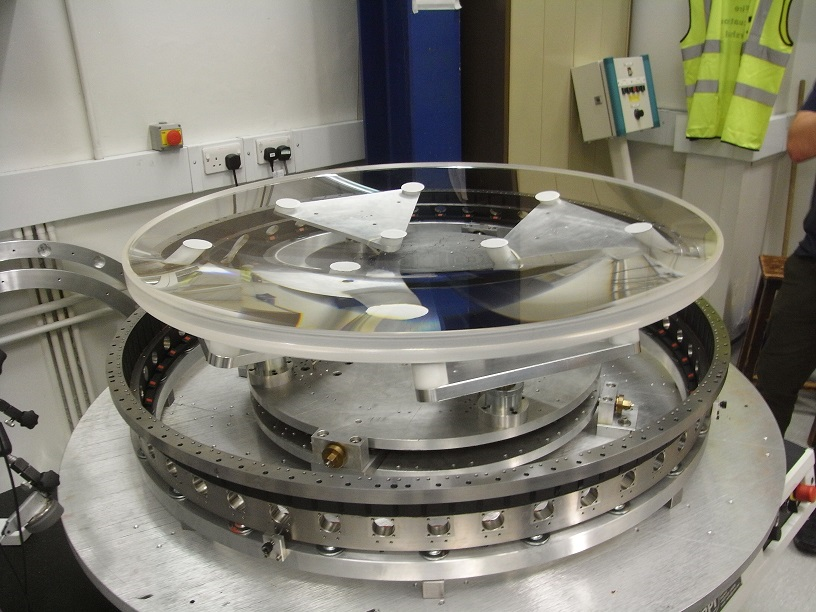}
\end{minipage}
\hspace{0.25in}
\begin{minipage}[b]{0.45\linewidth}
\includegraphics[width=\textwidth]{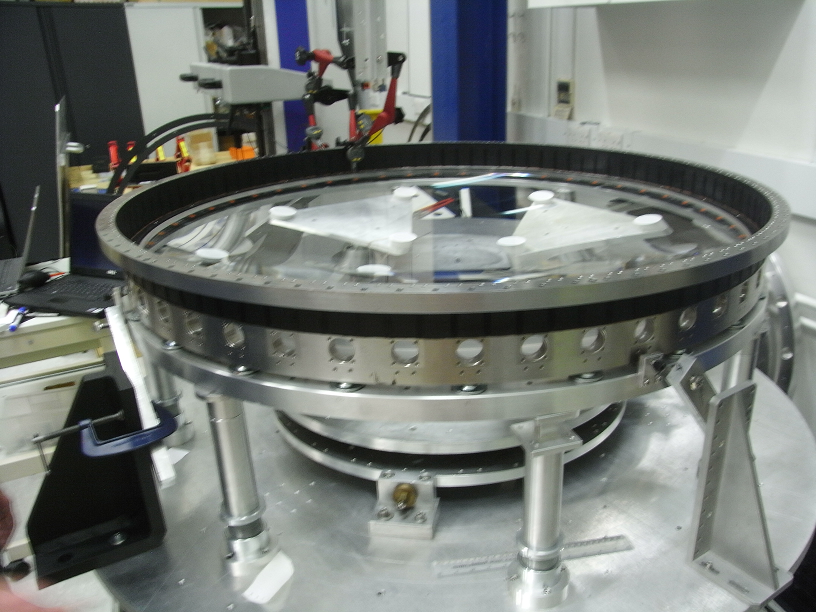}
\end{minipage}
\caption{Integration of the C1 lens and its cell.  Left: C1 lens and cell on alignment support structure.  Right: C1 lens mated with axial pads.}
\label{fig:C1onstandandinCell}
\end{figure}

The lenses were then installed in the cells using the following procedure. The lens cell was set up on the rotary table, along with the lens and cell alignment systems. The corresponding lens was then lifted from its crate using a custom shipping/handling frame. The shipping frames, designed and built by Berkeley, incorporated a lens flipping mechanism, necessary since all lenses except C3 were shipped upside-down from their installation orientation.  The lens was flipped if needed, and lowered onto a distributed support structure mounted on an XYZ adjustment system on the rotary table (Figure~\ref{fig:C1onstandandinCell}, left).

We next centered and levelled the lens using Sylvac digital dial gauge indicators, accurate to 1\,$\mu$m, mounted on a stable metrology tower.  We then raised the lens cell until it was within a few hundred micrometers of the lens, and levelled and centered the cell such that its x-y coordinate system center was on the rotary axis. Next we raised the cell carefully until the entire weight of the lens was transferred to the axial RTV pads installed on the cell.  Finally we measured the alignment of the lens to the cell using dial gauges and confirmed the contact between the axial pads and the lens surface (Figure~\ref{fig:C1onstandandinCell}, right).

\begin{figure}[h]
\centering
\includegraphics[height=2.5in]{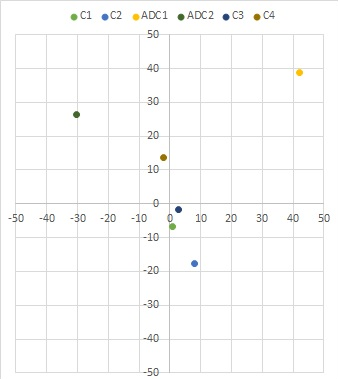}
\caption{Measured geometric centers of lens diameters.  Measurements are referenced to the barrel coordinate system in microns, after installation in cells. The measurements confirmed that the lenses remained well-centered after the radial pads were glued in place.}
\label{fig:LensCent}
\end{figure}

Next we installed the radial inserts and pads.  The radial inserts had previously been removed from the cells and their RTV pads attached to them using the same method as with the axial inserts. The RTV pads were produced by casting a rectangular sheet in a precision mold and cutting out the required circular pads with a stamp. The thickness of the pad was chosen to leave a small gap between the pad and the lens when the insert is installed, and this gap was filled with a thin glue layer of RTV when the insert and pad were installed in the cell. To minimize excess, the glue was applied using a doctor blade fixture: a roller was run over the glue while resting on thin guides surrounding the glue area, assuring a uniform, well-defined thickness. We rechecked the position of the lens relative to the cell after the gluing was complete and the cell retaining ring installed.  Figure~\ref{fig:LensCent} shows the final measurements of the lens centers.  Note that the ADC measurements were made for only one position of the cell; if the cell were rotated by the ADC mechanism then the center position would change. In this case the cell centers were aligned to the mechanism axis of rotation to within 25\,microns.  Next we compared the position of the lens centers to their corresponding cell mounting flange with a Faro gauge arm, and we used this measurement to determine the spacer thickness required to give the correct lens to lens spacing when installed in the barrel. Finally we painted the bare parts of the cell black, with the exception of the mounting flange. 

Almost all completed cells met their required lens-to-cell alignment tolerances.  The exception was the ADC2 lens which showed a higher tilt than specified.  This was due to the wedge angle of the lens being slightly smaller than nominal (though still within its manufacturing tolerance). This was expressed as an overall tilt of the lens of 120\,$\mu$rad. However, the combined tolerance of the lens tilt and wedge meet their higher-level requirement, and so DESI accepted this exception.

Once installation of the lenses into their cells was completed, the lens-cell assemblies were then installed in the barrel sections. The lowermost lens (closest to primary mirror) was installed first in each section. The barrel section was lowered until it was close to the lens cell assembly with its spacers in place. The fiducial dowels were then fitted into cell and barrel flanges. The lens-cell assembly was then carefully bolted to the barrel section. The position of the lens relative to the flanges of the barrel section were then measured using the Faro gauge arm. 

\begin{figure}
\centering
\includegraphics[height=3in]{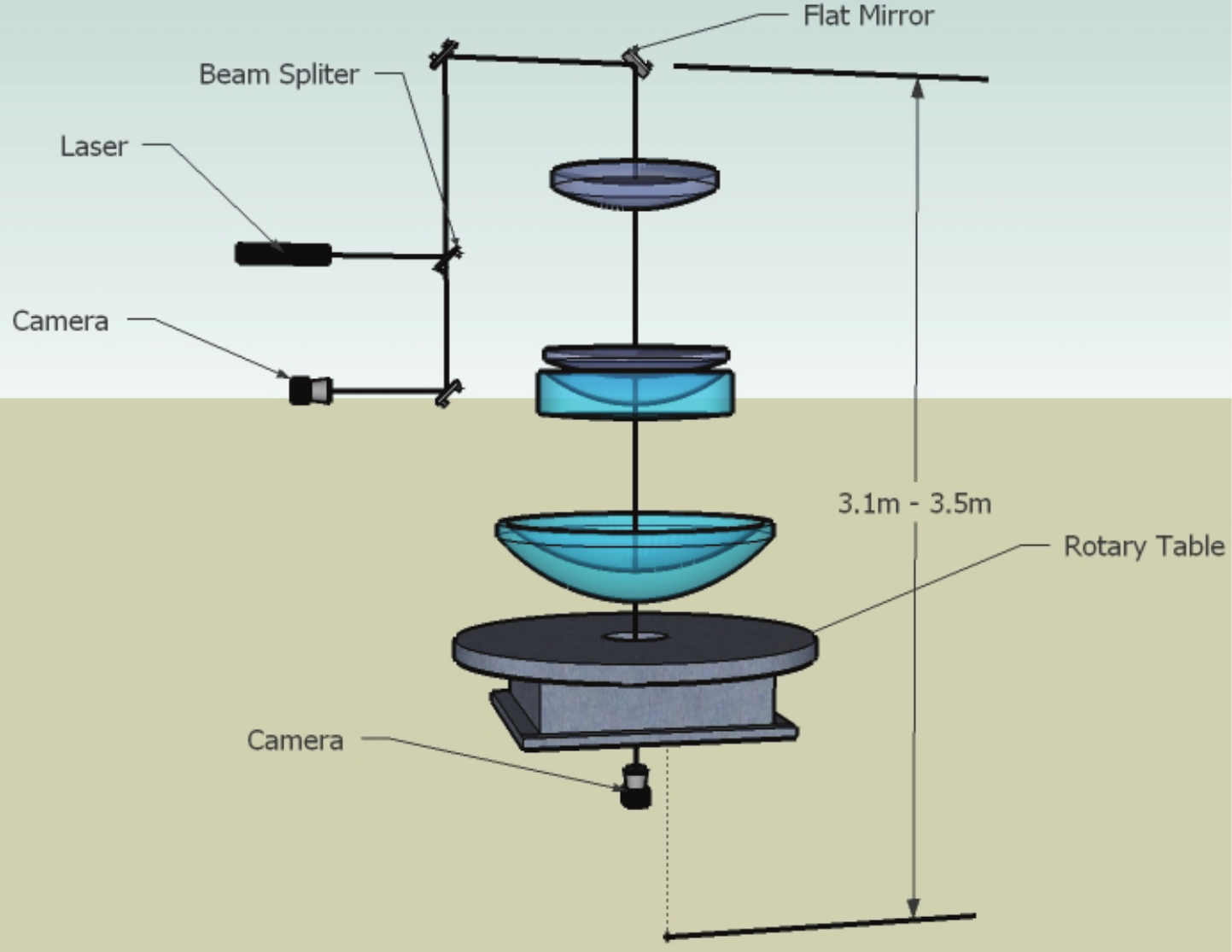}
\caption{Schematic of the pencil laser beam alignment system successfully used for the DECam corrector.  The DESI system was identical to this one. \citep[Image previously published in][]{brooks16}}
\label{fig:laseralign}
\end{figure}

The alignment of the lens was also measured using an optical laser pencil beam method, as shown in Figure~\ref{fig:laseralign}. We mounted the laser beam alignment system on a MiniTec modular aluminium frame system and co-aligned it to the rotary axis of the rotary table. The barrel section was aligned using previously measured external reference surfaces such that the chosen barrel axis position was on the rotary axis. By measuring the position of the return laser beam and the transmitted beam when the optic was rotated, we determined the tilt and decenter of the lens unambiguously.  We then installed the second lens-cell assembly into the section by lowering it with a crane onto the section, with spacers placed, without disturbing the laser beam. The fiducial dowels were inserted and the lens-cell pair bolted to the barrel. We then repeated the laser beam test along with Faro gauge arm and dial gauge measurements to establish the position of the second lens in the barrel section.

Once the lenses and cells were mounted in their respective barrel sections, we assembled the complete corrector.  The Front barrel section was placed on the rotary table and centered using the external reference surfaces measured at FNAL. The Mid section was then lowered until it was close to the Front section and the fiducial pins inserted. The sections were then bolted together.  Next we lowered the Aft section onto the Mid section and again inserted the fiducial pins and bolted the parts together. Finally we measured the relative centers of the three sections using the external reference surfaces on each barrel section. After confirming the alignment of the sections, we disassembled the corrector into its individual sections for shipping.

The assembly and alignment plan for the lenses, cells, and barrel are fully described in \citet{brooks16}.

\begin{figure}
\centering
\includegraphics[height=2.5in]{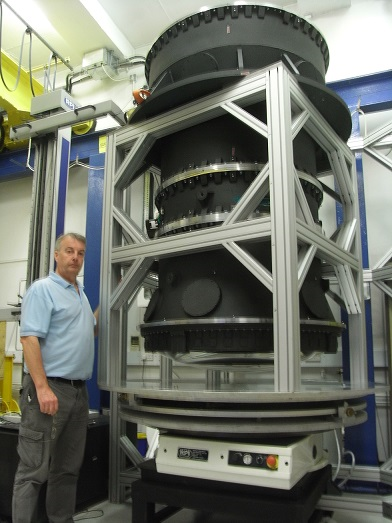}
\caption{The fully assembled corrector barrel assembly at UCL.  The assembly (black) is mounted on the UCL rotary table with its hexapod flange resting on mounting structure (gray).}
\label{fig:CorrectorUCL}
\end{figure}

\vspace{5mm}
\subsection{ADC Rotator}\label{sec:ADC_rotator}

\begin{figure}
\begin{center}
  \includegraphics[height=.45\textwidth]{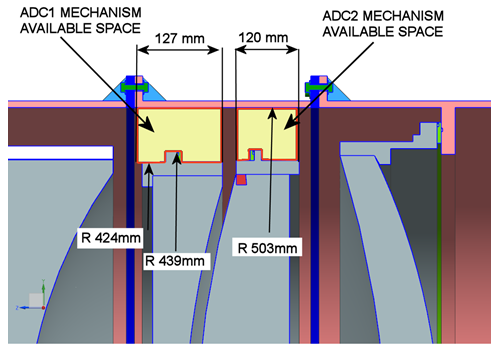}
  \caption{Available space to accommodate the ADC1 and ADC2 rotation mechanisms.}
\label{fig:adc-space}
\end{center}
\end{figure}

Table \ref{tab:rotator_reqs} lists the main requirements for the ADC rotator, the mechanism that rotates each ADC lens around the optical axis in order to correct for atmospheric dispersion.  The first three were DESI requirements for the ADC assembly, and the next four were allocated limits within the total ADC alignment tolerance budget.  The last requirement is the volume allocation for the rotator to fit within the ADC subassembly, shown in Figure \ref{fig:adc-space}.  The finished ADC rotator met all these requirements.  The values for the speed and ramp-up/ramp-down time requirements were flowed down from an analysis that showed that the ADC movement time given these values never dominated the total time needed for telescope reconfiguration between pointings.  The ADC movement time within the exposure sequence is described in more detail in \cite{abareshi2022}.

\begin{table}
\centering
\caption{ADC rotator requirements.}
\begin{tabular}{c|c} \hline 
Position accuracy	&	$\pm$ 0.1 degrees	\\
Maximum speed	&	5 degrees second$^{-1}$ \\
Ramp-up/ramp-down time	&	$<$ 3 seconds	\\
Static lateral tolerance	&	20 microns	\\
Static tilt tolerance	&	25 microrad	\\
Dynamic lateral tolerance	&	50 microns	\\
Dynamic tilt tolerance	&	80 microrad	\\
Maximum volume	&	Within allocated volume in lens cells	\\
\hline
\end{tabular}
\label{tab:rotator_reqs}
\end{table}

\begin{figure}
\begin{center}
  \includegraphics[height=.35\textwidth]{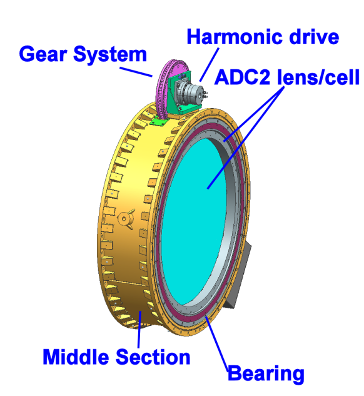}
  \hspace{4 mm}
  \includegraphics[height=.35\textwidth]{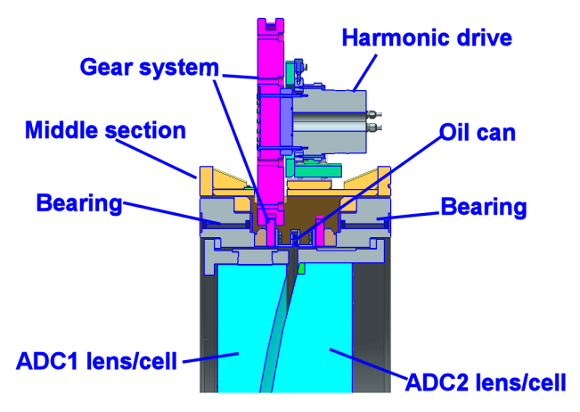}
  \caption{The ADC rotator.} 
\label{fig:ADC_solid_model}
\end{center}
\end{figure}

\begin{figure}[htb]
\begin{center}
  \includegraphics[height=.40\textwidth]{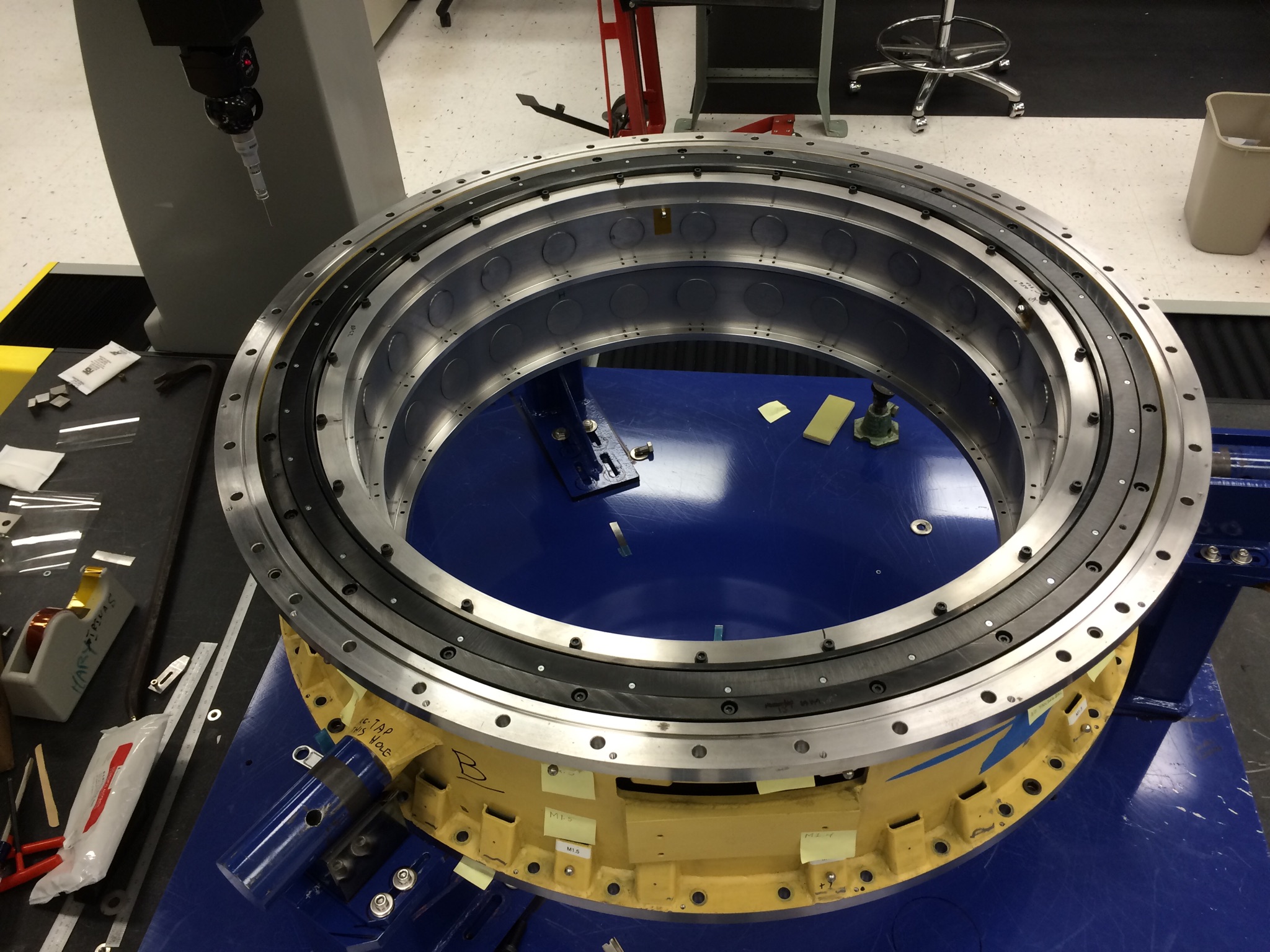}
  \caption{The ADC rotator.  The rotator is shown being aligned in the Middle section of the barrel using a CMM at FNAL.  The ADC1 and ADC2 lens cells are visible in the inner diameter.  One of the two custom bearings between the cells and barrel in visible in the lower left.}
\label{fig:ADC-picture_CMM}
\end{center}
\end{figure}

In this section we first describe the components of the ADC rotator, and then we show the performance of the entire system.  Figure \ref{fig:ADC_solid_model} shows two views of the ADC rotator.  The main components of the rotator are the two ADC1 and ADC2 lenses, the ADC1 and ADC2 lens cells housing the lenses, the custom bearings allowing for the rotations of the cells, the gear and pinion system used to rotate the bearings, and the rotary actuator that rotates the pinion.  The custom bearings are attached to the Middle section of the barrel.  Figure \ref{fig:ADC-picture_CMM} shows the ADC rotator installed in the Middle barrel section during alignment at FNAL.

The custom bearings were made by Kaydon.  Standard catalog bearings were considered, but a custom design turned out to be the simplest solution that met the requirements.  The bearing uses ceramic balls in order to minimize the amount of grease needed to lubricate the bearing.  The bearing preload was adjusted to make sure there would be no play in the bearings for any direction of the gravity vector.  

The bearings are driven by a gear and pinion system custom-made by Nexen, and was also designed to run with a minimal amount of lubrication.  Eight gear segments are attached to each bearing, which in turn are driven by an external pinion as shown in Figure \ref{fig:ADC_solid_model}.  The pinion makes contact with the gears attached to the bearings through a series of rollers positioned around the circumference of the pinion, which enables the system to run with little lubrication.  The system has no backlash, which is desirable to avoid any vibrations during operation.  To set the rotation of the lenses, we relied on the encoders in the harmonic drive motor for feedback, since there wasn't available volume to install external measuring devices inside the barrel's Middle section.  We aligned the eight gear segments onto each bearing at FNAL using custom tools provided by Nexen.  The distance between the pinion and the gears varied slightly after alignment, but we set an appropriate pressure for the pinion against the gears to avoid backlash.

Each pinion is driven by a rotary actuator motor, model SHA25A-SG from Harmonic Drive LLC.  The motor includes absolute rotational encoders that provide enough bits to allow the lenses to rotate by dozens of turns before having to reset the encoders.  The resolution of these encoders is much better than a millimeter across the circumference of the lenses.  Limit switches on the bearings provide a home position for each of the lenses with better than a millimeter precision.  
The rotary actuators provide the necessary gear reduction by use of a harmonic drive, which has the advantages of compactness, zero backlash, high repeatability, and high position accuracy.  In this mechanism an elliptical wave generator pushes a flexspline gear against a circular spline that has two more teeth than the flexspline; this difference in teeth provides the gear reduction.  

One potential disadvantage of this system is that irregularities, repeating every 180\,degrees because of the elliptical wave generator, have the potential to drive mechanical resonances present in the system.  Thus, mechanical resonances can make the system vibrate for certain actuator shaft rotational velocities.  This happened in fact during characterization of the actuators, when rotating the ADCs at 10\,degrees second$^{-1}$.  However, we identified the resonances in testing and confirmed they did not occur at the 5\,degrees second$^{-1}$ maximum speed requirement (see Table \ref{tab:rotator_reqs}).

\begin{figure}
\begin{center}
  \includegraphics[height=.3\textwidth]{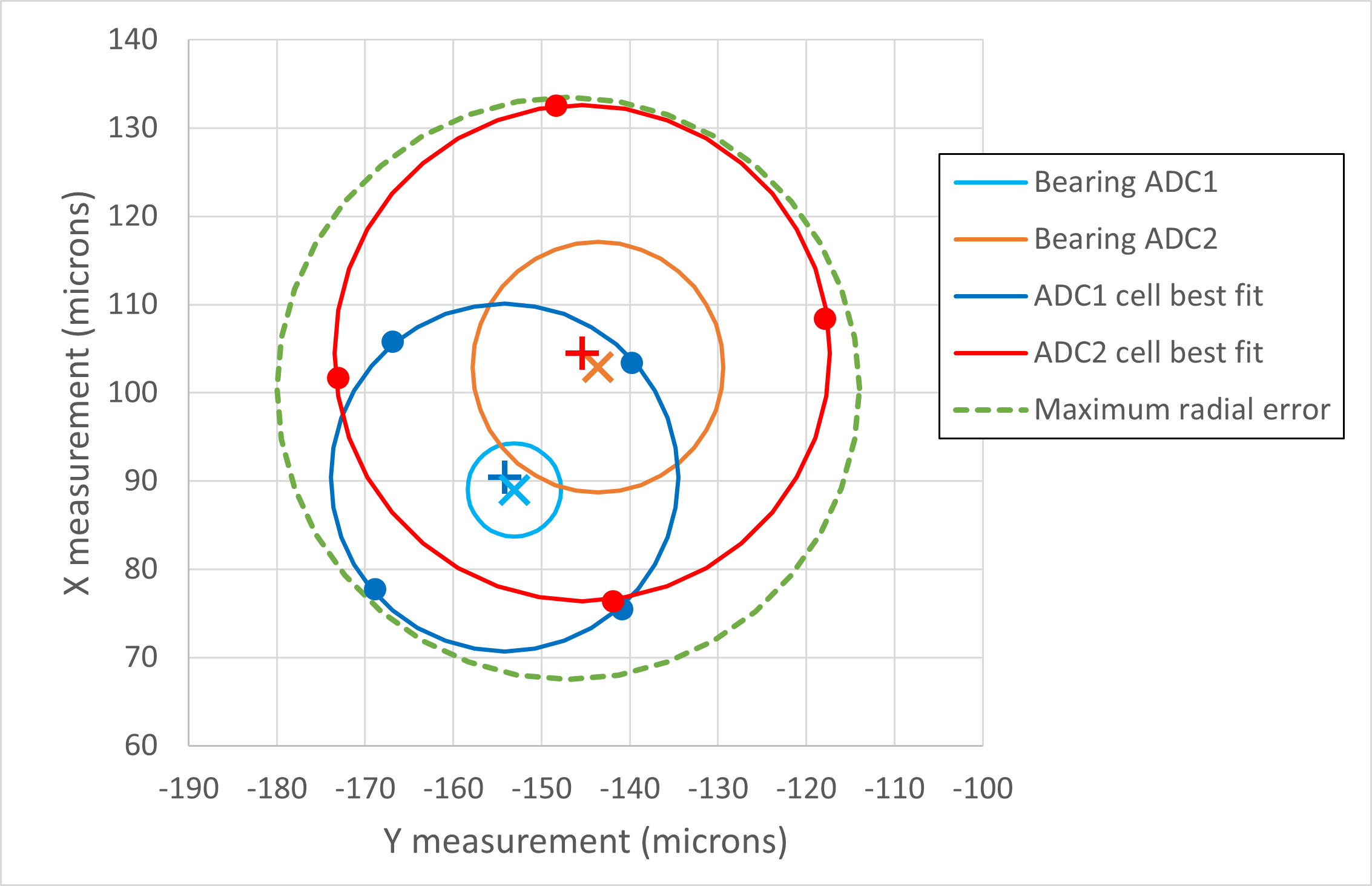}
  \hspace{6 mm}
  \includegraphics[height=.3\textwidth]{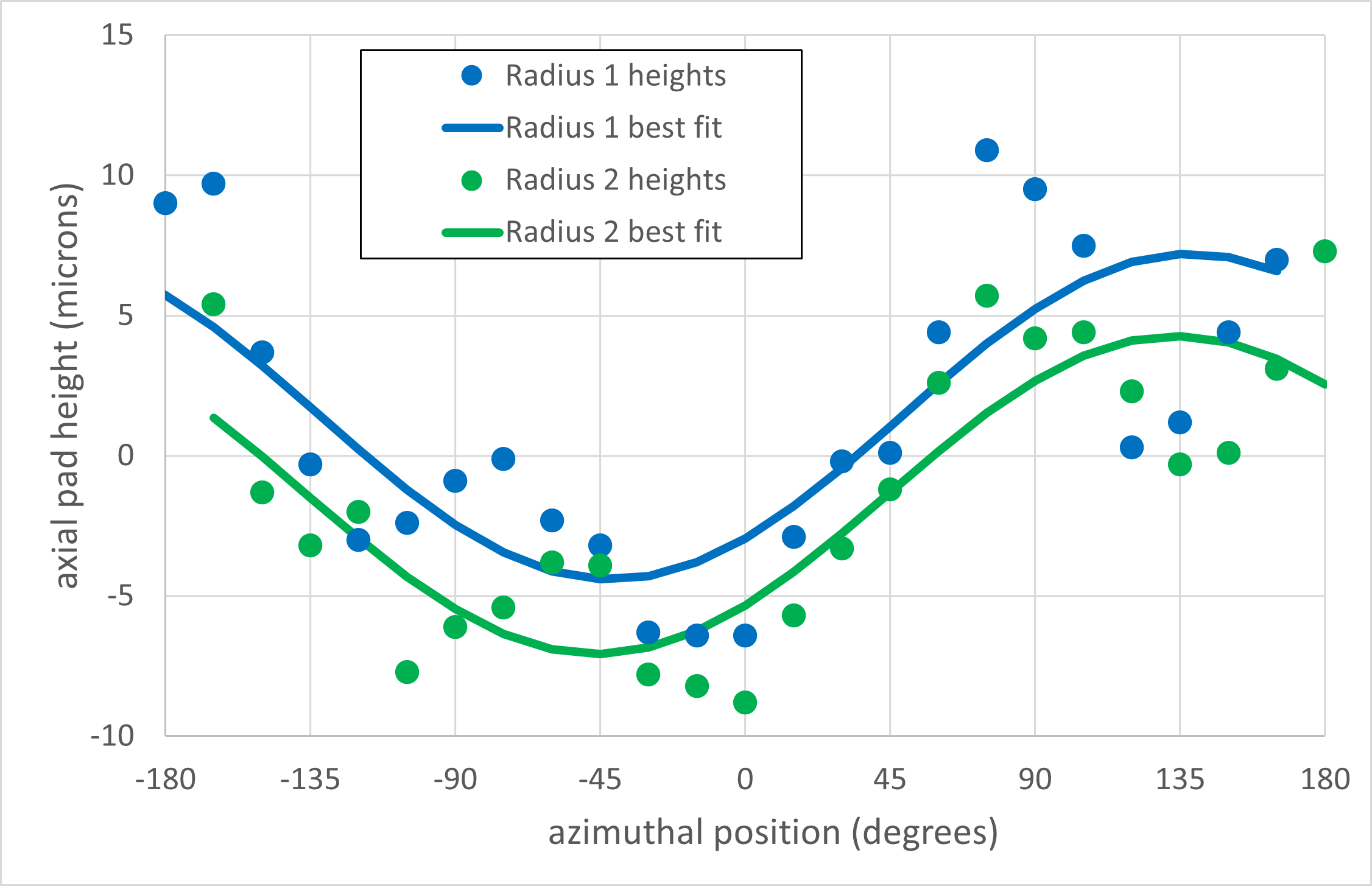}
  \caption{Final measurements after aligning the the ADC rotator bearings to the ADC lens cells.  Left: radial measurements for the bearings and the lens cells.  Right: an example of a tilt measurement for the ADC1 lens cell, showing the height of points on its axial pad surface in microns.}
\label{fig:adc-radial}
\end{center}
\end{figure}

To align the ADC bearings and the lens cells, we used a large CMM at FNAL to measure their centers.  The coordinate system for the measurements was defined to be centered on the barrel's Middle section, with the $xy$-plane parallel to the front face of the Middle section.  The center of rotation of the bearings was determined by measuring the inside surface of the ADC1 and ADC2 bearings over several rotation angles, and fitting a circle to the data.  The left plot in Figure \ref{fig:adc-radial} shows the results of this measurement; the data points represent the vector difference of the measured point from its nominal location.  The light blue and orange circles show a fit to the ADC1 and ADC2 bearing measurements, respectively.  The center of these circles, marked with crosses, determine the center of rotation of the bearings.  If the inside surface of the bearings had been perfectly centered with the bearing's center of rotation then the radius of these circles would be zero.

After installing and aligning the ADC1 and ADC2 lens cells, we performed similar measurements on the lens cells by measuring their radial pad locations.  The measurements and fitted circles are overlaid in Figure \ref{fig:adc-radial} (left), shown by the dark blue and red circles for ADC1 and ADC2 respectively.  The red and blue crosses at the center of those fitted circles determine the center of rotation of the cells.  Note that the centering of the cells is slightly worse than the centering of the bearings.  The green circle in the Figure encompasses all the decenterings of the bearings and cells, and has a radius of 33\,$\mu$m.  To this we can add the errors due to the loading of the bearings for different ADC rotator orientations; these were measured at Kaydon before the bearings were delivered to FNAL.  The final results are listed in Table \ref{tab:adc_radandtilt}.  The average of the center of rotations of the ADC1 and ADC2 bearings, i.e. the crosses in Figure \ref{fig:adc-radial} (left), was used to define the axis of the barrel, and all other lenses were aligned to this axis.

\begin{table}
\begin{center}
\caption{ADC rotator radial and tilt measurements.}
\label{tab:adc_radandtilt}
\begin{tabular}{c|ccc|ccc|cc|cc}
\hline
& \multicolumn{6}{|c|}{Radial} & \multicolumn{4}{|c}{Tilt} \\
& \multicolumn{6}{|c|}{(microns)} & \multicolumn{4}{|c}{(microradians)} \\
\hline
 & \multicolumn{3}{|c|}{Measurements} &  \multicolumn{3}{c|}{Requirements} & \multicolumn{2}{c|}{Measurements} & \multicolumn{2}{c}{Requirements} \\
\hline
 Cell & dynamic+static  & bearing & \textbf{total} & dynamic & static & \textbf{total} & \textbf{static}  & \textbf{dynamic}  & \textbf{static} & \textbf{dynamic} \\
\hline
ADC1 & $\le$ 33 & $\le$ 11 & $\le$ \textbf{44} & $\le$ 50 & $\le$ 20 & $\le$ \textbf{70} & \textbf{7}  & \textbf{10}  & \textbf{$\le$ 25} & \textbf{$\le$ 80} \\
ADC2 & $\le$ 33 & $\le$ 6 & $\le$ \textbf{39} & $\le$ 50 & $\le$ 20 & $\le$ \textbf{70} & \textbf{5-9}  & \textbf{4}  & \textbf{$\le$ 25} & \textbf{$\le$ 80} \\
\hline
\end{tabular}    
\tablecomments{Static radial requirements refer to the alignment of the cells and the bearings, and the dynamic radial requirements refer to the runout of the bearings. Their requirements were specified separately, but in the fully assembled system it is only possible to measure the combination of both errors. The static tilt values are the maximum measured tilt of the cell axial pad surface, and the dynamic tilt values are the maximum measured tilt of the bearings.  The total measured errors meet their requirements.}
\end{center}
\end{table}

To measure the tilt alignment of the lens cells, we measured the axial pad surfaces using the same CMM.  An example of these measurements is shown in the right plot in Figure \ref{fig:adc-radial}.  This shows two sets of measurements, each taken at different radii, of the axial pad surface height as a function of the azimuthal angle around the ADC1 lens cell.  A surface that is perfectly flat, and perpendicular to the axis of rotation, would show a line along the horizontal axis.  Sinusoidal deviations from this line show that the plane formed by the axial pad surface is tilted relative to a plane perfectly perpendicular to the barrel's axis.  By repeating these measurements for different rotations of the cell, we could then extract both the static and the dynamic tilts of the planes formed by the axial pad surface, and thus find the tilt of the ADC1 and ADC2 lenses.  Table \ref{tab:adc_radandtilt} shows the final calculated tilt measurements after the ADC rotator was aligned.  The measured tilts are all well below their requirements.

\subsection{Hexapod} \label{sec:hexapods}

The corrector barrel must be held in its nominal position with respect to the primary mirror while the telescope is slewed over the full range of Zenith angles and also while enduring the range of temperature variations in the Mayall dome.  Therefore a hexapod is necessary to adjust the barrel position in six degrees of freedom accurately and quickly.  To align the barrel to its best position during operation, the DESI Active Optics System (AOS) measures its misalignment error and calculates the needed actuator motions.  The error signal is measured by four sensors in the focal plane that record wavefront errors by observing out-of-focus stars in the field of view (Silber 2022, submitted).  The AOS is based on a similar algorithm used by DECam \citep{roodman14}.  

\begin{figure}[htb]
\begin{center}
  \includegraphics[width=.90\textwidth]{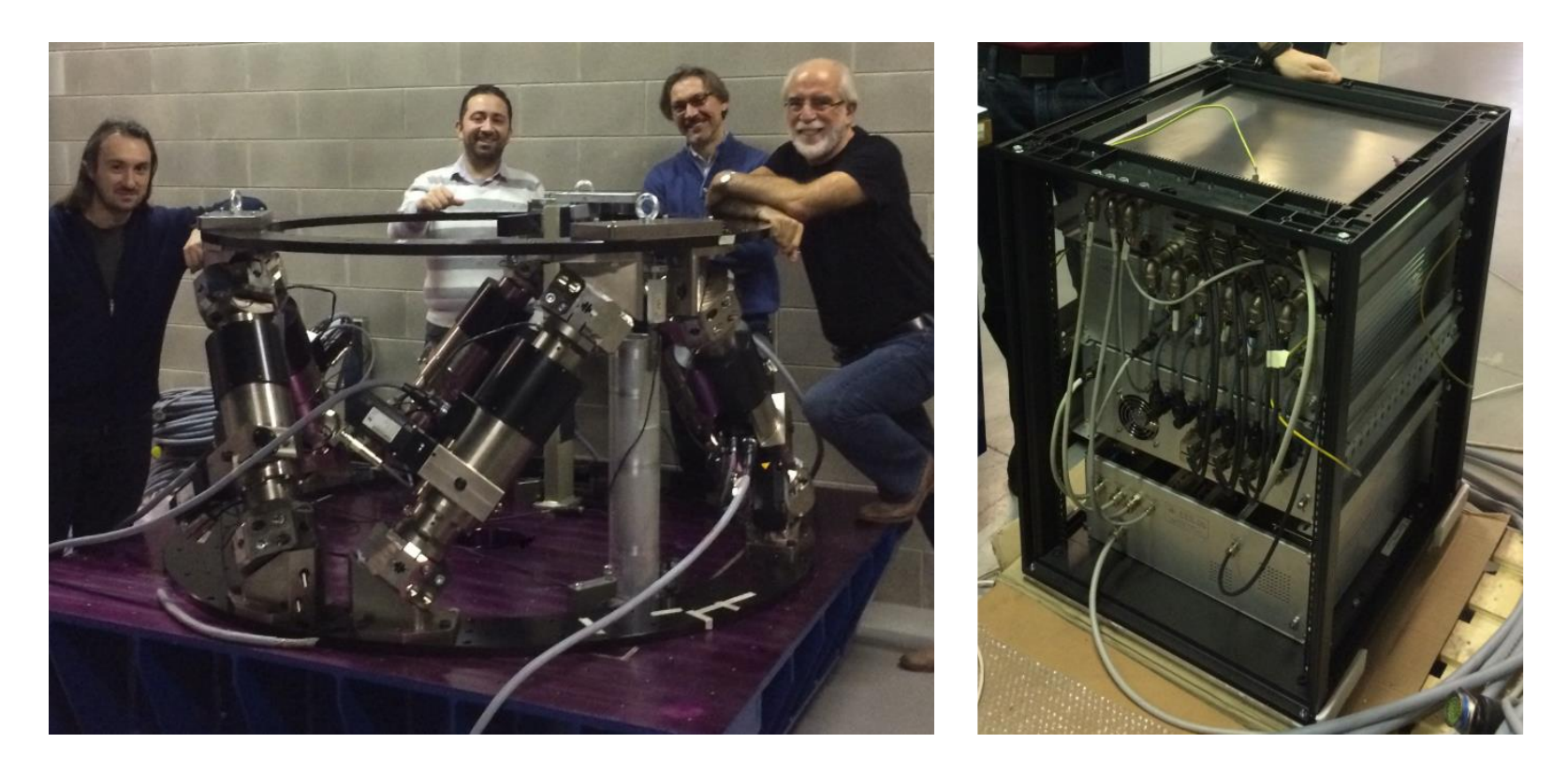}
  \caption{Left: the completed hexapod at A.D.S. International.  Right: the Hexapod Control Unit.}
\label{fig:hexapod_picture_ADS}
\end{center}
\end{figure}

The DESI hexapod was designed and manufactured by A.D.S. International S.r.l., as was the DECam hexapod.  Figure \ref{fig:hexapod_picture_ADS} shows the DESI hexapod and the Hexapod Control Unit (HCU) at A.D.S. before their delivery to DESI; the hexapod is also shown attached to the barrel in Figure \ref{fig:hex_oneside}.  The DESI hexapod is very similar in design and construction to the DECam hexapod, though the DESI version has a larger overall diameter, and a smaller vertical range which was exchanged for a larger range in tip and tilt.  

The hexapod system consists of six actuators, a mechanical joint with two degrees of freedom at both ends of each actuator, one stationary plate, a motion plate, and the HCU. The stationary plate bolts to the prime focus cage, and the motion plate bolts to the barrel. The six actuators are arranged in three triangles, with the apex of each triangle located where two actuators meet at the motion plate. The actuators at the base of each triangle are attached to the stationary plate that is fixed to the cage.  The hexapod effectively provides a three-point support to the motion plate.

The HCU controls all motions of the hexapod, and can be used in stand-alone manual operation mode or a remote operation mode. In remote operation mode, the controller moves the hexapod according to the position command it receives from the DESI Instrument Control System \citep[see e.g.][]{honscheid18}. It also returns hexapod status information to the control system in either mode. 

The flexure joints at the end of each actuator provide smooth, accurate motions without the stick-slip problems that, for example, a universal joint could have.  These flexures allow for the precise motion of the hexapod at the level of 1\,$\mu$m.  The measured stiffness of the seven DESI actuators (six installed in the hexapod plus one spare), with their corresponding flexures, ranged from 157 to 165\,N/$\mu$m.  This stiffness was necessary to minimize the deflections of the barrel with respect to the cage.  The hysteresis for the seven actuators plus flexures, for a $\pm$30\,kN maximum load, ranged from 3 to 8\,$\mu$m, and there was zero measured motion along the actuators during the transition from tension to compression.

\begin{table}
\begin{center}
\caption{Hexapod positioning requirements.}
\label{tab:corrmech}
\begin{tabular}{lc}
\hline
 Item & Requirement \\
\hline \hline
Hexapod resolution: & \\
\hspace{3mm} Lateral motion & 15~\micron \\
\hspace{3mm} Axial motion &  10~\micron \\
\hspace{3mm} Tip, tilt &  2 arcsec \\
\hspace{3mm} Roll &  3 arcsec \\
\hline
Hexapod range: & \\
\hspace{3mm} Lateral range & $\pm$\, 8 mm \\
\hspace{3mm} Axial range &  $\pm$\, 10 mm \\
\hspace{3mm} Tip, tilt range &  $\pm$\, 250 arcsec \\
\hline
\end{tabular}
\end{center}
\end{table}

\begin{figure}[htb]
\begin{center}
  \includegraphics[width=.90\textwidth]{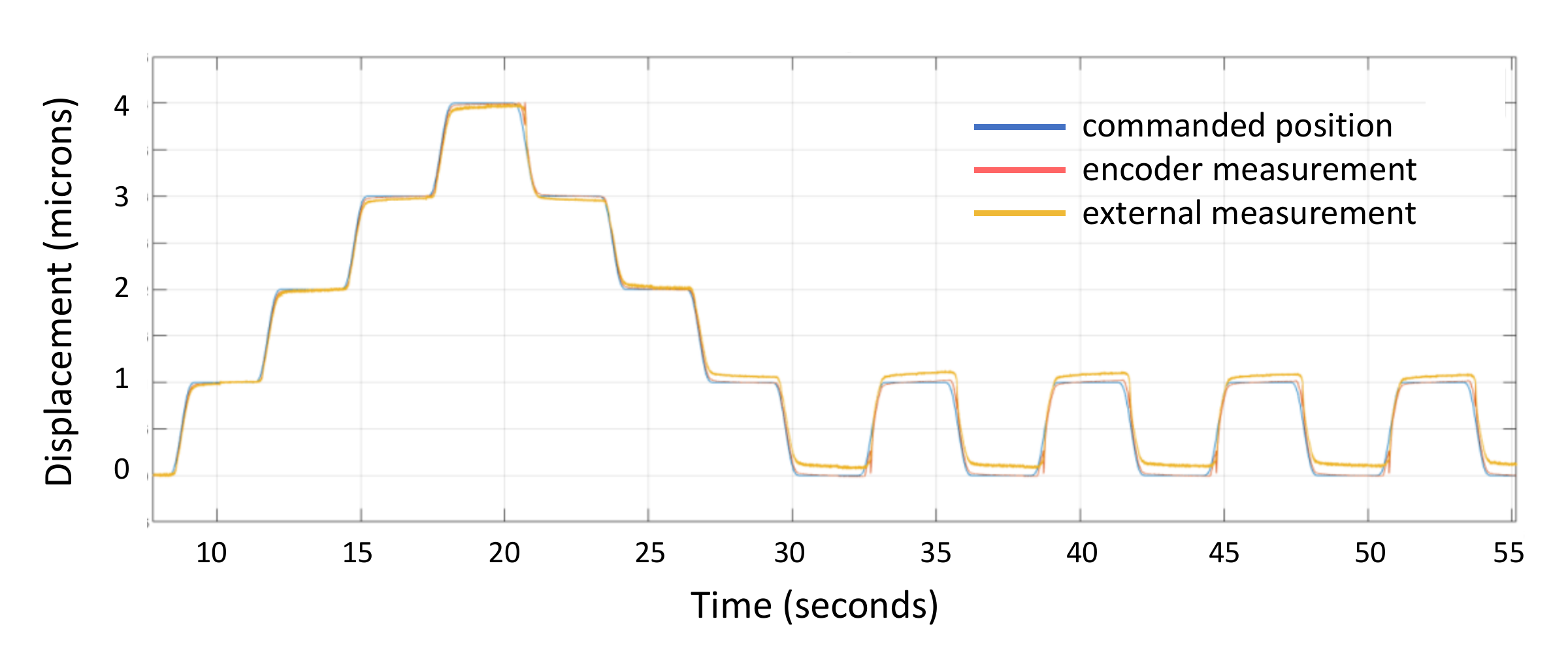}
  \caption{Hexapod actuator accuracy test.  The hexapod vendor tested the displacement of an actuator with flexures by commanding the actuator to a series of positions.  The plot compares the commanded position, the position measured by an internal encoder, and the position measured by an external device.  The differences are well within requirements.}
\label{fig:hexapod_actuator_test}
\end{center}
\end{figure}

The hexapod satisfied all the positioning requirements listed in Table \ref{tab:corrmech}.  Figure \ref{fig:hexapod_actuator_test} shows the displacement of one end of an actuator and flexure assembly, while the other end is held fixed, and the actuator was commanded to move in steps of 1\,$\mu$m.  Measurements were made using both the internal encoder of the actuator and an external device.  The difference between the expected and the actual motion is a small fraction of a micron.  When similar tests were done that covered the entire range of the actuators, the difference between the commanded and the measured displacement was only at the level of a few microns.  These differences were all well within requirements.

\subsection{Prime Focus Cage and Telescope Top End}\label{sec:cage_tele}

Table \ref{tab:cagereq} shows the main requirements that drove the prime focus cage and the telescope top end design.  Of these requirements, the sag, tilt and roll due to gravity and the decision to keep an open cage for easy access to the hexapod and ADC rotation mechanism drove most of the design decisions.  After a brief description of the main components of this subsystem we will concentrate on how these requirements influenced the design.

\begin{table}
\begin{center}
\caption{Requirements for the DESI ring and cage design.}
\label{tab:cagereq}
\begin{tabular}{lc}
\hline
\hline
Element  &  Requirement \\
\hline
\hline
Mass and dimensions  & \\
\hline
Mass of outer ring/spider vanes/cage/hexapods/barrel  & $\le$ 7,000 Kg \\
Cage outer diameter  &  $\le$ 1.8 m \\
Outer Ring ID  &  $\ge$ 4.6 m \\
Outer Ring OD  &  $\le$ 5.45 m \\
Vane thickness  &  $\le$ 25.4 mm \\
\hline
\hline
Spider vane adjustability  &  \\
\hline
Lateral and vertical motion  &  $\pm$ 50 mm \\
Tip/tilt  &  $\pm$ 1 degree \\
Step size  &  $\le$ 0.25 mm \\
\hline
\hline
Gravity sag, tilt and roll (HMP w.r.t. trusses)  & \\
\hline
Vertical sag with telescope at zenith  &  $\le$ 0.9 mm \\
Vertical sag with telescope horizontal  &  $\le$ 1.4 mm \\
Tilt  &  $\le$ 70 arcsec \\
Focal plane roll during a 20-minute observation  &  $\le$ 8 $\mu$m \\
\hline
\hline
Safety factor  & \\
\hline
Against yield, fracture or buckling  &  $\ge$ 4 \\
\hline
\hline
\end{tabular}
\end{center}
\end{table}

\begin{figure}[thb]
\begin{center}
    \includegraphics[height=.35\textwidth]{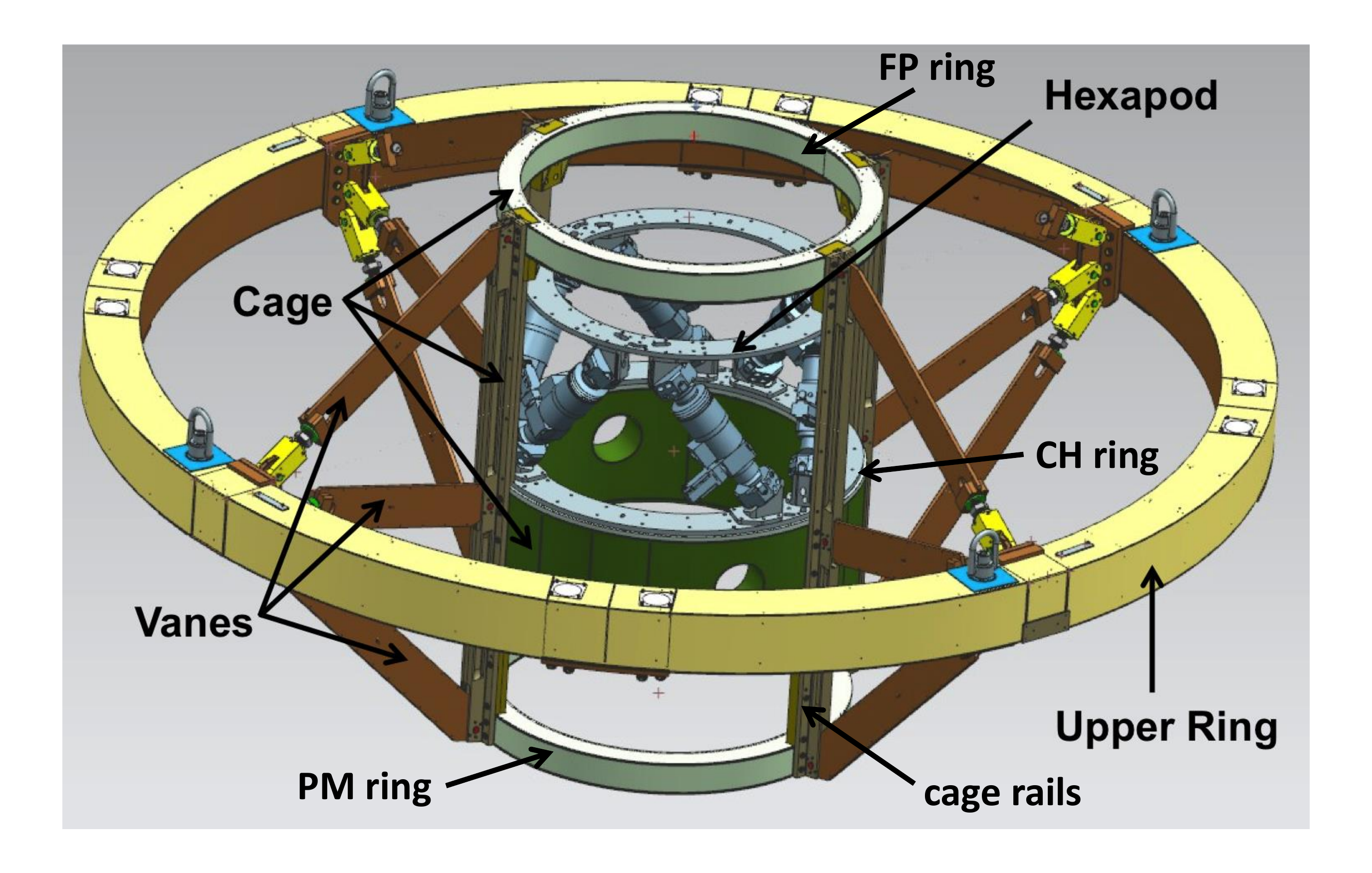}
    \hspace{3 mm}
    \includegraphics[height=.33\textwidth]{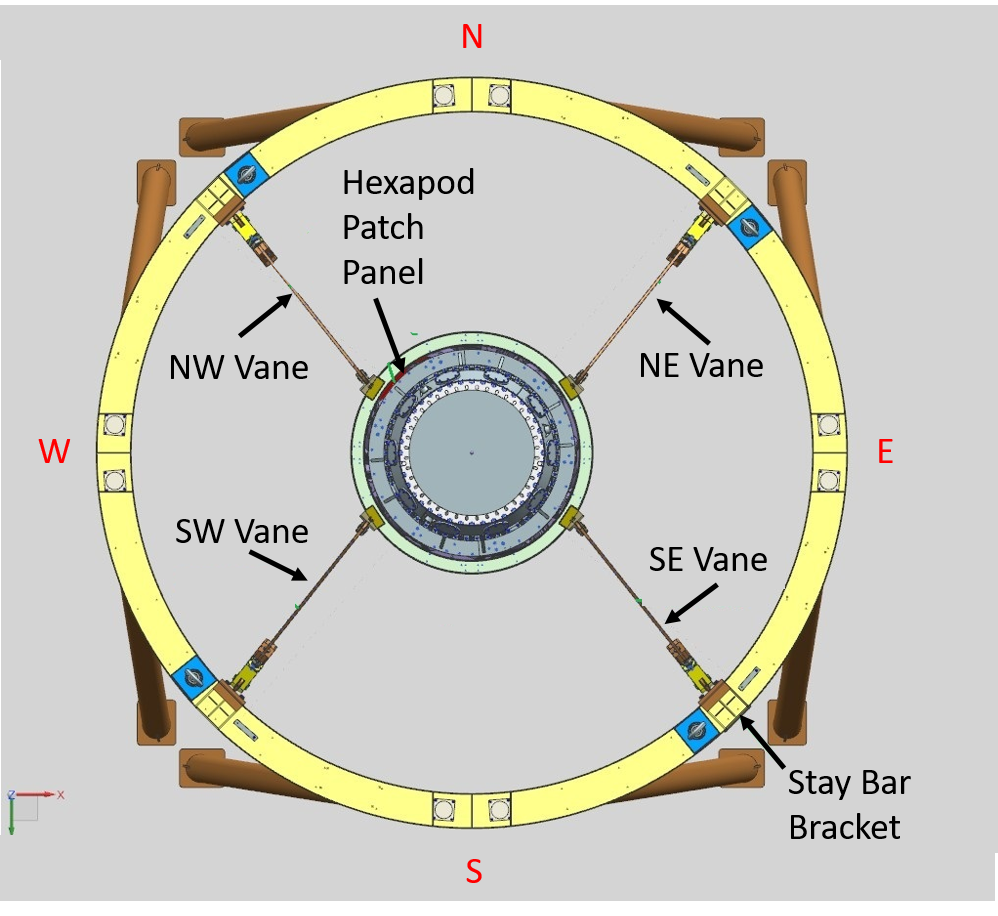}
  \caption{The Mayall top end.  Left: the top end includes the cage, the 12 vanes, and the upper ring.  The cage consists of the Primary Mirror (PM) ring, Cage Hexapod (CH) ring, and Focal Plane (FP) ring, all connected by four rails.  Right: end view showing the vane positions connecting the cage rails and upper ring. \citep[Left image previously published in][]{gutierrez18}}
\label{fig:cage3}
\end{center}
\end{figure}

Figure \ref{fig:cage3} shows the 10,700\,kg telescope top end.  The top end includes the prime focus cage, the spider vanes (or vanes, or fins) and the upper ring (aka the outer ring).  The cage consists of three rings connected by four thick rails: the Primary Mirror (PM) ring closest to the mirror, the Cage Hexapod (CH) ring to which the hexapod mounts, and the Focal Plane (FP) ring which supports the focal plane assembly thermal shield.  Three vanes connect each of the four cage rails to the upper ring, making a total of 12 vanes.  Each vane length is adjustable in order to center and align the cage with respect to the primary mirror.  The upper ring is used to mount the entire structure to the telescope Serrurier trusses.

\subsubsection{Prime Focus Cage} 

An open design, similar to DES, was selected for the DESI cage.  The main advantage provided by an open design is easy access to the hexapod and the ADC rotators, which is needed for maintenance and repairs.  A disadvantage of an open design (e.g. a closed design, like a tube, provides much smaller deflections) is that in order to meet the deflection requirements a middle vane is needed, which complicates the alignment of the cage and the tensioning of the vanes.  

When the telescope is in the horizontal position the maximum vertical sag of the Hexapod Mounting Plate (HMP) relative to the Serrurier trusses (see Table \ref{tab:cagereq}) is 1.4\,mm, or 17.5\% of the hexapod's range.  The maximum tilt of the HMP is 70\,arcsec, or 28\% of the hexapod's range.  The maximum vertical deflection when the telescope is at zenith is 0.9\,mm, or 9\% of the hexapod's range.  These motions leave ample margin in the hexapod ranges that can compensate for other effects such as telescope deflections and alignment errors.  

Three steps were taken to satisfy the requirements in the case when the telescope is in the horizontal position.  First, the PM ring was pushed as far out as possible to increase the effectiveness of the vanes to prevent cage rotations,  Second, within weight constraints, the hexapod ring was designed as stiff as possible to prevent barrel rotations.  Third, since the long unsupported part of the rails is the weakest link in terms of deflections, the middle vanes were placed inside the hexapod ring to provide backing to the part of the rail that has a slot; this slot allows for the vane insertion in the rail.  

\begin{figure}[thb]
\begin{center}
  \par \vspace{5 mm}
    \includegraphics[width=.28\textwidth]{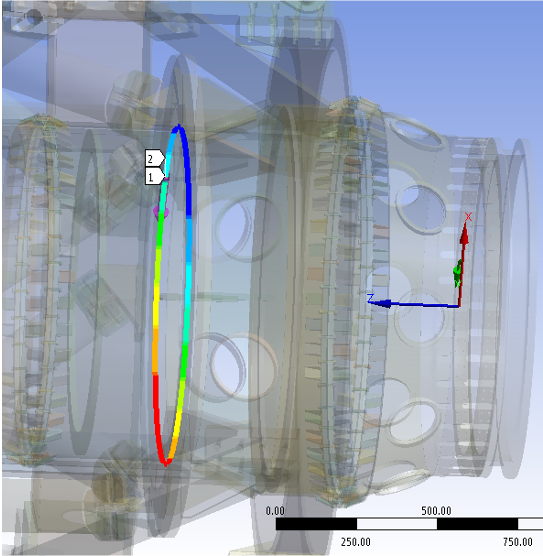}
    \includegraphics[width=.40\textwidth]{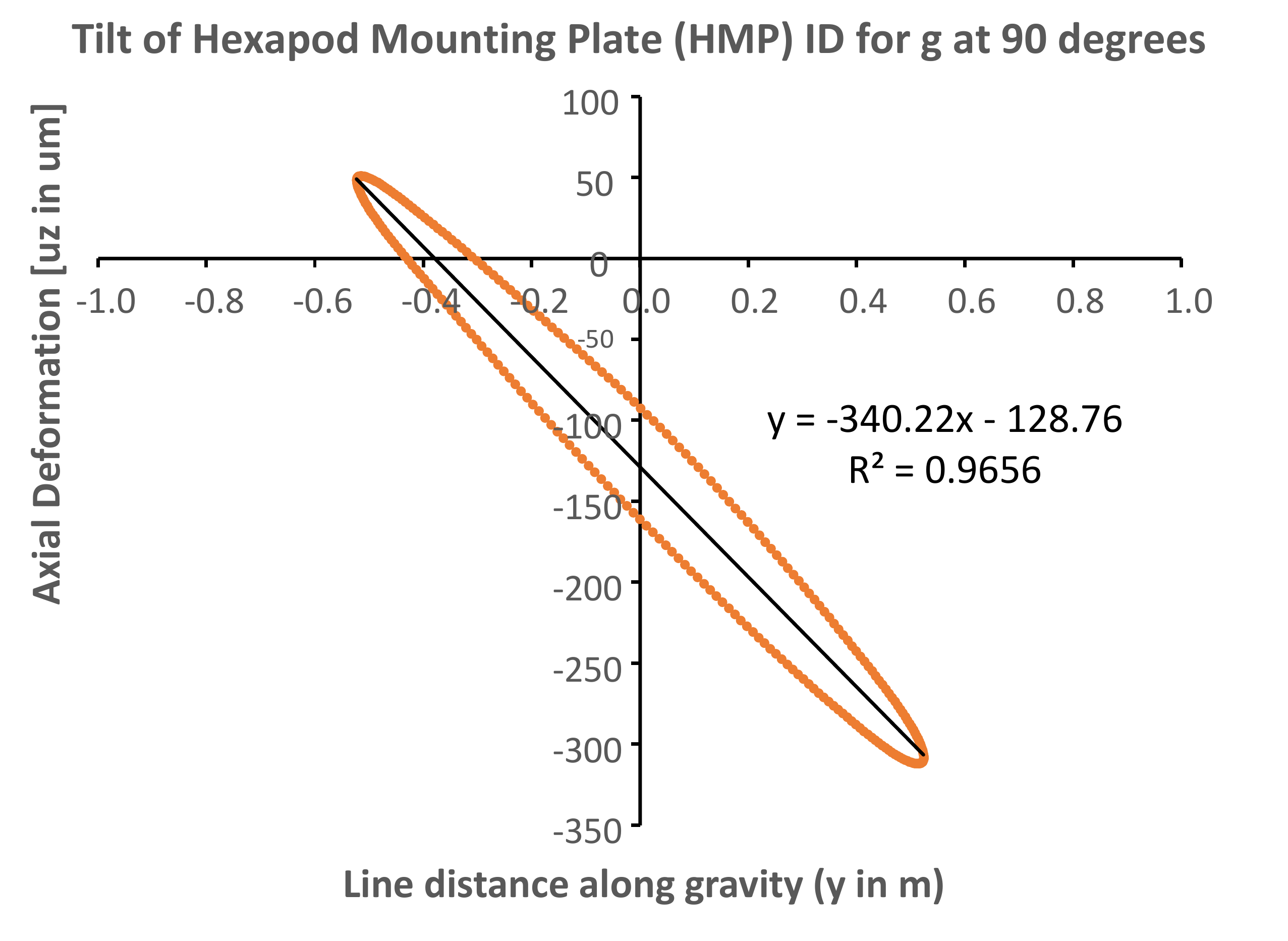}
    \includegraphics[width=.28\textwidth]{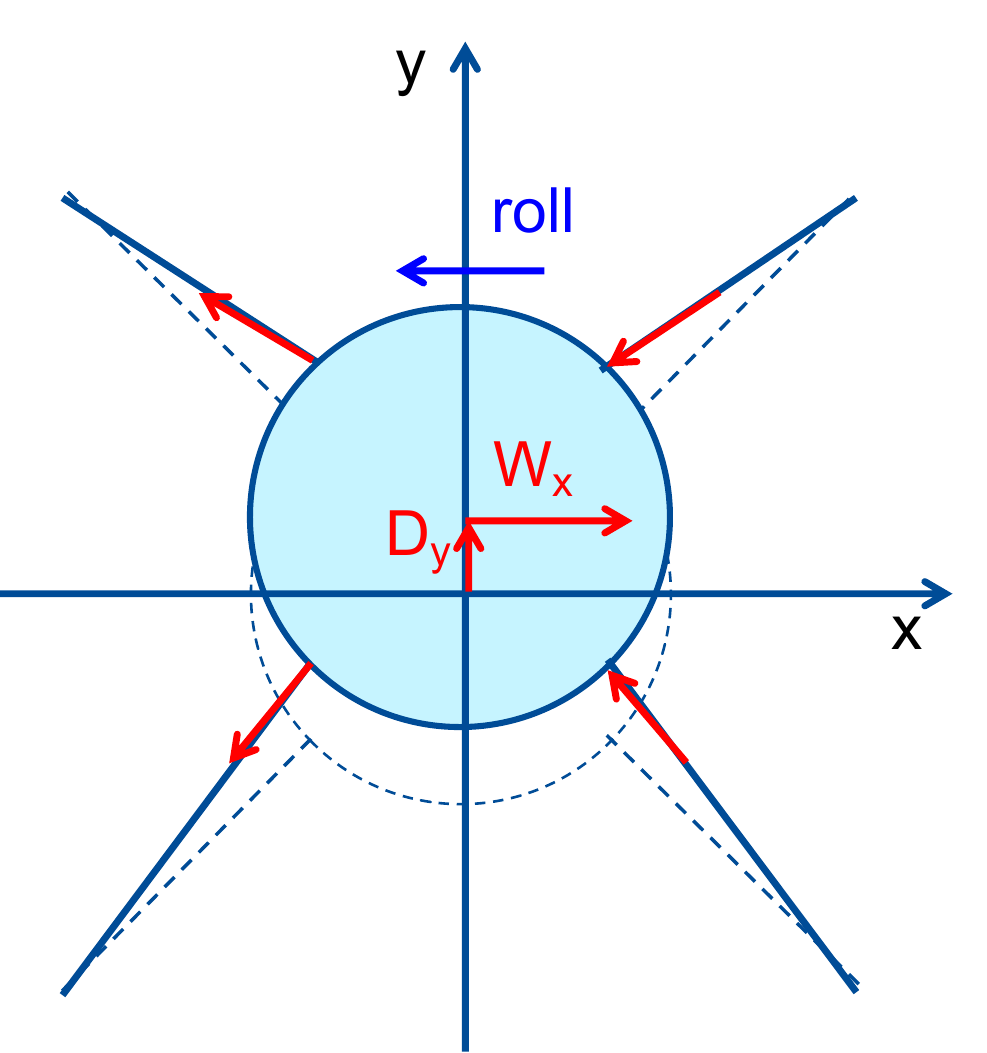}
  \par \vspace{2 mm}
  \caption{FEA of the cage design.  The telescope is oriented horizontally. 
 Left: the color ring shows the points on the inside of the HMP that were used to measure tilt.  Center: locations of the measured points.  The slope of the line fitted to the points is a measurement of the cage tilt.  Right: the cage can be made to roll by shifting the barrel away from the center of the outer ring.}
\label{fig:cage_FEA}
\end{center}
\end{figure}

Finite Element Analyses (FEAs) were used to determine if the cage met specifications.  To study sag and tilt of the HMP relative to the outer ring, an FEA was done with the telescope in the horizontal direction and gravity in between the vanes (labeled 0$^\circ$ and 90$^\circ$) and along the vanes (labeled as 45$^\circ$).  After subtracting 7\,arcsec of tilt for the outer ring the tilts for the HMP were 40, 49 and 63\,arcsec for 0$^\circ$, 45$^\circ$ and 90$^\circ$ respectively.  The left and center pictures in Figure \ref{fig:cage_FEA} show the FEA results for the 90$^\circ$ case.  The color ring in the left FEA picture shows the points on the HMP flange that were used to measure tilt.  The inner side of the flange was used because it is less subject to deformations by the hexapod actuators.  The axial deformations for these points as a function of the vertical coordinate are plotted in the center figure.  The slope of the line fit (70\,arcsec) is a measure of the tilt.  The total vertical sags of the HMP are 2.0, 2.3 and 2.6\,mm, with gravity at 0$^\circ$, 45$^\circ$ and 90$^\circ$ respectively.  Subtracting a 1.5\,mm sag for the Outer ring gives a maximum sag of 1.1\,mm.  The results for the FEA of the telescope in the vertical position shows that the deflection of the HMP relative to the outer ring is 0.8\,mm.  Thus the cage design meets all the gravity sag and tilt requirements.

DESI exposures are nominally 20\,minutes long, and the focal plane has a very large diameter, so preventing the roll of the cage during an exposure is very important.  This roll is required to be less than 8\,$\mu$m at the edge of the focal plane (see Table \ref{tab:cagereq}).  To provide a restoring force to prevent the cage from rolling, the vanes were not inserted symmetrically around the cage.  The right picture in Figure \ref{fig:cage3} shows how the vanes insert on the rails of the cage.  The insertion was displaced 11.5\,degrees from the 45\,degree line with pairs of vanes moving in opposite directions.  This generates enough restoring force to keep the roll of the cage within specifications.  The sketch on the right of Figure \ref{fig:cage_FEA} shows how decentering the cage relative to the OR produces roll.  In the sketch the weight $W$ of the cage is along the $x$-axis, then if the cage is centered the forces that the vanes exert on the cage (red arrows) will be symmetrical and there would be no torque around the axis of the cage.  If the cage is moved by an amount $D_y$ as indicated in the sketch (e.g. to center the cage relative to the primary mirror) then the upper forces become more tangential than the lower ones and a moment is generated that will make the cage rotate.  If the vanes are inserted at $\pm 45^\circ$ then the restoring force would be essentially zero (second order in the rotation angle) and as the gravity vector changes during long exposures the cage would roll by unacceptable amounts.  The decentering of the barrel by the hexapod produces an equivalent but opposite effect, because in this case the roll is produced by having the Center of Mass not centered on the cage.  As mentioned before, the symmetry was broken by displacing the vane's insertion points by 11.5\,degrees.  This number was selected such that a 12\,mm decentering of the cage combined with an 8\,mm barrel displacement of the hexapod would produce, during a 20\,minute observation, a maximum roll anywhere in the sky of less that 8\,$\mu$m at the edge of the focal plane.  The breaking of the vane's insertion symmetry in the cage also has the added advantage of stiffening a low-frequency resonant mode of the telescope in which the cage rotates around its axis.

Lastly, we reinforced the insertion points of the rails in the FP ring so that the entire system could be lifted from these points.  Figure \ref{fig:integ3}, right, shows them used for this purpose.

\subsubsection{Spider Vanes}

The spider vanes (or vanes, or fins) are 19\,mm thick and 200\,mm wide.  They are constructed in three different lengths, corresponding to the three different locations where they attach to the cage structure.  Each vane has a spherical bearing press fitted into the vane for attachment to the cage support rail and a threaded adapter at the opposite end for cage position adjustability.  This end attaches to an opposite-handed turnbuckle that provides 50\,mm of total adjustment.  The other end of the turnbuckle attaches to a flange in the outer ring through another spherical bearing such that at each end the vanes are free to rotate in any direction.  Each vane also has tapped holes for mounting cable trays.  These trays house the fiber cables, piping, and electrical cables for the focal plane assembly, as well as the power and control wiring for the hexapod and the ADC rotator.

All twelve vanes have strain gauges to allow for pretensioning of the vanes.  The pretension was calculated such that all twelve vanes will always be under tension independent of the position of the telescope.  The FEA calculated pretension for the vanes is 13.4\,kN for the PM-vanes (vanes closest to the primary mirror), 44.1\,kN for the C-vanes (the central vanes) and 45\,kN for the FP-vanes (vanes closest to the focal plane).  These pretension values are the zero-g tensions that meet the required minimum vane tension of 4.1\,kN for the PM vanes.  Since the vanes form an overly constrained system, only eight of the twelve vanes can be adjusted independently.  The pretension of the PM and C-vanes were selected for the adjustment because the PM-vanes have a much larger minimum tension that the other vanes.  With the telescope in the vertical position the goal was to achieve a total tension of 31\,kN for the PM-vanes and 49\,kN for the C-vanes, with a minimum of 30\,kN for the FP-vanes.  After the final alignment and pretensioning of the vanes, the minimum measured tension values of the vanes with the telescope in the vertical position were 33\,kN, 49\,kN and 44\,kN, for the PM, C and FP-vanes respectively; these values meet the required minimum with margin.

\subsubsection{Outer or Upper Ring}

The Mayall telescope is fully dedicated to DESI and there is no need to support operation of Cassegrain instruments, as was the case in the previous corrector design, and therefore the design eliminated the old "flipping" ring.  This allowed for a 2250\,kg saving in the total weight of the telescope top end.  The outer or upper ring sits atop the Serrurier trusses and supports the cage through the fins.  The bottom mounting holes that attach the ring to the trusses were drilled during installation.  An alignment “cross-wire” system, attached to the outer ring, was used to align the center of the ring to the center of the primary mirror.  To simulate the deformation of the outer ring by the forces exerted by the vanes, a fixture was designed to preload the ring to approximately the same tension as the net vector tension imposed by the final assembly of cage and vanes.  Once the preloaded outer ring was aligned to the primary mirror, by using an alignment telescope previously registered to the mirror's mechanical center, it was match drilled and pinned to the Serrurier trusses and then removed for the final assembly with the rest of the system.

\subsubsection{Materials and Coatings} 

The cage, vanes and outer ring are fabricated from A572 Grade 50 structural steel. Cage cover sections provide wind and thermal radiation protection for the camera/corrector.  Some surfaces of the top end assemblies are painted with an optical black paint as specified in the stray light analysis (see section \ref{sec:straylightmain}), specifically surfaces directly visible from the primary mirror such as the spider vanes and the ring underside.  The primer is Lord Corporation Aeroglaze 9947 and the top coat is Aeroglaze Z306, a suitable flat black coating with low outgassing properties.  Other surfaces facing outwards from the primary are painted white; the trusses were also left with their existing white paint, being deemed small contributors to stray light.  Mating contact surfaces for the rings and fins are not painted. These surfaces are protected from corrosion by a light coat of grease.  The cage, vanes and outer ring were built by CAID Industries in Tucson, Arizona.   Pictures of the fully assembled system, including the barrel and hexapod, are shown in Figures \ref{fig:integr2} and \ref{fig:integ3}.

\section{Stray Light} \label{sec:straylightmain}

Stray light is a concern with any optical instrument.  In the case of DESI, stray light that enters the fibers at the focal plane can be relayed into the spectrographs and eventually to the final images on the CCDs.  This degrades the instrument’s signal-to-noise ratio, and may produce artifacts in the measured spectra.  Stray light originating from one astronomical object can also contaminate the measured spectrum of another.  Therefore stray light must be studied and quantified, and also controlled if analysis shows it significantly affects the signal-to-noise ratio.

Our approach to stray light is summarized in \cite{miller16straylight}.  We contracted with Photon Engineering LLC (PE) in Tucson, AZ, to perform stray light analysis of the DESI front end, ending at the fibers at the focal surface.  PE created a stray light model of the front end, and we used it to identify the various physical contributors to stray light.  The model included the major subassemblies: the Mayall observatory and dome; the Mayall telescope including internal and external structure; the DESI corrector including the lenses, mechanical cells, and barrel structure; and the DESI focal plane assembly.  Surface properties were defined for all components that may be illuminated in the analysis, using reasonable estimates (AR coatings on glasses, black coatings on barrel surfaces, etc.)

Since stray light was expected to be an issue, we took a “best practices” approach from the start, implementing reasonable, commonplace, and cost-effective methods to block stray light without requiring a definite analysis to justify them.  Simple thin baffles were placed before and after each lens surface to block scatter from their edges.  A front baffle on the cage blocks wide-angle sources from entering the barrel directly.  Lenses were polished to achieve an excellent surface roughness spec, minimizing their scatter.  The primary mirror is cleaned at regular intervals, to reduce wide scatter from contamination buildup.  Most non-optical surfaces were treated to be highly non-reflective, e.g. black anodization or painting with Aeroglaze Z306.  The PE analysis started with these methods in place, and considered whether they were enough, or whether more was required.

Once the model was set up, PE ran several specific initial raytraces.  First, PE illuminated the model from the direction of the sky, and identified which surfaces were directly visible from the sky.  Second, PE illuminated the model from the direction of the focal surface, and in a similar fashion identified which surfaces are directly visible from the detector.  Then PE pinpointed which surfaces are seen by both: these were the specific surfaces that were most likely to be a stray light concern, since light from outside the observatory can scatter a single time from them and still reach the detector.  Thus PE could prioritize this short list of surfaces when setting up the large raytraces for later analysis, making for a more efficient raytrace.

To analyze stray light in the model, we relied upon Point Source Transmittance (PST) functions.  A PST is the amount of total irradiance that reaches the detector surface of an optical system (the focal surface in this case), given collimated input light over a range of incident angles.  The PST is normalized by the input irradiance, so it is a measure of the fraction of source light from a particular angle that reaches the focal surface.  Furthermore, a raytrace of a PST tracks the scattered paths of each ray and builds up statistics about where the significant scatter sources are in the model.

There are two significant sources of astronomical light that potentially create unacceptable scattered stray light in the DESI corrector.  The first is stars: their unidirectional incoming light can scatter inside the corrector, sending some light to the focal surface.  However, the scattered light that is spread across the entire focal surface is relatively low at each fiber, except in the case of the very brightest stars.
The second astronomical source is more substantial: the sky continuum background, which is atmospheric radiance falling upon the Mayall telescope from all directions.  In this case, the integrated light from all directions that reaches the focal surface can be large, and can be significantly more than the light from a single star.  To analyze this effect of a wide sky continuum, we integrated the PSTs over all input angles.  Table \ref{tab:SL2} shows the results of this integration.  The flux values are given as a fraction of the directly imaged sky background.  Prioritizing this list revealed which specific scattering sources can send unacceptable amounts of light to the focal surface, and thus suggested which contributors should be mitigated and which could be ignored.  However, this decision was also based on how difficult a mitigation would be: the difficulty is estimated in Table \ref{tab:SL2} as well.


\begin{table}
\centering
\footnotesize
\caption{Scattered light reaching the focal surface}
\label{tab:SL2}
\begin{tabular}{c|c|c|c|c|c|c}
\hline
	&		& \multicolumn{2}{c}{Integrated flux at FS}		&   &		&		\\
\multirow[b]{3}{8em}{\centering Scattering source} & 
\multirow[b]{3}{6em}{\centering Relevant input angles} & 
\multirow[b]{3}{6em}{\centering all flux reaching FS} &
\multirow[b]{3}{6em}{\centering accounting for fiber accept. angle} &
\multirow[b]{3}{5em}{\centering Ease of mitigation} &
\multirow[b]{3}{4.5em}{\centering Mitigate?} &
\multirow[b]{3}{5.5em}{\centering Effect when mitigated} \\ 
 & & & & & & \\
 & & & & & & \\ [3pt]
\hline													
M1 cell top	&	all	&	0.06616	&	0.00066	&	easy	&	Y	&	0.00013	\\
M1 cell inner walls	&	all	&	0.01879	&	0.00376	&	medium	&	Y	&	0.00188	\\
Corrector Lens surface scatter	&	$<$ 5 deg	&	0.01339	&	0.01339	&	very hard	&	N	&	0.01339	\\
Primary Mirror scatter	&	 $<$ 10 deg	&	0.00946	&	0.00946	&	hard	&	N	&	0.00946	\\
Barrel: Hexapod Mounting Flange	&	2-3 deg	&	0.00623	&	0.00208	&	easy	&	Y	&	0.00021	\\
Barrel: C4 mounting flange	&	2-3 deg	&		&		&	easy	&	Y	&		\\
Barrel: C4 flexure sneak path	&	2-3 deg	&		&		&	easy	&	Y	&		\\
Barrel: C2 mounting flange	&	4 deg	&		&		&	easy	&	Y	&		\\
M1 inner mask	&	all	&	0.00211	&	0.00211	&	medium	&	N	&	0.00211	\\
M1 cover petals, open	&	20-35 deg	&	0.00183	&	0.00002	&	medium	&	N	&	0.00002	\\
Upper structure: grazing off outer cage	&	all	&	0.00110	&	0.00110	&	medium	&	N	&	0.00110	\\
Upper structure: cage front ring	&	all	&		&		&	medium	&	N	&		\\
FPS: grazing off outer cage	&	all	&	0.00068	&	0.00068	&	medium	&	N	&	0.00068	\\
Cell: C1 scatter	&	0-5 deg	&	0.00025	&	0.00013	&	easy	&	Y	&	0.00003	\\
Cell: C2 scatter	&	0-4 deg	&	0.00082	&	0.00041	&	easy	&	Y	&	0.00008	\\
Cell: C3 scatter	&	0-3 deg	&	0.00001	&	0.00001	&	medium	&	N	&	0.00001	\\
Cell: C4 scatter	&	2 deg	&	0.00005	&	0.00002	&	medium	&	N	&	0.00002	\\
Cell: ADC1 scatter	&	0-3 deg	&	0.00006	&	0.00003	&	medium	&	N	&	0.00003	\\
Cell: ADC2 scatter	&	0-3 deg	&	0.00006	&	0.00003	&	easy	&	Y	&	0.00001	\\
cage front baffle	&	0-9 deg	&	0.00074	&	0.00074	&	easy	&	N	&	0.00074	\\
FVC assembly	&	all	&	0.00003	&	0.00003	&	medium	&	N	&	0.00003	\\
Building	&	all	&	0.00001	&	0.00001	&	hard	&	N	&	0.00001	\\
Ghosts	&	0-4 deg	&	\multicolumn{2}{c}{separate analysis}		&	very hard	&	N	&		\\	
GFA reflection	&	2 deg	&	\multicolumn{2}{c}{separate analysis}	&	medium	&	N	&		\\	
FPA backplane	&	0-2 deg	&	\multicolumn{2}{c}{separate analysis}	&	medium	&	Y	&	\\ [6pt]
\textbf{Total:}	&		&	\textbf{0.122}	&	\textbf{0.035}	&		&		&	\textbf{0.030}	\\
\hline
\end{tabular}
\tablecomments{Table lists the scattering sources that contribute to the total scattered light reaching the focal surface (FS). The scattering sources are listed roughly in order of magnitude.  Flux at the focal surface is calculated by integrating the PST data over all input angles, and is displayed as a fraction of the incident scene light. \citep[Previously published in][]{miller16straylight}}
\end{table}


Note an important caveat to the initial analysis: at the output of the fibers in the DESI spectrographs, a pupil stop blocks rays that exit at too large of an angle.  Thus there is an effective acceptance angle at the fiber input.  This acceptance angle limits the incident stray light at the fiber that makes it all the way through the system.  Table \ref{tab:SL2} includes a column that estimates the integrated flux, including the effect of this acceptance angle cutoff.  For some scattering sources, the magnitude of stray light is substantially reduced by this effect.  For example, the non-black top of the Mayall telescope center section (the primary mirror cell and surrounding structure) creates substantial scattered light that reaches the focal surface fibers, but very little of it passes all the way through the spectrographs.  Thus some scattering sources were no longer deemed to cause significant stray light, and the list of scattering sources that required mitigation was reduced.

\begin{figure}
\centering 
\includegraphics[width=.95\textwidth]{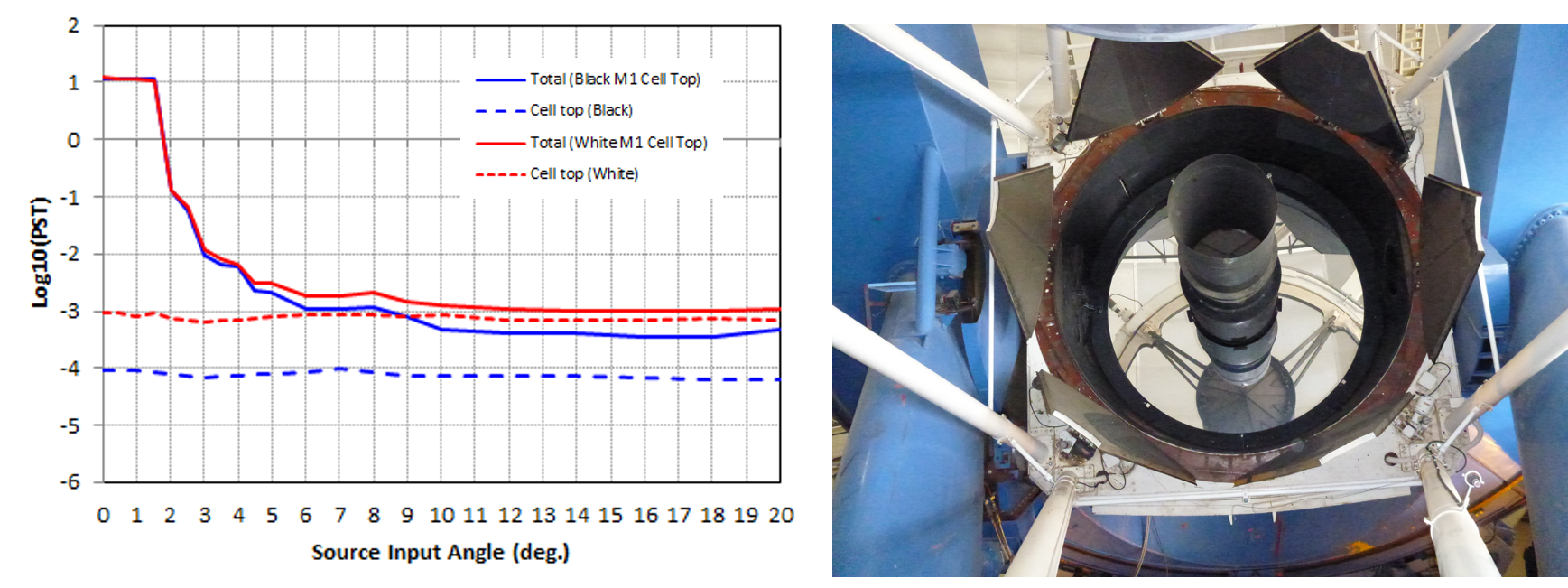}
\caption{Example of stray light analysis and mitigation.  Left: PST comparing the scatter from non-black paints on the center section to black paints.  Right: the Mayall telescope center section before painting any part black.  Repainting the center section black reduces its contribution to scattered light by an order of magnitude.  This in turn reduces the total stray light across all input angles in the left figure: the reduction in stray light integrated over all angles is significant. \citep[Images previously published in][]{miller16straylight}}
\label{fig:SLex1} 
\end{figure}

The rightmost three columns of Table \ref{tab:SL2} address possible mitigation of the various scattering sources.  Some were deemed relatively easy to mitigate, and some would take substantial expense.  Therefore DESI chose to mitigate only some of them, specifically ones that were relatively easy fixes, or else reduced stray light substantially.  The rightmost column shows the predicted stray light after mitigation; while the reduction is not drastic - the fraction of incident light reaching the focal plane changes from 0.035 to 0.030 (3.5\% to 3.0\%) - the fixes were relatively easy, or were already in line with best engineering practices.  Some examples include:

1.  Non-black paint on the telescope center section, shown in Figure~\ref{fig:SLex1}, can cause diffuse scatter over wide angular range.  Therefore this area of the telescope was painted black, well before the corrector integration, reducing scatter by about a factor of 10.  Figure~\ref{fig:SLex1} also shows the PST before and after this change.  Some of this change becomes irrelevant when the fiber acceptance angle is considered, but applying black paint was an easy precaution.

\begin{figure}
\centering 
\includegraphics[width=.95\textwidth]{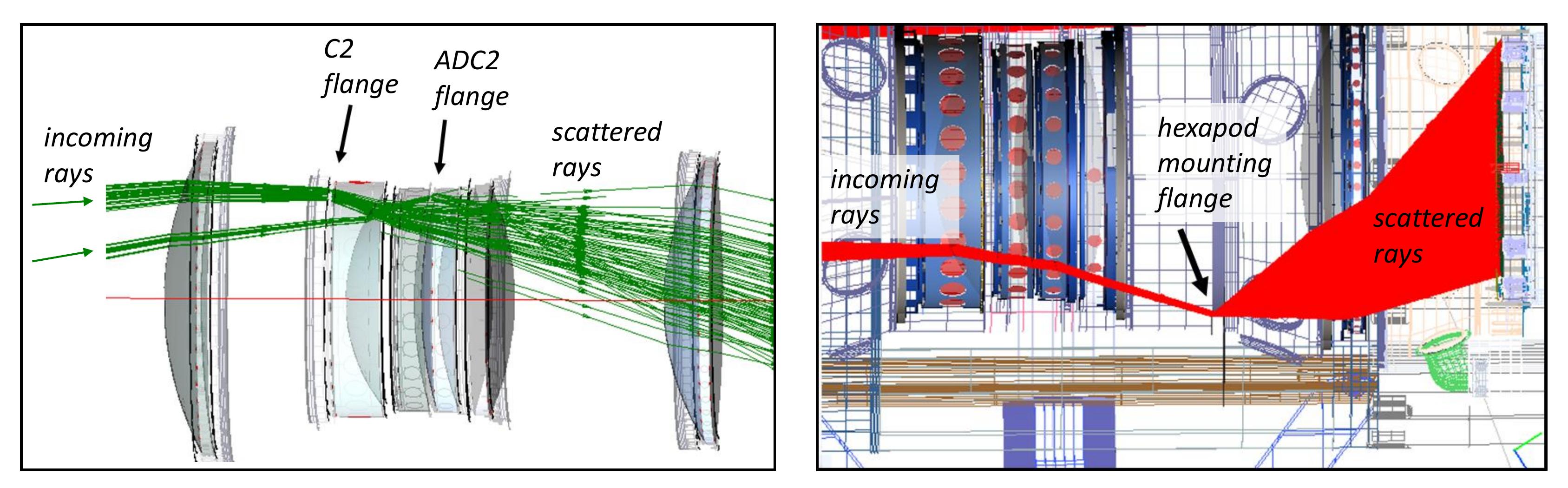}
\caption{Examples of scattering inside the barrel.  Left: Model of scatter from flat surfaces on C2 and ADC2 lens cells.  For a particular input angle, the incoming rays (from the left) were scattered at the lens edges (labeled by the black arrows), sending diffuse light towards the focal surface.  These scatter sources were mitigated by adding fine grooves to the flange surfaces.  Right: Model of scatter from the hexapod mounting flange inside the corrector barrel.  For a particular input angle, the incoming rays (from the left) were scattered by the flange (labeled by the black arrow), sending diffuse light towards the focal surface.  Low baffles immediately past the flange were added to block the scatter.  \citep[Images previously published in][]{miller16straylight}}
\label{fig:SLex2and3} 
\end{figure}

2.  Flat flanges on several lens cell designs can cause specular glints for slightly out-of-field light.  Therefore fine grooves were added to these flange surfaces, in order to reduce grazing angle scatter.  Figure~\ref{fig:SLex2and3}, left, illustrates several glints in the stray light model at the C2 and ADC2 lens cells, before the grooves were added.

3.  Several internal barrel flanges can scatter light widely.  Therefore low baffles were added around these flanges to block the scatter path.  Figure~\ref{fig:SLex2and3}, right, shows the scatter off of the hexapod mounting flange, the largest of the barrel flange scatterers, before the low baffles were added.


Our goal for maximum stray light contamination in the corrector was consistent with the requirement for ghosts, as discussed in section \ref{sec:ghostsection}: less than 1\% of the field of view should be contaminated by non-science light that is brighter than 1\% of the sky background already in the field of view.  However, Table \ref{tab:SL2} shows that the total sky background that reaches the focal plane from scattering was predicted to be 3\% of the sky background after putting mitigations in place, which was above the goal.  This extra background light adds to the measurement noise, decreases the instrument SNR, and increases the instrument exposure time by 6\%.  During the design phase, we decided to accept this prediction because the survey time budget included 23\% margin, part of which could be allocated to this particular effect.

In fact, during the main survey the average sky background measured by the spectrographs was almost exactly as predicted by the throughput budget, given a roughly 10\% variation from night to night.  That is, no extra background light due to the scattering was observed, or at least it was well within the nightly variations.  It is worth noting that some scattering estimates were conservative, such as the mirror surface cleanliness and lens surface roughnesses, and so the prediction of background due to stray light may have been high.

\section{Integration}

\subsection{Corrector Integration Approach}

The excellent performance of the corrector requires precise integration and alignment.  Typically a lens assembly would be tested optically after assembly as a verification that the complete assembly was built correctly.  This is often implemented as a full-aperture, double-pass interferometry test.  However, this is a very demanding test: it is difficult to find facilities that can host an optical test for such a large lens assembly; it would require substantial new test equipment; it would cost many months in the tight project schedule.  After a trade study, we decided not to perform an optical verification test on the entire corrector.  Instead, extra effort was invested into the mechanical verification of the corrector.  As described above, final alignment of the lenses was assured by careful mechanical measurements with dial indicators and a large CMM, and also verified by a laser beam test.  Furthermore, extra care was taken to verify the individual lenses were fabricated correctly.  Major parameters of the lenses were verified using multiple measurement techniques.  For example, as described in section \ref{sec:2p7} we sent the C2 lens to NASA's Goddard Space Flight Center to confirm its early figure progress on a large CMM, since its aspheric convex surface is particularly challenging to verify.  Finally, the complete assembly was optically modeled with all tolerances and uncertainties to verify by analysis that the assembly would meet its requirements.

\subsection{Shipping}

The corrector components were manufactured in a wide range of locations and vendors, and the parts were therefore shipped extensively, both individually and in assemblies.  For example, each lens traveled from the blank vendor, to the polishing vendor, to the coating vendor, to UCL for corrector integration, then finally to Kitt Peak for telescope integration.  Each leg of the trip was tracked carefully, with pre- and post-shipping inspections, and usually with a devoted shipper delivering the lens door-to-door.  The extra cost of this was considered well worthwhile considering the accumulated cost and uniqueness of a lens. 

\begin{figure}
\centering 
\includegraphics[width=.8\textwidth]{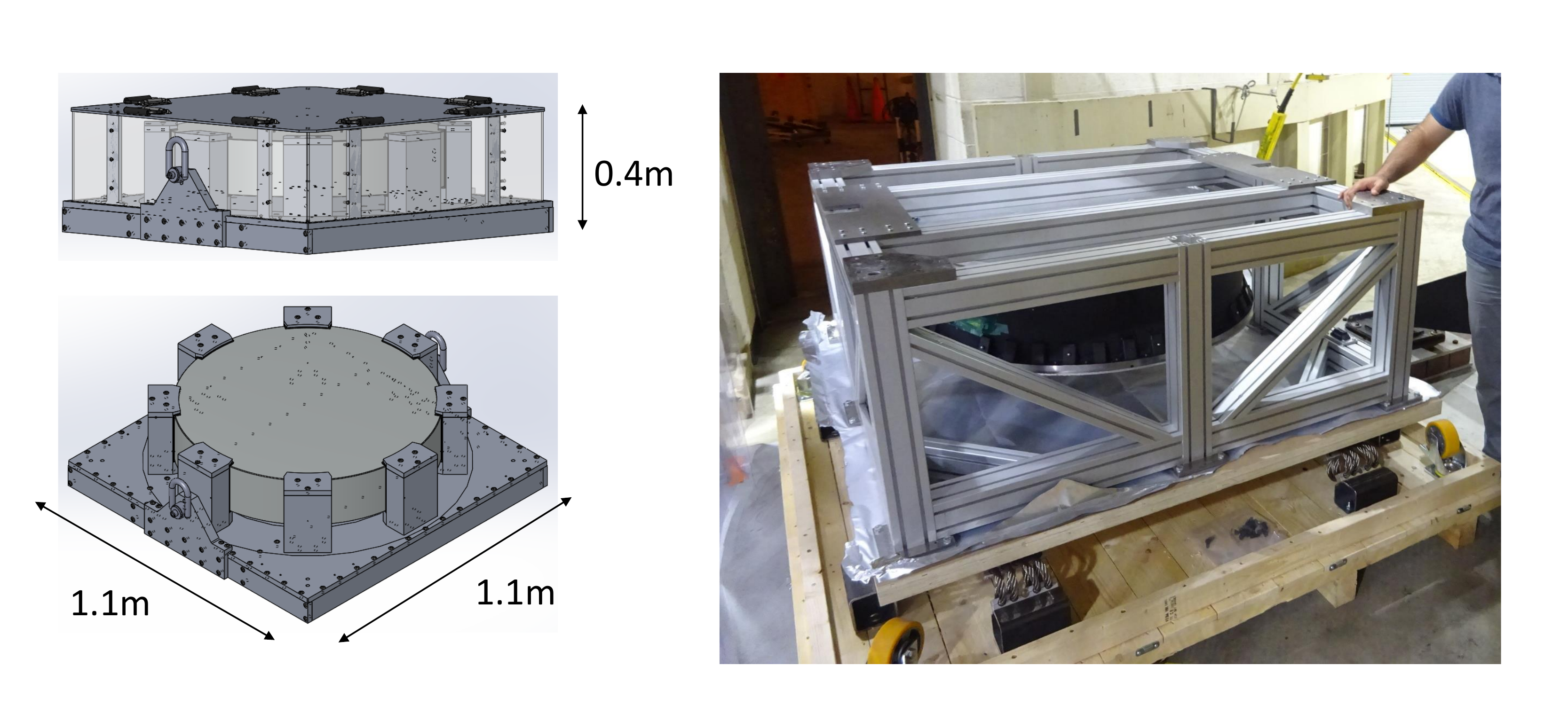}
\caption{Shipping fixtures.  Left: Model of C2 lens in its devoted shipping fixture.  Right: the Mid barrel segment being unloaded at Kitt Peak; the segment is secured in an aluminum frame, and rests on wirecoil isolators on its crate base.}
\label{fig:shippingfixtures} 
\end{figure}

The blanks were shipped from the glass manufacturers to the polishing vendors in standard wooden crates, packed securely in foam.  DESI designed and built individual aluminum shipping fixtures for the polished lenses.  These carried the finished polished lenses to the coating vendor, and the coated lenses to UCL for integration.  Figure \ref{fig:shippingfixtures} shows the C2 shipping fixture as an example.  Each lens was held securely around its circumference, clamped by Neoprene rubber pads and supported on aluminum stands.  The fixtures doubled as lifting and flipping fixtures for handling the lenses safely; their design made sure to account for how all parties would use them in their facilities.  For example, the center of mass, including the lens, is near the axis between the lift rings, to allow for easy flipping.

The stackup of misalignment errors in the lenses to their cells, the cells to the barrel segments, and segments to segments needed to be kept small in total to meet the requirements of Table \ref{tab:lens_place}.  Since the barrels and cells were contracted and measured in different places (FNAL and UCL, respectively) it was important to assure that the parts would be sufficiently well-aligned when integrated.  To achieve this, the cells were shipped after their fabrication from UCL to FNAL, where they were aligned by CMM in the as-built barrel sections, and doweled to establish repeatability, before returning to UCL for lens integration.  All along, careful mechanical measurements of centering and tilt of the barrels and cells were made during their builds, in order to verify that the error stackup was within requirement.  Therefore DESI was assured that the complete assembly would be well-aligned when the barrel segments were shipped to UCL for integration.

After the lens cells and barrels were integrated at UCL, and the alignment fully verified as described above, the assembly was dismantled into three barrel sections that were shipped from UCL to Kitt Peak in individual wooden crates.  To assure the carefully-aligned assemblies were not jolted during shipment, they were mounted to custom aluminum frames, and mounted on wirecoil isolators in the crate (Figure \ref{fig:shippingfixtures}).  Shock sensors were mounted on the frames for each section.  The UCL rotary table along with a metrology test frame and equipment were also shipped to the Mayall.
The crates containing the barrel sections and ancillary test equipment were flown from the UK to Tucson on a chartered airfreight flight.  The sections were then taken by truck from Tucson to the Mayall telescope.
The barrel cage, fins, and upper telescope ring were trucked directly from CAID in Tucson to Kitt Peak.  The hexapod was shipped from ADS in Italy to Kitt Peak.

\subsection{On-site Integration and Installation} \label{sec:onsiteint}

All corrector subassemblies were delivered to the Mayall telescope by July 2018, for the beginning of their on-site installation.  The three barrel sections were first unpacked from their crates, and the rotary table and metrology framework installed on the ground floor of the Mayall telescope. The Front barrel section was then mounted on the rotary table and the lens alignment checked using the laser pencil beam system described earlier. This was repeated for the Aft section by itself, and then the Mid section by itself. The results were compared to the same measurements taken during assembly at UCL and found to be unchanged.

\begin{figure}
\centering 
\includegraphics[width=.8\textwidth]{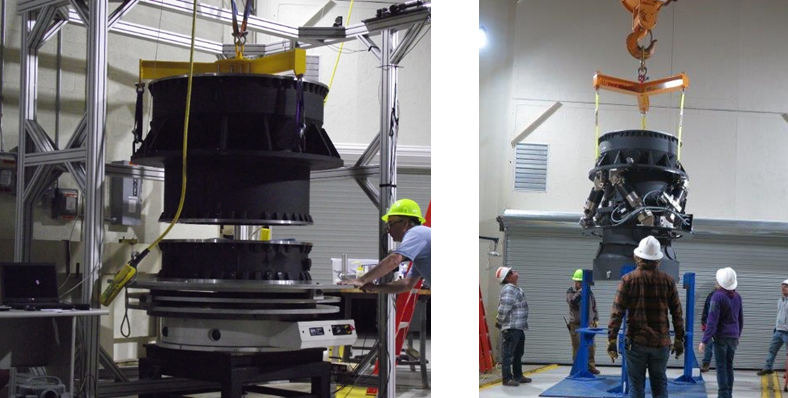}
\caption{Corrector assembly at Kitt Peak.  Left: The Aft and Middle barrel sections being reassembled on the UCL precision turntable on the Mayall ground floor for post-shipment alignment verification.  Right: the fully-assembled corrector barrel and hexapod being lifted by the Mayall dome crane from the integration stand on the ground floor to the cage/vanes/ring assembly on the C floor in the Mayall dome. \citep[Images previously published in][]{besuner20}}
\label{fig:integr1} 
\end{figure}

The full corrector barrel was then assembled using the procedure outlined in Section \ref{sec:alignment} above.  The relative alignment of each section was measured on the rotary table, using a MicroEpsilon OptoNCDT distance-measuring laser on external reference surfaces, to confirm that the entire assembly was unchanged compared to pre-shipping measurements taken at UCL (Figure \ref{fig:integr1}, left). At this point, the mechanical measurements assured that the assembled corrector was still within specifications after shipping, and the integration into the Mayall telescope could proceed.  

Meanwhile on the floor next to the Mayall telescope itself, the cage was installed into the upper ring by the spider vanes.  Then the cage was aligned to the ring by measuring references on the cage and ring with a laser tracker, and by adjusting the spider vane lengths.  This located the cage to within a few millimeters of nominal.

\begin{figure}
\centering 
\includegraphics[width=.8\textwidth]{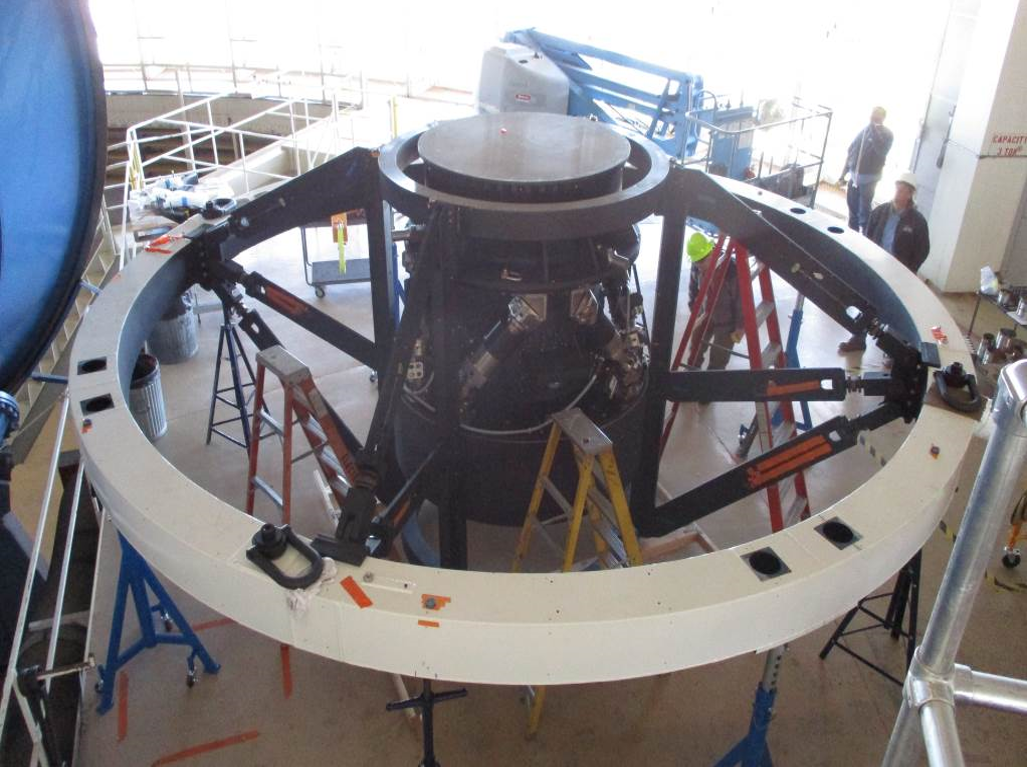}
\caption{The corrector and hexapod installed in the cage and upper ring located in the Mayall dome.  The corrector has been centered relative to the upper ring. \citep[Image previously published in][]{besuner20}}
\label{fig:integr2} 
\end{figure}

\begin{figure}[thb]
\begin{center}
  \par \vspace{5 mm}
    \includegraphics[height=.34\textwidth]{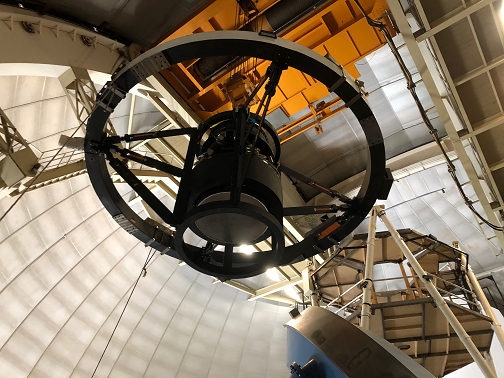}
    \includegraphics[height=.34\textwidth]{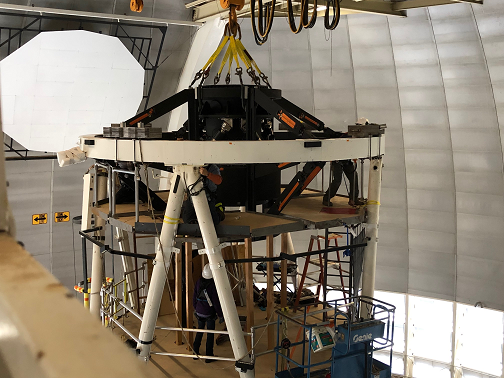}
  \par \vspace{2 mm}
  \caption{Integration of the top end.  Left: top end being listed by crane.  Right: top end installed on Mayall telescope trusses. \citep[Right image previously published in][]{besuner20}}
\label{fig:integ3}
\end{center}
\end{figure}

Next the hexapod was installed onto the corrector barrel assembly.  The Aft section was temporarily removed for the installation, since the hexapod assembly is smaller than the Front and Aft barrels, and then replaced using the pins.  The Mayall crane then lifted the corrector and hexapod up to the Mayall floor where it was installed into the cage and upper ring (Figure \ref{fig:integr1}, right, and Figure \ref{fig:integr2}).  Finally the entire top end assembly was installed onto the telescope trusses (Figure \ref{fig:integ3}).

After these steps, the laser tracker was again used to guide adjustment of the spider vanes to center the corrector, this time with respect to the primary mirror.  This was done for the case of the telescope pointing at zenith, and with the hexapod set to the middle of its adjustment range.  The laser tracker proved to be a convenient, reliable, and direct measurement method for alignment.

The laser tracker also proved useful as an independent check of assumed parameters.  We had measured the primary mirror shape with a laser tracker in 2014, and found differences from the published value in both the radius of curvature and in the location of the optical vertex with respect to the mechanical center.  We theorized that the differences to be due to measurement errors during mirror fabrication in the 1960s.  In the case of the radius, the updated value was 21.318\,meters, 18\,mm shorter than the published value of 21.336\,meters.  The new values were confirmed in 2018 during the installation of DESI at the Mayall telescope, when we adjusted the hexapod to determine the location of both best focus and best centration of the corrector with respect to the primary mirror.  The required shift of the hexapod to achieve best focus was -9\,mm away from what was expected from the published value, exactly corresponding to the -18\,mm update of the primary mirror radius.  The position of the cage was adjusted iteratively after installation on the telescope, with the final adjustments performed after initial on-sky testing, to ensure best focus and centration with the hexapod at the center of its ranges.  The final adjustments were verified with on-sky measurements using out-of-focus star images, and also by laser tracker.

Integration of all of the DESI subassemblies at the Mayall is fully described in \citet{besuner20}.

\section{Corrector Performance}

Initial on-sky testing of the DESI corrector began in April 2019 \citep{meisner20} with the DESI Commissioning Instrument \citep{ross18}. The Commissioning Instrument (CI) acted as a substitute focal surface for verifying the DESI front end performance in advance of the eventual DESI robotic focal plane assembly.  The CI included five commercial CCD detectors, each of which imaged a 24\,arcmin$^2$ field of view through a Sloan Digital Sky Survey (SDSS) $r$-band filter.  One of these CCDs was positioned at the center of the field and the remaining four approximately 400mm off axis in the four cardinal directions. The CI also included 22 illuminated fiducials that provided position references for calibrating the optical distortion patterns of the corrector.  The design, assembly, and verification of the CI was led by the Ohio State University. More details are available in \citet{ross18} and \citet{coles18}.

\begin{figure}
\centering 
\includegraphics[width=.8\textwidth]{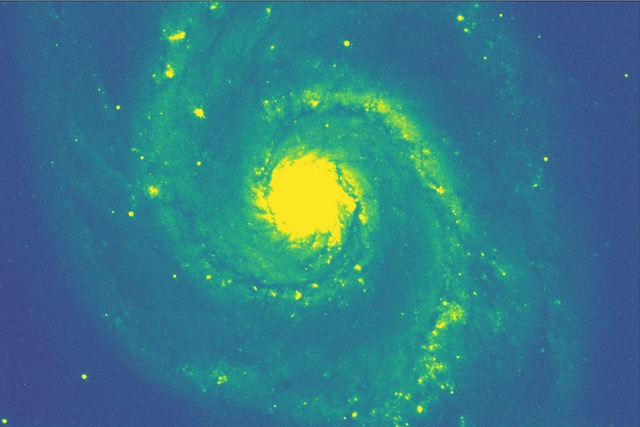}
\caption{Image of the galaxy M51 galaxy obtained on 2019 April 1, the first night of observations with the corrector. This image was obtained with the DESI Commissioning Instrument through an $r-$band filter with the CCD camera at the center of the field of view. The image quality in this observation was approximately $0.65''$ FWHM. The field of view of one of the CI CCDs is approximately $6.7' \times 4.4'$. North is up and East is to the left in the image.}
\label{fig:M51} 
\end{figure}

\begin{figure}
\centering 
\includegraphics[width=.8\textwidth]{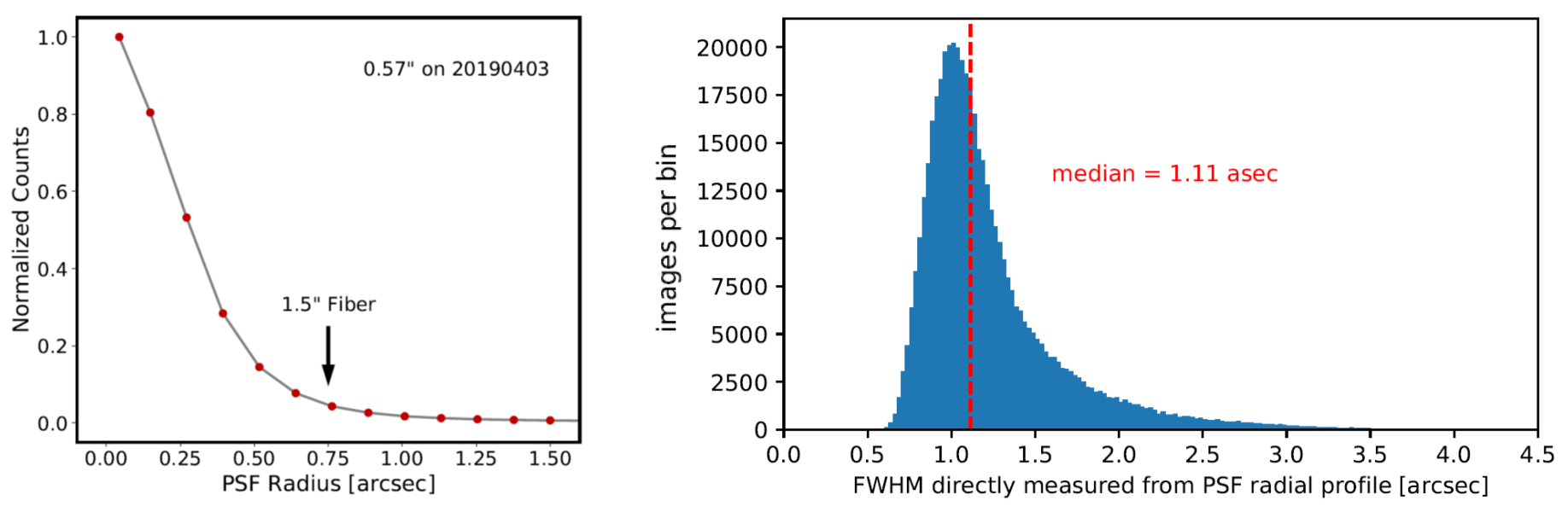}
\caption{Measurement of the delivered image quality of the corrector.  Left: radial profile of star on a night with superb seeing ($0.57''$ on 2019 April 3).  Right: histogram of FWHM measurements from approximately 480,000\,$r-$band images obtained during the DESI Commissioning phase, October 2019 to March 2020. The median FWHM is $1.11''$, consistent with the expected performance of the corrector, and with the pre-DESI performance of the MOSAIC prime focus camera. \citep[Right image previously published in][]{meisner20}}
\label{fig:measuredPSF} 
\end{figure}

\begin{figure}
\centering 
\includegraphics[width=.8\textwidth]{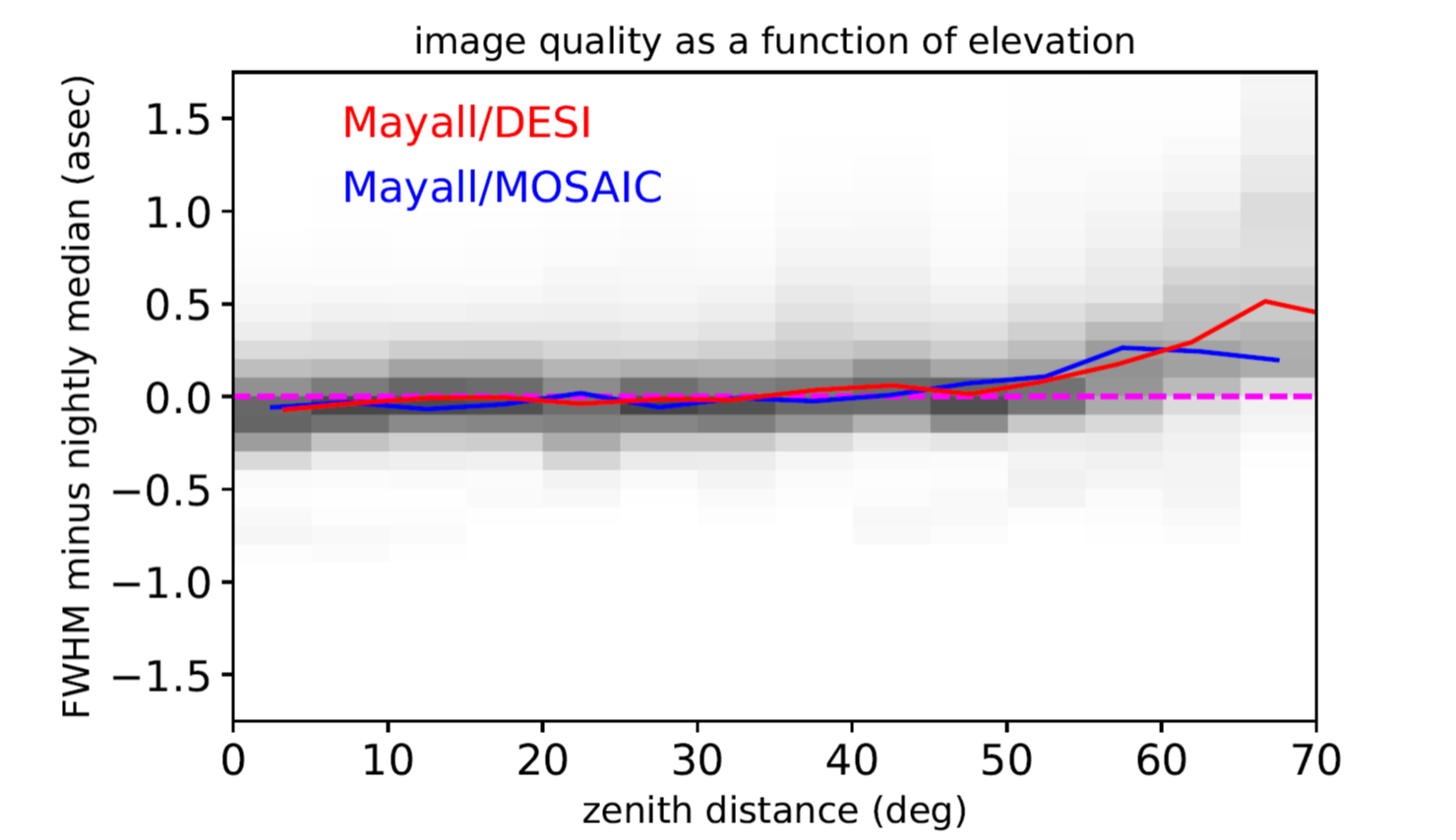}
\caption{Measured PSF FWHM image quality as a function of zenith angle. Approximately 480,000\,guider camera images contribute to this plot; the grayscale areas show the density of data points binned by zenith angle and difference between each observation's FWHM and the corresponding nightly median FWHM. The moving median trend of Mayall/DESI image quality versus zenith distance (red line) remains quite flat until around $50-60^\circ$, as expected from the optical design. The corresponding Mayall/MOSAIC trend (blue line) is very similar. The dashed magenta line denotes average image quality independent of zenith distance. \citep[Image previously published in][]{meisner20}}
\label{fig:IQvselevation} 
\end{figure}

The CI provided the first verification of the excellent imaging quality of the DESI corrector after integration.  Figure \ref{fig:M51} shows an image of the galaxy M51 recorded with the central CCD detector on 2019 April 1, the first night of observations. The CI recorded star images with a FWHM qqq of approximately 0.65\,arcseconds that night, a particularly good night for seeing. The left plot of Figure \ref{fig:measuredPSF} shows the PSF radial profile taken with CI observations on April 3, another night with excellent seeing.  The right plot in the Figure shows a histogram of all PSF width measurements derived from stellar sources during the entire Commissioning phase.  The median FWHM value is 1.11\,arcseconds; this is primarily due to seeing and is expected from the delivered image quality of the Mayall that was achieved using the MOSAIC prime focus camera before DESI \citep{dey14}.  However, this PSF width is also consistent with the image quality of the DESI corrector expected from modeling simulations and budgeted error terms.  These data show that the image quality remains constant for targets over the full design range of Zenith angles, thus demonstrating the success of the ADC system. Figure~\ref{fig:IQvselevation} shows the image quality as a function of Zenith angle after subtraction of the nightly median image quality, again calculated using all the images from the Commissioning phase. There is a modest degradation at large angles, above about $50-60^\circ$, which is expected from the optical design at the outermost fields where the guider CCDs are located (see Figure~\ref{fig:blur}).

\begin{figure}
\centering 
\includegraphics[width=.8\textwidth]{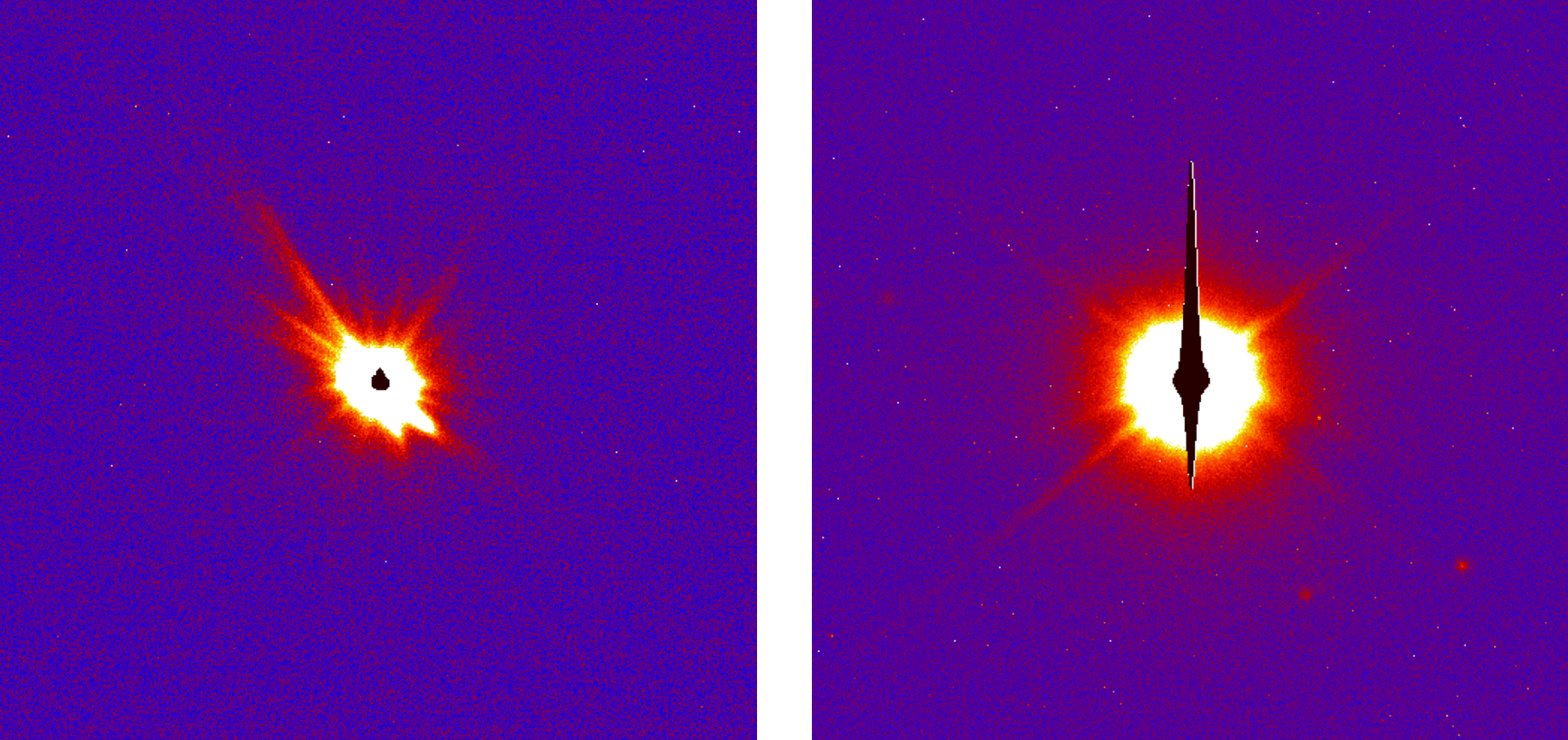}
\caption{Image of a star before (left) and after (right) mirror edge mask installation.  The mask blocks the outer edge of the primary mirror where there are polishing errors, and thus removes smear in the image.  The image was captured on a CI CCD and is overexposed to show the dimmer outer region of the PSF; the bright central core is blacked out in the image.}
\label{fig:beforeAfter} 
\end{figure}

The initial images with the CI were obtained without the primary mirror outer mask in place (described in \ref{sec:optdesign}).  Without the mask, images were smeared due to known polishing errors around the mirror edge. This smear was eliminated when we re-installed the primary mirror mask.  Figure \ref{fig:beforeAfter} shows an image of a bright star before and after the re-installation of the primary mirror mask. 

We validated the ADC capability using the Fiber View Camera (FVC), a diagnostic camera located in the primary mirror hole that observes the focal plane through the corrector lenses \citep{silber2022}.  To do this, we illuminated the CI fiducials and rotated each ADC lens through 360\,degrees, confirming that the FVC images shifted exactly as expected in both amplitude and direction.

We used the CI to check the corrector throughput with measurements of stars on nights with excellent conditions. This calculation included the standard atmospheric extinction model for Kitt Peak, the area of the primary mirror (with the mask in place), the expected vignetting due to the corrector, cage, and vanes, the coating and transparency of each corrector lens, the nominal throughput and bandpass of the $r$-band filter, and the quantum efficiency of the CI CCDs. The predicted and measured flux agreed to better than 10\%, which provided excellent validation of the corrector performance, as well as many of the model parameters used for the DESI survey design.

The CI allowed for the first validation of the AOS algorithm (see Section \ref{sec:hexapods}) using DESI hardware.  After a series of checks and calibrations, the AOS successfully actuated the corrector barrel so that it was well-aligned to the primary mirror, and the corrector achieved the excellent imaging performance described above.

The CI was removed in May 2019, and in its place the DESI Focal Plane Assembly (FPA) was integrated into the telescope and connected to the fiber system and spectographs.  The DESI commissioning phase began in October 2019, and other aspects of the corrector could be validated using the new hardware.

Scattering within the corrector system redistributes object flux and increases the background level in fields with, or near, bright sources.  Early simulations for the DESI survey assumed that the scattered component would increase the background level by no more than 1\% over what was expected in nominal conditions.  To check this assumption, we empirically estimated the scattered light contribution with a deep spectroscopic observation with the bright star Sirius positioned in the center of the DESI field of view where there are no fiber positioners.  In this configuration, scattered light within the corrector system has the characteristic spectral energy distribution of Sirius, and this light is seen in spectra obtained by the fibers distributed across the focal surface.  We fit the Sirius light to the spectral data and made a map of the magnitude of Sirius light that was scattered across the focal plane.  We then calculated that the integrated flux in this scattered component was on the order of 1\% of the total expected flux from the star, thus validating our earlier assumption.

We further validated the ADC capability, this time using the FPA and spectrographs.  To do this, we steered the telescope to a high Zenith angle, imaged a field of bright stars onto an array of fibers, rotated the ADC lenses to their expected orientations for that Zenith angle, and then moved the fibers in small steps around its nominal location (''dithered'').  By measuring the spectra in each slightly-different location we could identify where various wavelengths of a given star image were being focused at the focal plane.  We found that they were all coincident to a few microns thus validating the ADC operation.  If the ADC were not working then the focused rays would be separated by about 70\,microns over the bandpass.

Finally, we calibrated the corrector distortion model.  The corrector introduces distortion into the mapping of the sky to the focal plane, with both static and variable components, the latter depending on the ADC motions.  We developed a distortion model based on spin-weighted Zernike polynomials that included 12 orthogonal terms (of which only 10 are significant) and accounted for both radial and non-axisymmetric contributions \citep{kent18}.  The distortion model is critical to be able to align fibers to targets accurately. Validation and calibration of the distortion model was done by viewing the back-illuminated fiducials in the FPA using the FVC.
Additionally, the dithering operation described above and in \cite{DESIcollab2022_overview} allowed the fiber positions to be mapped accurately to the sky and correct for any residual distortion introduced by the FVC.

\section{Lessons Learned}

Like any project of this scale, the extensive design and build of DESI over many years included its share of challenges and missteps along with its many successes, and it is worth sharing what can be learned to benefit future similar projects.  Here is a list of some “lessons learned” from the DESI experience.

\begin{itemize}
    \item Consider optical design within the overall framework of the corrector performance, and perform trades early.  We performed extensive optical design and analysis of the corrector lenses early on, and were able to strike a good balance between design simplicity (number of lenses, aspheres) and reasonable fabrication and alignment tolerances.  We consulted with potential optics vendors during this early stage to make sure the lens specifications were within their fabrication and test capabilities.
    
    \item Iterate optical design with the mechanical design early on.  We were able to balance the difficulty of the requirements on the optics and mechanics, for example by prototyping the fiber positioners to convince ourselves that we could allow an optical design with a strong aspheric curvature on the focal surface.
    
    \item Start the procurement process early for long-lead items or items that are challenging to manufacture, and include ample contingency.  The large corrector lenses were identified as long-lead items that would require procurement of the glass blanks, polishing, and coating.  Furthermore the lenses had some associated schedule risk due to their size and challenging specs.  DESI was fortunately able to begin the procurement process for these lenses before project-level reviews would typically authorize, thanks to private funding.  Even then, some lenses took longer than expected and used up their schedule contingency.
    
    \item The vendors are absolutely critical to the success of a project, so thoroughly vet all relevant capabilities of potential vendors.  We chose vendors carefully for the lens blanks, the polishing, and the coating.  For each step a pool of vendors was narrowed down by a competitive bid process that used a best-value approach.  This approach gauged each vendor using weighted factors based on: quoted cost and schedule; their relevant experience (successes) at the scale being bid; their fabrication equipment in-house; their verification equipment in-house; their proposed work flow plans (especially relative to their other existing customers); their contingency plans/approaches in case of problems; the capability of any proposed subcontractors.  The vendor’s willingness to communicate openly, directly, and on a regular basis is also very important.
    
    \item Devote resources to logistics.  Shipping lenses, building fixtures, etc. may not seem like challenging problems worthy of engineering time, but making sure all the transportation details are covered is well worth the cost of avoiding costly delays late in the program.  DESI held reviews with all vendors, integration facilities, and the Mayall telescope, specifically to review their handling equipment and facilities, and uncovered many potential issues that could be fixed ahead of time.
    
    \item Devote resources up front to implement safety protections.  The cost is relatively small, but the payoff can be large.  For example, a heavy lifting strap fell onto the C1 lens during assembly of the barrel; fortunately we had built and installed a sturdy lens cover to protect C1, and so the lens was safe.
    
    \item Problems will always arise.  Be ready to engage closely with vendors when they do, and actively manage the problem with the vendor to prevent cost/schedule spiraling.
    
    \item Have relevant expertise available when problems arise, either on-project or at partner institutions.  We also found it valuable to invest in the capability to evaluate the performance impacts of non-conformances, e.g. with well-maintained models.
    
    \item Consider the value of appropriate testing.  We traded whether to perform a full assembly-level optical test on the corrector, and successfully saved cost and schedule by directing effort into component-level testing and modeling instead.
    
\end{itemize}

\section{Summary}

DESI is a highly complex instrument that was designed to achieve cutting-edge science, and thus it requires a high-performing prime focus corrector.  In this paper we provided a description of the requirements, the design, and implementation of the DESI corrector and the corrector support system.

The optical design meets all of its challenging requirements.  The corrector converts the F/2.8 beam from the primary mirror to F/3.9, corrects the converging beam over a wide 3.2-degree field of view and a wide 360-980\,nm bandpass, as well as adjusts for atmospheric dispersion over a range of Zenith angles from 0 to 60\,degrees.  The design provides high-throughput, high-quality imaging for the DESI instrument.

The heart of the corrector is the six lenses, the largest of which is 1.1\,meters in diameter and the heaviest of which is 237\,kg.  They were made from high-homogeneity fused silica and borosilicate material.  They were polished to exacting surface figures, with some taking up to 3 years to complete.  The lenses are coated with a broadband anti-reflection coating that allows transmission of $\geq 99.0$\% averaged over the defined bandpass, thus helping DESI to achieve its overall throughput requirement.  The counter-rotating pair of ADC lenses successfully compensates for atmospheric dispersion from Zenith angles up to 60\,degrees.  While other ADCs have been built before, DESI is the first corrector to use two separate single wedged lenses to perform the compensation.

The lenses are mounted in individual cells, the cells are mounted in barrel segments, and segments are mounted together into an overall barrel.  Lenses, cells, and barrel segments all have stringent requirements on their mutual alignment, as low as tens of microns, and all elements were fabricated and integrated using precise metrology to achieve those requirements.  All elements were designed with thermal stability in mind: they include RTV expansion pads and flexures to mitigate the thermal stresses between elements with materials of different CTEs.  The barrel was designed with the goal of keeping the corrector mass and diameter the same as in the case of the previous corrector installed on the Mayall telescope, despite DESI using larger diameter lenses.  We achieved this by implementing a single shell barrel design with no reinforcing ribs.  FEA showed that increasing the shell thickness was the most effective way to keep the barrel rigidity acceptably high.  The barrel consists of only three segments to hold all lenses and support structure, plus the front shroud and FPD, and it connects to the hexapod through a single flange.  The hexapod has up to $\pm10$\,mm of adjustment range and precisely readjusts the barrel position relative to the primary mirror to correct for dynamic changes due to changing Zenith angle and temperature changes in the observatory.

We fabricated a new cage, vanes, and upper ring to support the large mass of the corrector.  Their design was primarily driven by requirements on acceptable sag, tilt, and roll of the corrector.  The steel cage consists of three rings and four rails; it uses an open design to allow access to the hexapod and ADC rotators for maintenance and repairs.  The twelve vanes connect the cage to the upper ring and can be adjusted in length to align the corrector within the telescope.  The steel upper ring supports the entire top end and mates to the existing Serrurier trusses on the Mayall telescope.

We analyzed the sources of stray light, including ghosts, that could contaminate the measured spectra, and made design decisions to mitigate the top sources.  This included painting the primary mirror cell section black where it is visible to the focal plane, adding grooves to grazing-angle cell surfaces that could scatter towards the focal plane, painting all non-mating surfaces in the cell and barrel segments with black Aeroglaze paint, and adding various baffles before and after each lens.  These mitigations lowered the stray light potentially reaching the focal plane fibers to an acceptably small fraction of the science light.

The corrector sections were delivered to the Mayall telescope, and the barrel segments and lenses were measured to verify their alignment before their integration with the hexapod, cage, vanes, outer ring, and telescope.  Laser tracker tests further verified the correct alignment after integration.

During the commissioning phase, the corrector successfully demonstrated that it could achieve its expected excellent image quality and throughput, using a temporary Commissioning Instrument with five CCDs in place of the eventual focal plane assembly.  The FWHM imaging was measured to be 0.57\,arcseconds on the first night of imaging, demonstrating the corrector's superb optical performance.  The median FWHM imaging measurement over the 2-month commissioning campaign was 1.11\,arcseconds which is consistent with the expected seeing at Kitt Peak.  This included observations at Zenith angles up to 60\,degrees, thus validating the design and operation of the ADC system.  The measured throughput of the corrector was within 10\% of the expected value, providing another validation of the corrector build and integration.

DESI embarked on its 5-year main survey in May 2021, and has already collected a tremendous amount of science data, typically observing at least 80,000\,galaxies and quasars on clear nights.  The corrector is a critical element that supports DESI in its science mission of unprecedented spectroscopic measurements of the cosmos.


\acknowledgments


The authors wish to thank Marty Valente and Jim Burge (Arizona Optical Systems), David Anderson and Vilma Anderson (Rayleigh Optical Corporation), and George Gardopee and Andrew Clarkson (L3 Brashear) for their helpful review of the lens polishing sections.  We also wish to thank Dr. Charles Kennemore III (Viavi Solutions) for his extensive documentation of the coating work that supported this paper.  We also thank Patrick Jelinsky for his advice and guidance with the final drafts of this paper.

The DESI collaboration appreciates Trish Dobson and American Cargoservice Inc. for managing the extensive logistics required to transport the corrector barrel sections safely from the UK to Kitt Peak by air and truck.

The authors acknowledge the support of the Gordon and Betty Moore Foundation, grant number 3581, and the Heising-Simons Foundation, grant number 2014-91, for their help in procuring long-lead components that enabled the corrector to meet its schedule.


Authors P.D and D.B acknowledge the support of the UK Science and Technology Facilities Council (STFC), grant number ST/M00287X/1.
The Fermiab authors acknowledge support from Fermi Research Alliance, LLC (FRA), acting under Contract No. DE-AC02-07CH11359 with the U.S. Department of Energy.

This material is based upon work supported by the U.S. Department of Energy (DOE), Office of Science, Office of High-Energy Physics, under Contract No. DE–AC02–05CH11231, and by the National Energy Research Scientific Computing Center, a DOE Office of Science User Facility under the same contract. Additional support for DESI was provided by the U.S. National Science Foundation (NSF), Division of Astronomical Sciences under Contract No. AST-0950945 to the NSF’s National Optical-Infrared Astronomy Research Laboratory; the Science and Technologies Facilities Council of the United Kingdom; the Gordon and Betty Moore Foundation; the Heising-Simons Foundation; the French Alternative Energies and Atomic Energy Commission (CEA); the National Council of Science and Technology of Mexico (CONACYT); the Ministry of Science and Innovation of Spain (MICINN), and by the DESI Member Institutions: \url{https://www.desi.lbl.gov/collaborating-institutions}. Any opinions, findings, and conclusions or recommendations expressed in this material are those of the author(s) and do not necessarily reflect the views of the U. S. National Science Foundation, the U. S. Department of Energy, or any of the listed funding agencies.

The authors are honored to be permitted to conduct scientific research on Iolkam Du’ag (Kitt Peak), a mountain with particular significance to the Tohono O’odham Nation.

For more information, visit \url{https://desi.lbl.gov}

Source data files for all plots are archived at \url{https://doi.org/10.5281/zenodo.7844288}

\vspace{5mm}
\facility{Mayall (DESI)}


\bibliography{references}{}
\bibliographystyle{aasjournal}

\end{document}